\RequirePackage{fix-cm} 
\documentclass[a4paper, twoside, reqno, 12pt]{amsart}
\usepackage{fixltx2e}   

\usepackage{etex}

\usepackage[latin1]{inputenc}
\usepackage[T1]{fontenc}

\usepackage{esint}
\usepackage{dsfont}
\usepackage{xspace}
\usepackage{amsgen}
\usepackage{amsthm}
\usepackage{amssymb}
\usepackage{amsmath}
\usepackage{wasysym}
\usepackage{upgreek}
\usepackage{amsfonts}
\usepackage{stmaryrd}
\usepackage{nicefrac}
\usepackage{mathtools}

\usepackage[mathcal, mathscr]{euscript}

\usepackage{mathrsfs}
\DeclareMathAlphabet{\mathscrbf}{OMS}{mdugm}{b}{n}

\usepackage[lined, boxed, linesnumbered, english]{algorithm2e}

\usepackage{multicol}
\usepackage{multirow}

\usepackage{a4wide}

\usepackage[dvipsnames, table]{xcolor}
\definecolor{bckg}{RGB}{20.8, 20.8, 20.8}
\definecolor{oneblue}{rgb}{0.0, 0.0, 0.85}
\definecolor{Lightblue}{RGB}{214, 214, 214}
\definecolor{bluepigment}{rgb}{0.2, 0.2, 0.6}
\definecolor{charcoal}{rgb}{0.21, 0.27, 0.31}
\definecolor{denimblue}{rgb}{0.08, 0.38, 0.74}
\definecolor{Lightgray}{rgb}{0.89, 0.89, 0.89}
\definecolor{darkgrey}{rgb}{0.273, 0.281, 0.30}
\definecolor{darkelectricblue}{rgb}{0.33, 0.41, 0.47}

\usepackage[sort&compress, comma, square, numbers]{natbib}

\usepackage{psfrag}
\usepackage{graphicx}
\usepackage{subfigure}
\usepackage{morefloats}
\usepackage{indentfirst}

\usepackage{acronym}
\usepackage{microtype}
\usepackage[labelsep=period,%
            labelfont={bf,sf,color=bluepigment},%
            justification=raggedright]{caption}

\usepackage[perpage, symbol]{footmisc}

\usepackage[usenames, pdf]{pstricks}
\usepackage{epsfig}
\usepackage{pst-grad} 
\usepackage{pst-plot} 

\usepackage{url}
\usepackage[colorlinks,
           urlcolor=oneblue,
           linkcolor=denimblue,
           citecolor=NavyBlue,
           bookmarksopen=false,
           pdfpagemode=UseNone,
           pagebackref]{hyperref}

\usepackage[explicit]{titlesec}

\titleformat{\section}[block]
  {\color{NavyBlue}\Large\sffamily\bfseries}
  {}
  {0.0em}
  {\colorbox{bckg!5}{\strut\parbox{\dimexpr\linewidth-2\fboxsep\relax}{\thesection. #1}}}
  [\vspace*{0.33em}]

\titleformat{name=\section,numberless}[block]
  {\color{NavyBlue}\Large\sffamily\bfseries}
  {}
  {0.0em}
  {\colorbox{bckg!5}{\strut\parbox{\dimexpr\linewidth-2\fboxsep\relax}{#1}}}
  [\vspace*{0.33em}]

\titleformat{\subsection}
  {\color{NavyBlue}\large\sffamily\bfseries}
  {}
  {0.0em}
  {\colorbox{bckg!5}{\parbox{\dimexpr\linewidth-2\fboxsep\relax}{\thesubsection. #1}}}
  [\vspace*{0.33em}]

\titleformat{name=\subsection,numberless}
  {\color{NavyBlue}\Large\sffamily\bfseries}
  {}
  {0em}
  {\colorbox{bckg!5}{\parbox{\dimexpr\linewidth-2\fboxsep\relax}{#1}}}
  [\vspace*{0.33em}]

\titleformat{\subsubsection}
  {\color{bluepigment}\sffamily\normalsize\bfseries}
  {\thesubsubsection}
  {0.5em}
  {#1}
  [\vspace*{0.33em}]

\titleformat{\paragraph}[runin]
  {\color{bluepigment}\sffamily\small\bfseries}
  {}
  {0em}
  {#1}

\titlespacing{\section}{0.0em}{1.5em plus 2pt minus 2pt}%
{1.0em plus 2pt minus 2pt}[0em]
\titlespacing{\subsection}{0.5em}{1.5em plus 2pt minus 2pt}%
{1.0em}[0em]
\titlespacing{\subsubsection}{0.5em}{1.5em plus 2pt minus 2pt}%
{1.0em plus 2pt minus 2pt}[0em]

\usepackage{titletoc}

\contentsmargin{0.5em}
\newlength{\tocsep} 
\titlecontents{section}[\tocsep]
  {\addvspace{10pt}\bfseries\sffamily}
  {\contentslabel[\thecontentslabel]{\tocsep}}
  {}
  {\ \titlerule*[0.75pc]{.}\ \thecontentspage}
  []
\titlecontents{subsection}[\tocsep]
  {\addvspace{8pt}\sffamily}
  {\contentslabel[\thecontentslabel]{\tocsep}}
  {}
  {\ \titlerule*[0.5pc]{.}\ \thecontentspage}
  []
\titlecontents*{subsubsection}[\tocsep]
  {\addvspace{2pt}\footnotesize\sffamily}
  {}
  {}
  {\ \titlerule*[0.35pc]{.}\ \thecontentspage}
  [\\*]

\makeatletter
\def\@setauthors{%
  \begingroup
  \def\thanks{\protect\thanks@warning}%
  \trivlist
  \centering\footnotesize \@topsep30\p@\relax
  \advance\@topsep by -\baselineskip
  \item\relax
  \author@andify\authors
  \def\\{\protect\linebreak}%
  \textsc{\normalsize\textcolor{darkelectricblue}{\authors}}%
  \ifx\@empty\contribs
  \else
    ,\penalty-3 \space \@setcontribs
    \@closetoccontribs
  \fi
  \endtrivlist
  \endgroup
}
\def\@settitle{\begin{center}%
  \baselineskip14\p@\relax
    \bfseries
    \textsc{\Large\textcolor{charcoal}{\@title}}
  \end{center}%
}
\makeatother

\usepackage{enumitem}
\setlist[description]{%
  topsep=30pt,               
  itemsep=5pt,               
  font={\bfseries\sffamily\color{NavyBlue}}, 
}

\usepackage{fancyhdr}
\usepackage{lastpage}

\newcommand*\Title{\textcolor{bluepigment}{Spectral Methods --- Part 1}}
\newcommand*\Authors{\textcolor{bluepigment}{S.~Gasparin, J.~Berger, D.~Dutykh \& N.~Mendes}}
\newcommand*{\plogo}{\textcolor{gray}{{\texttt{arXiv.org} / \textsc{hal}}}} 

\numberwithin{equation}{section}

\newcommand{\ie}{\emph{i.e.}\xspace}
\newcommand{\eg}{\emph{e.g.}\xspace}

\newcommand*\egal{\ = \ }
\newcommand*\plus{\ + \ }
\newcommand*\moins{\ - \ }
\newcommand*\egalb{\, = \, }

\renewcommand{\O}{\mathcal{O}}
\newcommand*{\Ox}{\Omega_{\, x}}

\renewcommand{\b}{\mathrm{b \,}}

\newcommand{\Mat}{\mathrm{Mat}\,}

\newcommand{\Pvref}{P_{\,v}^{\,0}}
\newcommand{\BivL}{\mathrm{Bi}_{\,v,\,\mathrm{L}}}
\newcommand{\BivR}{\mathrm{Bi}_{\,v,\,\mathrm{R}}}
\newcommand{\cm}{c_{\,m}}
\newcommand{\cms}{c_{\,m}^{\,\star}}
\newcommand{\dm}{d_{\,m}}
\newcommand{\dmref}{d_{\,m}^{\,0}}
\newcommand{\dms}{d_{\,m}^{\,\star}}
\newcommand{\gl}{g_{\,l}}
\newcommand{\glL}{g_{\,l,\,\mathrm{L}}}
\newcommand{\glR}{g_{\,l,\,\mathrm{R}}}
\newcommand{\glsL}{g_{\,l,\,\mathrm{L}}^{\,\star}}
\newcommand{\glsR}{g_{\,l,\, \mathrm{R}}^{\,\star}}
\newcommand{\hv}{h_{\,v}}
\newcommand{\hvL}{h_{\,v,\,\mathrm{L}}}
\newcommand{\hvR}{h_{\,v,\,\mathrm{R}}}
\newcommand{\kl}{k_{\,l}}
\newcommand{\kv}{k_{\,v}}
\newcommand{\Pc}{P_{\,c}}
\newcommand{\Ps}{P_{\,s}}
\newcommand{\Pv}{P_{\,v}}
\newcommand{\Pvi}{P_{\,v}^{\,i}}
\newcommand{\PvL}{P_{\,v,\,\mathrm{L}}}
\newcommand{\PvR}{P_{\,v, \,\mathrm{R}}}
\newcommand{\Rv}{R_{\,v}}
\newcommand{\tref}{t^{\,0}}
\newcommand{\uL}{u_{\,\mathrm{L}}}
\newcommand{\uR}{u_{\,\mathrm{R}}}
\newcommand{\rholv}{\rho_{\,l+v}}
\newcommand{\xs}{x^{\,\star}}
\newcommand{\ts}{t^{\,\star}}
\newcommand{\M}{\mathcal{M}}
\newcommand{\A}{\mathcal{A}}
\newcommand{\R}{\mathds{R}}

\newcommand{\T}{\mathsf{T}}
\newcommand{\Bi}{\mathrm{Bi}}
\newcommand{\gls}{g_{\,l}^{\,\star}}

\newcommand*\pd[2]{\frac{\partial #1}{\partial #2}}

\newcommand{\eqdef}{\mathop{\stackrel{\ \mathrm{def}}{:=}\ }}

\newcommand{\dix}[1]{ \times 10^{\,#1}}

\newcommand{\Eu}{\textsc{Euler}}
\newcommand{\CN}{\textsc{Crank}--\textsc{Nicolson}}

\begin{document}

\title[\Title]{Solving nonlinear diffusive problems in buildings by means of a Spectral Reduced--Order Model}

\author[S.~Gasparin]{Suelen Gasparin$^*$}
\address{\textbf{S.~Gasparin:} LAMA, UMR 5127 CNRS, Universit\'e Savoie Mont Blanc, Campus Scientifique, F-73376 Le Bourget-du-Lac Cedex, France and Thermal Systems Laboratory, Mechanical Engineering Graduate Program, Pontifical Catholic University of Paran\'a, Rua Imaculada Concei\c{c}\~{a}o, 1155, CEP: 80215-901, Curitiba -- Paran\'a, Brazil}
\email{suelengasparin@hotmail.com}
\urladdr{https://www.researchgate.net/profile/Suelen\_Gasparin/}
\thanks{$^*$ Corresponding author}

\author[J.~Berger]{Julien Berger}
\address{\textbf{J.~Berger:} LOCIE, UMR 5271 CNRS, Universit\'e Savoie Mont Blanc, Campus Scientifique, F-73376 Le Bourget-du-Lac Cedex, France}
\email{Berger.Julien@univ-smb.fr}
\urladdr{https://www.researchgate.net/profile/Julien\_Berger3/}

\author[D.~Dutykh]{Denys Dutykh}
\address{\textbf{D.~Dutykh:} Univ. Grenoble Alpes, Univ. Savoie Mont Blanc, CNRS, LAMA, 73000 Chamb\'ery, France and LAMA, UMR 5127 CNRS, Universit\'e Savoie Mont Blanc, Campus Scientifique, F-73376 Le Bourget-du-Lac Cedex, France}
\email{Denys.Dutykh@univ-smb.fr}
\urladdr{http://www.denys-dutykh.com/}

\author[N.~Mendes]{Nathan Mendes}
\address{\textbf{N.~Mendes:} Thermal Systems Laboratory, Mechanical Engineering Graduate Program, Pontifical Catholic University of Paran\'a, Rua Imaculada Concei\c{c}\~{a}o, 1155, CEP: 80215-901, Curitiba -- Paran\'a, Brazil}
\email{Nathan.Mendes@pucpr.edu.br}
\urladdr{https://www.researchgate.net/profile/Nathan\_Mendes/}


\begin{titlepage}
\thispagestyle{empty} 
\noindent
{\Large Suelen \textsc{Gasparin}}\\
{\it\textcolor{gray}{Pontifical Catholic University of Paran\'a, Brazil}}\\
{\it\textcolor{gray}{LAMA--CNRS, Universit\'e Savoie Mont Blanc, France}}
\\[0.02\textheight]
{\Large Julien \textsc{Berger}}\\
{\it\textcolor{gray}{LOCIE--CNRS, Universit\'e Savoie Mont Blanc, France}}
\\[0.02\textheight]
{\Large Denys \textsc{Dutykh}}\\
{\it\textcolor{gray}{LAMA--CNRS, Universit\'e Savoie Mont Blanc, France}}
\\[0.02\textheight]
{\Large Nathan \textsc{Mendes}}\\
{\it\textcolor{gray}{Pontifical Catholic University of Paran\'a, Brazil}}
\\[0.10\textheight]

\colorbox{Lightblue}{
  \parbox[t]{1.0\textwidth}{
    \centering\huge\sc
    \vspace*{0.7cm}
    
    \textcolor{bluepigment}{Solving nonlinear diffusive problems in buildings by means of a Spectral Reduced--Order Model}

    \vspace*{0.7cm}
  }
}

\vfill 

\raggedleft     
{\large \plogo} 
\end{titlepage}


\newpage
\thispagestyle{empty} 
\par\vspace*{\fill}   
\begin{flushright} 
{\textcolor{denimblue}{\textsc{Last modified:}} \today}
\end{flushright}


\newpage
\maketitle
\thispagestyle{empty}


\begin{abstract}

This paper proposes the use of a Spectral method to simulate diffusive moisture transfer through porous materials as a Reduced--Order Model (ROM). The Spectral approach is an \textit{a priori} method assuming a separated representation of the solution. The method is compared with both classical \textsc{Euler} implicit and \textsc{Crank--Nicolson} schemes, considered as large original models. Their performance --- in terms of accuracy, complexity reduction and CPU time reduction --- are discussed for linear and nonlinear cases of moisture diffusive transfer through single and multi-layered one-dimensional domains, considering highly moisture-dependent properties. Results show that the Spectral reduced-order model approach enables to simulate accurately the field of interest. Furthermore, numerical gains become particularly interesting for nonlinear cases since the proposed method can drastically reduce the computer run time, by a factor of $100\,$, when compared to the traditional \textsc{Crank--Nicolson} scheme for one-dimensional applications.


\bigskip
\noindent \textbf{\keywordsname:} Spectral methods; \textsc{Chebyshev} polynomials; \textsc{Tau-Galerkin} method; numerical simulation;  diffusive phenomena; reduced-order modelling \\


\end{abstract}


\newpage
\tableofcontents
\thispagestyle{empty}


\newpage
\section{Introduction}

Moisture transfer through porous materials is a matter of concern in many areas, such as building physics, food engineering, hydrology, agriculture, geophysics, environmental engineering, energy systems, among others, where the transient evolution of moisture may play a role of paramount importance. Particularly, in the area of building physics, moisture transfer process through the porous envelope, roofing systems and ground may strongly affect energy and hygrothermal performance of those elements and, at the same time, it can influence the building occupants' health, the material's durability and the energy consumption and demand of the edifice.

The mechanisms that control the transport of moisture in porous building materials occurs simultaneously in its different phases. In the vapour phase, the moisture transfer is mostly governed by diffusive and convective transport while in the liquid phase it is governed mainly by capillarity, which is strongly influenced by weather conditions \cite{DeFreitas1996}.

Over the last decades, several models were proposed in the literature to mathematically describe the moisture transport as described in \cite{Mendes2017} and for the assessment of moisture effects, numerical tools have been developed to accurately simulate the processes of moisture transfer in building materials \cite{Barreira2010}. Since the $1990$'s, many computer-based tools for the prediction of the hygrothermal performance were developed, such as \texttt{DELPHIN} \cite{BauklimatikDresden2011}, \texttt{MATCH} \cite{Rode2003}, \texttt{MOIST} \cite{Burch1993}, \texttt{WUFI} \cite{IBP2005} and \texttt{UMIDUS} \cite{PUCPR2016, Mendes1999}. Moisture models have also been implemented in whole-building simulation tools and tested in the frame of the International Energy Agency Annex~$41$, which reported on most of the detailed models and their successful applications for accurate assessment of hygrothermal transfer in buildings \cite{Woloszyn2008}.

As building material properties are temperature- and moisture-dependent and the boundary conditions are driven by weather variables, the models included in those tools are based on numerical approaches using discrete representations of the continuous equations. To compute the solution, standard discretisation and incremental techniques are applied such as the \textsc{Euler} implicit scheme in \cite{Mendes2005, IBP2005, BauklimatikDresden2011, Steeman2009, Rouchier2013, Janssen2014, Janssen2007} to solve large systems of equations (of an order of $10^{\,6}$ for three-dimensional problems). Furthermore, when dealing with nonlinearities, hygrothermal properties of porous materials have to be updated as a function of the temperature and moisture content fields at each iteration. The difficulties to compute the solution increase, particularly when using implicit schemes that require sub-iterations to treat those issues. In the literature, the important numerical costs of simulation tools \cite{DosSantos2006, Dalgliesh2005, Mortensen2007, Abuku2009} is also mentioned and it is a matter of concern due to the great scale of buildings, where heat and moisture transfer phenomena have to be simulated. For those reasons, innovative and efficient ways of numerical simulation are worthy of further investigation and model reduction techniques can be an interesting alternative approach to deal with this problem.

The intent in constructing reduced-order models (ROMs) is to provide accurate description of the physical phenomena by decreasing the degrees of freedom, while retaining the model's fidelity, at a computational cost much lower than the large original model \cite{Reddy2017}. In recent years, reduced-order modeling techniques have proven to be powerful tools for solving various problems. Important efforts have been dedicated to developing reduced-order models that can provide accurate predictions while dramatically reducing computational time, for a wide range of applications, covering different fields such as fluid mechanics, heat transfer, structural dynamics among others \cite{Lucia2004, Herzet2018, Bai2002}. Examples with finite-element and finite-volume applications can be found in \cite{Ullmann2016} and \cite{Stabile2018}, respectively. In their work, they apply reduced-order models to build accurate solutions with less computational effort than the large original model. A careful attention must be paid regarding the definition of ROMs since sometimes it is related to degradation of the physical model \cite{Schilders2008}, which is not the case of the present work.

Reduced-order models can be classified as \emph{a priori} or \emph{a posteriori} methods. The \emph{a posteriori} approaches need a preliminary computed (or even experimental) solution data of the large original problem to build the reduced one. Whereas the \emph{a priori} ones do not need preliminary information on the studied problem. The reduced-order model is unknown \emph{a priori} and is directly built. Since the $2000$th, aiming to reduce the computational cost, reduction model techniques started to take place in the context of heat and moisture transfer for building physics applications, as an alternative to traditional methods. Different kinds of approaches can be considered, such as the \emph{a posteriori} Proper Orthogonal Decomposition (POD), the Modal Basis Reduction (MBR) and the \emph{a priori} Proper Generalized Decomposition (PGD), which has shown a relevant reduction of the computational cost for successful applications in the building physics area \cite{Berger2016c}.

Spectral methods are successfully applied in studies of wave propagation,  meteorology, computational fluid dynamics, quantum mechanics and several other fields \cite{Canuto2006}. Some works on the transport phenomena can be found in literature involving diffusive \cite{Guo2012, Wang2016}, convective \cite{Chen2016, RamReddy2015} and radiative \cite{Li2008a, Chen2015b, Ma2014} heat transfer. Spectral techniques applied in these works are varied, adopted according to the geometry, boundary conditions and field of application. In recent works, researchers have implemented spectral methods for solving heat and moisture transfer in food engineering \cite{Pasban2017} and on fluid flow \cite{Motsa2015}. According to the authors' knowledge, there is no research in the literature so far regarding the application of spectral methods for solving diffusive moisture transfer in building physics.

Therefore, the scope of this work is to present an innovative approach, applied for the first time in the context of building physics, \ie, the \emph{a priori} Spectral reduced-order model technique. In this work, the method is used to compute one-dimensional moisture diffusion in porous materials. The objective is to significantly reduce the computational cost while maintaining high fidelity solutions. This technique assumes separated tensorial representation of the solution by a finite sum of function products. It fixes a set of spatial basis functions to be the \textsc{Chebyshev} polynomials and then, a system of ordinary differential equations is built to compute the temporal coefficients of the solution using the \textsc{Tau--Galerkin} method. In this work, aiming at proposing the use of a Spectral method to simulate a complex and time-consuming phenomenon of diffusive moisture transfer, the temperature effect has been disregarded, but it must be the next step of the investigation on a highly efficient numerical method to simulate combined heat and moisture transport through porous building elements.

The efficiency of the Spectral approach will be analyzed/proven/demonstrated for simple and multilayered domains with highly nonlinear properties with sharp boundary conditions and profiles of solutions. For this purpose, the manuscript is organized as follows. First, the description of the physical phenomena is presented (Section~\ref{sec:phys_equat}). Then, the Spectral technique is described (Section~\ref{sec:spectral_linear}). In the sequence, the proposed method is applied to four different cases in one dimension. The first one considers linear transfer (Section~\ref{sec:case_linear}) to validate the method. The second one focuses on a weak nonlinear transfer (Section~\ref{sec:case_weak_nonlinear}), in which some simplifications are considered, while the third one presents a strongly nonlinear transfer case with moisture-dependent material properties (Section~\ref{sec:spectral_nonlinear}). Finally, the last case study (Section~\ref{sec:case_multilayer}) considers a multilayered wall with important interface conditions imposed.

It is important to mention that the case studies of this article are directly related to two recent papers, focusing on the establishment of efficient numerical models for nonlinear moisture transfer in terms of accuracy and reduced computational effort. The first work \cite{Janssen2014} discusses the choice of the physical potential for the formulation of the physical problem, while the second one \cite{Gasparin2017} provides a discussion of the numerical methods enabling to build a large original model to solve a nonlinear moisture diffusion problem. 


\section{Moisture transfer in porous materials}
\label{sec:phys_equat}

The physical problem involves one-dimensional moisture diffusion through a porous material defined by the spatial domain $\Ox \,=\, [\, 0, \, L \,]\,$. The moisture transfer occurs according to liquid and vapour diffusion processes. The physical problem can be formulated as \cite{Luikov1966, Abahri2016}: 
\begin{align}\label{eq:moisture_equation_1D}
  & \pd{\rholv}{t} \egal \pd{}{x} \left( \, \kl \, \pd{\Pc}{x} \plus \kv \, \pd{\Pv}{x} \, \right) \,,
\end{align}
where $\rholv$ is the volumetric moisture content of the material and $\kv$ and $\kl\,$, the vapour and liquid permeabilities.

Eq.~\eqref{eq:moisture_equation_1D} can be written using the vapour pressure $\Pv$ as the driving potential. For this, we consider the physical relation, known as the \textsc{Kelvin} equation, between $\Pv$ and $\Pc\,$:
\begin{align*}
  \Pc & \egal \Rv \, T \, \rho_{\,l} \, \ln\left(\frac{\Pv}{\Ps\,(\,T\,)}\right)\,,
\end{align*}
where $\phi \egalb \nicefrac{\Pv}{\Ps\,(\,T\,)}$ is the relative humidity. Thus, the derivative of $\Pc$ regarding $\Pv$ is expressed as:
\begin{align*}
  \pd{\Pc}{\Pv} & \egal \frac{R_{\,v} \, T\, \rho_{\,l}}{\Pv} \,,
\end{align*}
and the derivative regarding $x$ can be written as:
\begin{align}\label{eq:diff_Pc_x}
  \pd{\Pc}{x} \egal \underbrace{\pd{\Pc}{\Pv}}_{= \frac{R_{v} T \rho_{l}}{\Pv}} \cdot\ \pd{\Pv}{x} \plus \pd{\Pc}{T} \cdot \pd{T}{x} \,.
\end{align}
In addition, the left-hand term of Eq.~\eqref{eq:moisture_equation_1D} can also be expressed in terms of $\Pv$ and $T\,$:
\begin{align}\label{eq:diff_rho_t}
  \pd{\rholv}{t} \egal \pd{\rholv}{\phi} \cdot \pd{\phi}{\Pv} \cdot \pd{\Pv}{t} \plus \pd{\rholv}{T} \cdot \pd{T}{t} \,.
\end{align}
As the problem has been assumed isothermal, the temperature derivatives vanishes so that Eqs.~\eqref{eq:diff_Pc_x} and \eqref{eq:diff_rho_t} can be written as: 
\begin{align*}
  \pd{\Pc}{x} &\egal  \frac{\Rv \, T\, \rho_{\,l}}{\Pv} \cdot \pd{\Pv}{x} \,, \\
  \pd{\rholv}{t} &\egal \ \pd{\rholv}{\phi} \cdot \pd{\phi}{\Pv} \cdot \pd{\Pv}{t} \,.
\end{align*}

Considering the relation $\rholv \egal f\, (\,\phi\,) \egal f\, (\, \Pv \,,T \,)$, obtained from material properties and from the relation $\phi \egalb \nicefrac{\Pv}{\Ps(T)}$ between the vapour pressure $\Pv$ and the relative humidity $\phi$, we get: 
\begin{align*}
  \pd{\rholv}{t} \egal f^{\,\prime}\,(\,\phi\,) \, \frac{1}{\Ps} \cdot \pd{\Pv}{t} \,.
\end{align*}

Eq.~\eqref{eq:moisture_equation_1D} can be therefore rewritten as:
\begin{align}\label{eq:moisture_equation_1D_v2}
  & f^{\,\prime}\,(\,\phi\,) \, \frac{1}{\Ps} \cdot \pd{\Pv}{t} \egal \pd{}{x} \biggl[ \, \biggl( \, \kl \, \frac{\Rv \, T\, \rho_{\,l}}{\Pv} \plus \kv \, \biggr) \cdot \pd{\Pv}{x} \, \biggr] \,.
\end{align}
The material properties $f\,$, $\kl$ and $\kv$ depend on the vapour pressure $\Pv$. Therefore, we denote $\dm \eqdef \kl \, \dfrac{\Rv \, T\, \rho_{\,l}}{\Pv} \plus \kv $ as the global moisture transport coefficient and $\cm \eqdef f^{\,\prime}\,(\,\phi\,) \; \dfrac{1}{\Ps}$ as the moisture storage coefficient. Thus, considering the previous notation, Eq.~\eqref{eq:moisture_equation_1D_v2} can be written as:
\begin{align*}
  \cm\, \pd{\Pv}{t} \egal \pd{}{x} \biggl[ \, \dm \, \pd{\Pv}{x} \, \biggr] \,.
\end{align*}

At the material bounding surfaces, \textsc{Robin}-type boundary conditions are considered:
\begin{align*}
  \dm \, \pd{\Pv}{x} &\egal \hvL \cdot \left( \, \Pv \moins \PvL \, \right) \moins \glL \, , && x \egal 0 \,, \\
  -\ \dm \, \pd{\Pv}{x} &\egal \hvR \cdot \left( \, \Pv \moins \PvR \, \right) \moins \glR \, ,&& x \egal L \,,
\end{align*}
where $\PvL$ and $\PvR$ are the vapour pressure of the ambient air, $\glL$ and $\glR\,$, the liquid water flow (driving rain) and $R$ and $L$ stand for the right and left bounding surfaces. For the initial condition, the vapour pressure distribution is written in function of the space:
\begin{align*}
  \Pv\, (\,x\,,\,t=0\,) \egal \Pvi\,(x) \,.
\end{align*}
The initial condition can either have a uniform distribution or a profile more appropriated to the boundary conditions to reduce a warm-up simulation period, which can be very significant (at the order of years) depending up on the material hygrothermal properties and on the thickness of the building component.

It is important to obtain a unitless formulation of governing equations while performing mathematical and numerical analysis of given practical problems, due to a certain number of reasons already discussed in \cite{Gasparin2017}. Therefore, we define the following dimensionless parameters:
\begin{align*}
  & u \egal \frac{\Pv}{\Pvref} \,,
  && \xs \egal \frac{x}{L} \,, 
  && \ts \egal \frac{t}{\tref} \,,
  && \cms \egal \frac{\cm \cdot L^{\,2}}{\dmref \cdot \tref} \,,\\[3pt]
  & \dms \egal \frac{\dm}{\dmref} \,,
  && \Bi \egal \frac{\hv \cdot L}{\dmref} \,,
  && \gls \egal \frac{\gl \cdot L}{\dmref \cdot \Pvref}\,.
\end{align*}
where the superscript $0$ represents a reference value, chosen according to the application problem and the superscript $\star$ represents a dimensionless quantity of the same variable. In this way, the dimensionless problem is written as:
\begin{subequations}\label{eq:moisture_dimensionlesspb_1D}
  \begin{align}
   \cms \, \pd{u}{\ts} &\egal \pd{}{\xs} \left( \, \dms \, \pd{u}{\xs} \, \right) \,,
  & \ts & \ > \ 0\,, \;&  \xs & \ \in \ \big[ \, 0, \, 1 \, \big] \,, \label{eq:moisture_governing_dimensionless}\\[3pt]
   \dms \, \pd{u}{\xs} &\egal \BivL \cdot \left( \, u \moins \uL \, \right) \moins \glsL \,,
  & \ts & \ > \ 0\,, \,&  \xs & \egal 0 \,, \\[3pt]
   \moins \dms \, \pd{u}{\xs} &\egal \BivR \cdot \left( \, u \moins \uR \, \right) \moins \glsR \,,
  & \ts & \ > \ 0\,, \,&   \xs & \egal 1 \,, \\[3pt]
   u &\egal u_{\,0}\, (\,\xs) \,,
  & \ts & \egal 0\,, \,&  \xs & \ \in \ \big[ \, 0, \, 1 \, \big] \,.
  \end{align}
\end{subequations}

Finally, this is the problem of interest considered here for resolution. Now, the procedure of the Spectral method will be described to propose a Reduce Order Model for the solution of this problem. 


\section{Spectral reduced-order model for linear transfer}
\label{sec:spectral_linear}

While finite-difference and finite-element methods are based on a local representation of functions, using low-order approximations, Spectral methods consider a global representation of the solution, which yields beyond all orders approximations \cite{Boyd2000}. In the global representation approach, the value of the derivative at a certain spatial location depends on the solution on the entire domain and not only on its neighbours. Spectral methods consider a sum of polynomials that suit for this whole domain, almost like an analytical solution, providing a high approximation of the solution. As its error decreases exponentially it is possible to have the same accuracy of other methods but with a lower number of modes, which makes this method memory usage minimized, allowing to store and operate a lower number of degrees of freedom \cite{Trefethen1996}. The Spectral methods used in this work are the \textsc{Chebyshev} polynomials on the basis function and the \textsc{Tau--Galerkin} method to compute the temporal coefficients.


\subsection{Method description}

For the sake of simplicity and without loosing the generality, this method is first explained considering the dimensionless coefficients $\dms$ and $\cms$ as constants, noting $\nu\, \eqdef\, \dfrac{\dms}{\cms}$ and thus, considering the linear diffusion equation:
\begin{align}\label{eq:heat1d}
\pd{u}{t} \egal \nu \,  \pd{^{\,2} u}{x^{\,2}}  \,,
\end{align}
for $t \, > \, 0\, $ and $x \, \in \, \big[-1\,,  1 \, \big]\,$; the $\star$ symbol was dropped for the purpose of conciseness to explain the method. A special attention must be given to the spatial domain, because the \textsc{Chebyshev} Spectral method we use is described between the interval $\big[ -1\,,  1 \, \big]\,$. Thus, if the dimensionless interval is not in this interval, a change of variables (domain transformation) must be performed for the computational domain.

The boundary conditions are written as:
\begin{subequations}\label{eq:bc_heateq}
  \begin{align}
  \pd{u}{x}  &\egal \BivL \cdot \Bigl(u \moins \uL \bigl(\, t \, \bigr) \Bigr) \,, &  x & \egal -\ 1 \,, \\
  \moins \pd{u}{x} &\egal \BivR \cdot \Bigl(u \moins \uR \bigl(\, t \,\bigr) \Bigr)\,, &  x & \egal 1 \,.
  \end{align}
\end{subequations}

The Spectral method assumes that the unknown $u\,(\,x,\,t\,)$ from Eq.~\eqref{eq:heat1d} can be accurately represented as a finite sum \cite{Mendes2017}:
\begin{equation}\label{eq:series_ap}
  u\, (\,x,\, t\,) \egal \sum_{i\, =\, 0}^{\infty} \, a_{\,i}\, (\,t\,)\, \varphi_{\,i}\, (\,x\,)\egal   \underbrace{\sum_{i\, =\, 0}^{n} \, a_{\,i}\, (\,t\,)\, \varphi_{\,i}\, (\,x\,)}_{\egal u_{\,n}\, (\,x,\, t\,)} \plus \underbrace{\sum_{i\, =\, n+1}^{\infty} \, a_{\,i}\, (\,t\,)\, \varphi_{\,i}\, (\,x\,)}_{\ll 1}  \,.
\end{equation}
Here, $\{\varphi_{\,i}\, (\,x\,)\}_{\,i\, =\, 0}^{\,n}$ is a set of basis functions that remains constant in time, $\{a_{\,i}\, (\,t\,)\}_{\,i\, =\, 0}^{\,n}$ are the corresponding time-dependent spectral coefficients and $n$ represents the number of degrees of freedom of the solution. Eq.~\eqref{eq:series_ap} can be seen as a series truncation after $N \, = \, n\, +\, 1$ modes. The \textsc{Chebyshev} polynomials are chosen as the basis functions since they are optimal in $\mathcal{L}_{\infty}$ approximation norm \cite{Gautschi2004}. It should be observed that other bases can be used, such as the \textsc{Fourier} and \textsc{Legendre} polynomials. Therefore, we have:
\begin{align*}
  \varphi_{\,i}\,(\,x\,)\ \equiv\ \T_{\,i}\,(\,x\,)\,.
\end{align*}

The first \textsc{Chebyshev} polynomials are:
\begin{align*}
  &\T_{\,0}\,(\,x\,) \egal 1\,, & & \T_{\,1}\,(\,x\,) \egal x\,, & & \T_{\,2}\,(\,x\,) \egal 2\, x^{\,2} \moins 1\,, && \T_{\,3}\,(\,x\,) \egal 4\, x^{\,3} \moins 3\, x\,, \, \ldots
\end{align*}
and, higher order polynomials can be constructed using a recursive relation \cite{Peyret2002}: 
\begin{align*}
  \T_{\,i+1}\,(\,x\,) \egal 2\, x\, \T_{\,i}\,(\,x\,) \moins \T_{\,i-1}\,(\,x\,)\,.
\end{align*}

As we have chosen the basis functions, now we can write the derivatives:
\begin{subequations}\label{eq:derivatives}
  \begin{align}
  \pd{u_{\,n}}{x} &\egal \sum_{i\, =\, 0}^n \, a_{\,i}\,(\,t\,)\, \pd{\T_{\,i}}{x}\,(\,x\,)\egal \sum_{i\, =\, 0}^n \tilde{a}_{\,i}\,(\,t\,)\, \T_{\,i}\,(\,x\,)\,,\label{eq:derivative1}\\
  \pd{^{\,2} u_{\,n}}{x^{\,2}} &\egal \sum_{i\, =\, 0}^n \, a_{\,i}\,(\,t\,)\, \pd{^{\,2} \T_{\,i}}{x^{\,2}}\,(\,x\,)\egal \sum_{i\, =\, 0}^n \Tilde{\Tilde{a}}_{\,i}\,(\,t\,)\, \T_{\,i}\,(\,x\,) \,, \label{eq:derivative2}\\
  \pd{u_{\,n}}{t} &\egal \sum_{i\, =\, 0}^n \, \dot{a}_{\,i}\,(\,t\,)\, \T_{\,i}\,(\,x\,)\,,\label{eq:derivative3} 
  \end{align}
\end{subequations}
where the dot denotes $\dot{a}_{\,i}\, (\,t\,) \eqdef \dfrac{\mathrm{d}a\, (\,t\,) }{\mathrm{d}t} $ according to \textsc{Newton} notation. Note that the derivatives are re-expanded in the same \textsc{Chebyshev} basis function. As a result, coefficients $\{\tilde{a}_{\,i}\, (\,t\,)\}$ and $\{\Tilde{\Tilde{a}}_{\,i}\, (\,t\,)\}$ must be re-expressed in terms of coefficients $\{a_{\,i}\, (\,t\,)\}$. The connection is given explicitly from the recurrence relation of the \textsc{Chebyshev} polynomial derivatives \cite{Peyret2002}:
\begin{align*}
  & \tilde{a}_{\,i} \egal \dfrac{2}{c_{\,i}} \sum_{\substack{p\, =\,i\,+\,1 \\ p\,+\,i\; \text{odd}}}^{\,n} \, p \, a_{\,p}\, , & 
  i \egal 0,\ldots,n-1, 
  \\[3pt]
  & \Tilde{a}_{\,n} \ \equiv \ 0 \,, \\[3pt]
  & \Tilde{\Tilde{a}}_{\,i} \egal \dfrac{1}{c_{\,i}} \sum_{\substack{p\, =\, i\,+\,2 \\ p\,+\,i\; \text{even}}}^{\,n}\, p\,\Bigl(\,p^{\,2} \moins i^{\,2}\,\Bigr)\, a_{\,p}\, , & 
  i \egal 0,\ldots,n-2,  \\[3pt] 
  & \Tilde{\Tilde{a}}_{\,n-1} \ \equiv \  \Tilde{\Tilde{a}}_{\,n} \ \equiv \ 0 \,,
\end{align*}
with,
\begin{align*}
  c_{\,i} \egal \left\lbrace 
  \begin{matrix}
  2 \,, & \text{if} & i \egal 0\,,\\
  1 \,, & \text{if} & i \ >\ 0\,.
  \end{matrix} \right.
\end{align*}

Using the expression of the derivatives provided by Eqs.~\eqref{eq:derivative2} and \eqref{eq:derivative3}, the residual of the diffusion equation~\eqref{eq:heat1d} is: 
\begin{align}\label{eq:heat_eq_residual}
  R\,(\,x\,,\,t \,) \egal \sum_{i\, =\, 0}^n \, \Bigl[\, \dot{a}_{\,i}\, (\,t\,) \moins \nu \ \Tilde{\Tilde{a}}_{\,i}\, (\,t\,)\, \Bigr]\, \T_{\,i}\, (\,x\,)\,,
\end{align}
which is considered a misfit of the approximate solution. The purpose is to minimize the residual:
\begin{align*}
  \Bigl\Vert\;  R\,(\,x\,,\,t \,)\; \Bigr\Vert_{\,2} \ \longrightarrow\ \min \,,
\end{align*}
which is realized via the \textsc{Tau}--\textsc{Galerkin} method, which requires Eq.~\eqref{eq:heat_eq_residual} to be orthogonal to the \textsc{Chebyshev} basis functions $\langle\,R\,,\T_{\,i} \,\rangle \,=\, 0\,$:
\begin{align*}
  \langle\,R\,,\T_{\,i} \,\rangle \egal \int_{-1}^{1}\, \dfrac{R\,(\,x\,,t\,)\,\T_{\,i}\,(\,x\,)}{\sqrt{1 \moins x^{\,2}}}\, \mathrm{d} x \egal 0\,,
\end{align*}
namely,
\begin{align}\label{eq:inner_product}
  \int_{-1}^{1}\, \Biggl[\sum_{i\, =\, 0}^n \, \Bigl(\, \dot{a}_{\,i}\, (\,t\,) \moins \nu \ \Tilde{\Tilde{a}}_{\,i}\, (\,t\,)\, \Bigr)\, \dfrac{\T_{\,i}\, (\,x\,) \,\T_{\,j}\,(\,x\,)}{\sqrt{1 \moins x^{\,2}}}\,\Biggr] \mathrm{d} x \egal 0\,.
\end{align}
Then, taking the orthogonality property of the \textsc{Chebyshev} polynomials into account \cite{Peyret2002},  it leads to the following relation among the spectral coefficients:
\begin{align*}
  \dot{a}_{\,i}\, (\,t\,) \moins \nu \, \Tilde{\Tilde{a}}_{\,i}\, (\,t\,) \egal 0\,, & & 
  i \egal 0, \,1, \,\ldots,\, n-2 \,.
\end{align*}

Finally, after the projection and expansion of the residual, the result is a system of Ordinary Differential Equations (ODEs), with $n\, -\, 2$ equations to be solved as a function of time. The two extra coefficients are obtained by substituting the derivative~\eqref{eq:derivative1} into the boundary conditions~\eqref{eq:bc_heateq}:
\begin{subequations}
  \begin{align}
  \sum_{i\, =\, 0}^{n} \, \tilde{a}_{\,i}\, (\,t\,)\, \T_{\,i}\, (-1) \moins \BivL \, \sum_{i\, =\, 0}^n \, a_{\,i}\,(\,t\,)\, \T_{\,i}\,(-1) \plus \BivL \, \uL\, (\,t\,)  &\egal 0\,, \label{eq:bc1_aproxi} \\
  \moins \sum_{i\, =\, 0}^n \, \tilde{a}_{\,i}\, (\,t\,)\, \T_{\,i}\, (\, 1\,) \moins \BivR \, \sum_{i\, =\, 0}^n \, a_{\,i}\, (\,t\,)\, \T_{\,i}\, (\, 1\,) \plus \BivR \,\uR\, (\,t\,) &\egal 0\,, \label{eq:bc2_aproxi}
  \end{align}
\end{subequations}
with $\T_{\,i}\,(-1)\, =\, (-1)^{\,i}$ and $\T_{\,i}\,(\, 1\,)\, \equiv\, 1$ (see \cite{Peyret2002}). Eqs.~\eqref{eq:bc1_aproxi} and \eqref{eq:bc2_aproxi} are written in an explicit way, with coefficients $a_{\,n}$ and $a_{\, n-1}$ expressed in terms of all the other coefficients.

Therefore, the original partial differential equation~\eqref{eq:heat1d} is reduced to a system of ODEs plus two algebraic expressions. For linear problems, the system of ODEs is explicitly built. Moreover, the reduced system of ordinary differential equations has the following form:
\begin{align}\label{eq:system_ODE}
  \dot{a}_{\,i}\, (\,t\,) \egal \A \, a_{\,i}\,(\,t\,) + \b\, (\,t\,)\,,  & &
  i \egal 0, \,1, \,\ldots,\, n-2 \,,
\end{align}
where, $\A \in \Mat_{(n-2)\times (n-2)}(\R )\,$, with constant coefficients and with $n \simeq\,\O\,(\, 10\,)\, \,$. Besides, $\b(\,t\,) \in \R^{\,(n-2)}$ is a vector coming usually from boundary conditions.

Initial values of the coefficients $\{a_{\,i}\,(t\, =\, 0)\}$ are calculated by the \textsc{Galerkin} projection of the initial condition \cite{Canuto2006}:
\begin{align}\label{eq:system_ODE_int}
  a_{\,0,\,i}\ \equiv\ a_{\,i}\,(\, 0\,) \egal \dfrac{2}{\pi\, c_{\,i}}\, \int_{-1}^{\,1}\, \dfrac{u_{\, 0}\,(\,x\,)\, \T_{\,i}\,(\,x\,)}{\sqrt{1 \moins x^{\,2}}}\, \mathrm{d}x\,, & &
  i \egal 0, \,1, \,\ldots,\, n-2 \,,
\end{align}
where, $u_{\, 0}\,(\,x\,)$, is the dimensionless initial condition. After solving the \emph{reduced} system of ODEs (Eqs.~\eqref{eq:system_ODE} and \eqref{eq:system_ODE_int}), it is possible to compose the solution along with the \textsc{Chebyshev} polynomial.

Thus, by using the Spectral--ROM approach to build the reduced-order model, the time-dependent coefficients $\{a_{\,i}\,(\,t\,) \}$ are computed by solving the following system:
\begin{equation}\label{eq:system_ode_general}
  \left\{ \begin{array}{rcl}
    \dot{a}\, (\,t\,) & \egal& \A\, a\, (\,t\,) \plus \b\, (\,t\,) \,, \\
    a\, (\, 0\,)& \egal& a_{\,0} \,,
  \end{array}\right.
\end{equation}
remembering that $\A \in \Mat_{s\times s}\,(\R )$ is a constant coefficient matrix, $\b(\,t\,) \in \R^s$ is a vector coming from the boundary conditions and $a_{\,0}$ is the vector of initial spectral coefficients. The main advantage of a Spectral--ROM is that $s \ll p$, where $p$ is the number of degrees of freedom needed to solve problem~\eqref{eq:bc_heateq} by means of conventional methods (finite-differences, finite-elements and finite-volumes). We note that the matrix $\A$ and the vector $\b(\,t\,)$ might depend on problem parameters, such as the diffusion coefficient $\nu\,$:
\begin{align*}
  \A \egal \A \, (\, t \,; \nu\,)\,, \ \ \text{and} \ \ \b \egal \b(\, t \,; \nu\,) \,.
\end{align*}
Different approaches can be used to solve the system of ODEs~\eqref{eq:system_ode_general}, depending on the cases considered. The most straightforward way to use the Spectral--ROM from Eq.~\eqref{eq:system_ode_general} is to apply a numerical integration scheme, \eg, an adaptive \textsc{Runge}--\textsc{Kutta} with moderate accuracy, since Eq.~\eqref{eq:system_ode_general} is just a ROM. So, with an embedded error control and not so stringent tolerances, it can be done very efficiently. In this study, we shall employ ODE solvers for simplicity, since we are interested in the whole trajectory.


\subsection{Validation of the numerical solution}

To compare and validate the proposed method, the error between a solutions obtained by one of the numerical methods $u^{\, \mathrm{num}}\, (\, x,\, t\,)$, and the reference solution $u^{\, \mathrm{ref}} \, (\, x,\, t\,)$, is computed as a function of $x$ by the following formulation:
\begin{align*}
  \varepsilon_{\,2}\, (\,x\,)\ &\eqdef\ \sqrt{\,\frac{1}{N_{\,t}} \, \sum_{j\, =\, 1}^{N_{\,t}} \, \Bigl( \, u_{\, j}^{\, \mathrm{num}}\, (\, x\,, t\,) \moins u_{\, j}^{\mathrm{\, ref}}\, (\, x\,, t\,) \, \Bigr)^{\,2}}\,,
\end{align*}
where $N_{\,t}$ is the number of temporal steps. The global error $\varepsilon_{\, \infty}$ is given by the maximum value of $\varepsilon_{\,2}\, (\,x\,)$: 
\begin{align*}
  \varepsilon_{\, \infty}\ &\eqdef\ \sup_{x \ \in \ \bigl[\, 0 \,,\, L \,\bigr]} \, \varepsilon_{\,2}\, (\,x\,) \,.
\end{align*}
The computation of the reference solution $u^{\, \mathrm{ref}}\,(\,x\,, t\,)$ is detailed in Sections~\ref{sec:case_linear}, \ref{sec:case_weak_nonlinear} and \ref{sec:case_strong_nonlinear}.


\section{Numerical application} 

\subsection{Linear case}
\label{sec:case_linear}

\begin{figure}
\centering
\includegraphics[width=0.47\textwidth]{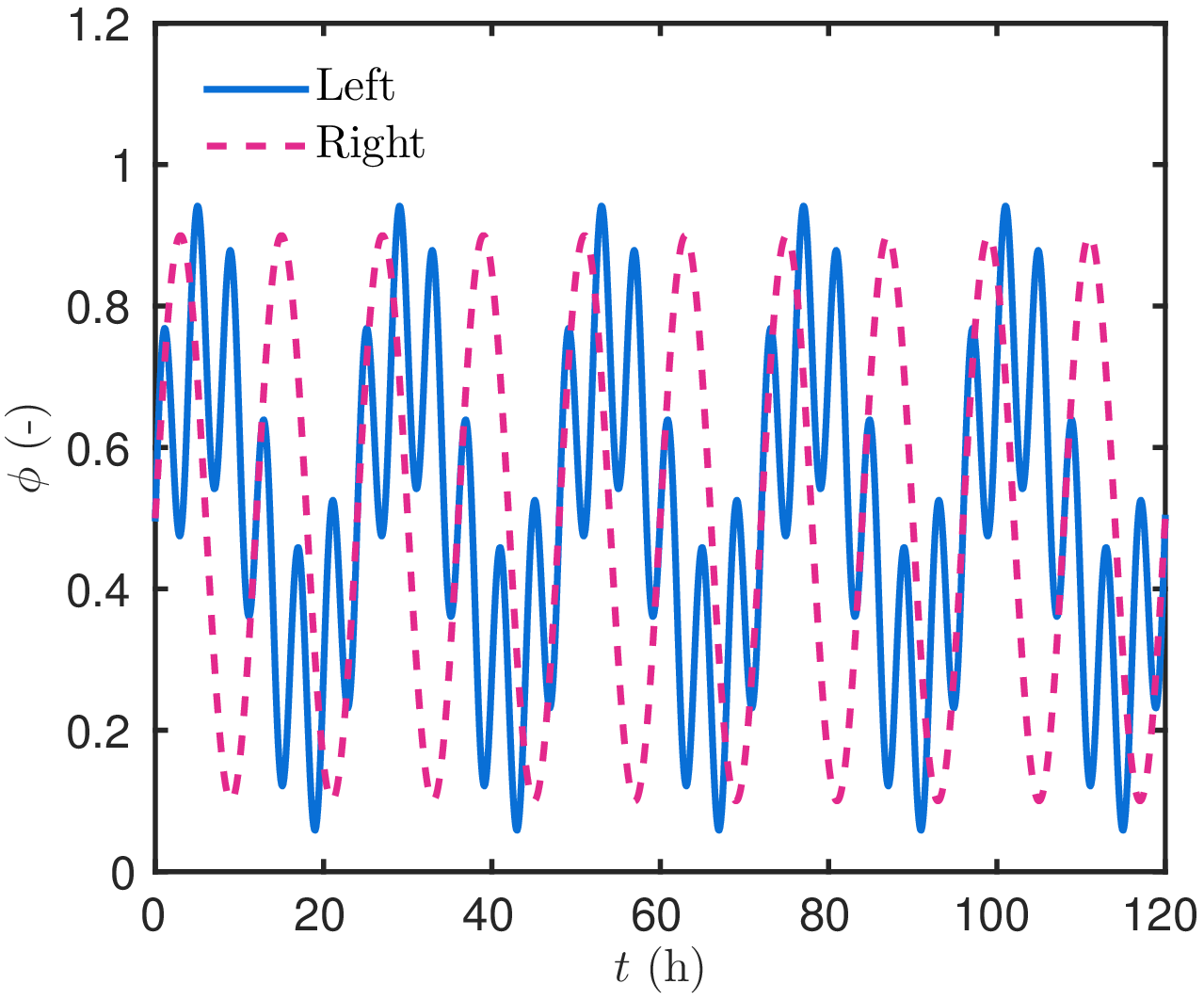}
\caption{\small\em Boundary conditions at the left side $(x \egal 0 \, \mathsf{m})$ and at the right side $(x \egal 0.1 \, \mathsf{m})\,$.}
\label{fig_AN1:BC}
\end{figure}

The first case considers linear moisture transfer in a material with $0.1 \, \mathsf{m}$ of length. The moisture transport coefficient has a value of $\dm \,=\, 1.97 \cdot 10^{-10}\, \mathsf{s}$ and the moisture storage a value of $\cm \,=\, 7.09 \cdot 10^{-3}\, \mathsf{kg/m^3/Pa}$ \cite{Gasparin2017}. The initial vapour pressure across the material is considered to be uniform as $\Pvi \,=\, 1.16 \cdot 10^{\,3}\, \mathsf{Pa \,}$, corresponding to a relative humidity of $50\, \%$ and to a temperature of $20^{\, \circ}\mathsf{C}$. Simulations are performed for a total time of $120\, \mathsf{h}\,$. The boundary conditions, represented by the relative humidity $\phi$ are given in Figure~\ref{fig_AN1:BC}. The sinusoidal variations oscillate between dry and moist states during the total simulation time. The convective vapour transfer coefficients are set to $\hvL \, =\, 2 \cdot 10^{-7}\, \mathsf{s/m}$ and $\hvR \, =\, 3 \cdot 10^{-8}\, \mathsf{s/m}$ for the left and right boundaries, respectively. As the readers may be interested in simulating the proposed case, dimensionless values are provided in Appendix~\ref{annexe:dimensionless}.

This case study is performed with the Spectral--ROM using $N \,=\, 6$ modes and with two central finite-difference approximations schemes: \emph{(i)} the \Eu ~implicit and \emph{(ii)} the \CN . The reference solution is computed using the \texttt{Matlab} open source toolbox \texttt{Chebfun} \cite{Driscoll2014}.

The reduced system of ODEs is implemented in \texttt{Matlab} and the spectral coefficients $\{a_{\,n}\,(\,t\,)\}$ are calculated for any intermediate time instant by the solver \texttt{ODE45}. The solver is set with an absolute and relative tolerance of $\mathsf{tol}\, =\,10^{\,-4}\,$. The integration in time is based on an explicit \textsc{Runge--Kutta} formula for \texttt{ODE45}. The inputs are the initial time, the final time and the time step (optional) and the solver supplies the integration at the given time. One should recall that computations of the Spectral solution are performed for the reference spatial domain of $[\, -1,\, 1\, ]$ and then transformed to the interested one.

\begin{figure}
\centering
\subfigure[a][\label{fig_AN1:x0time}]{\includegraphics[width=0.45\textwidth]{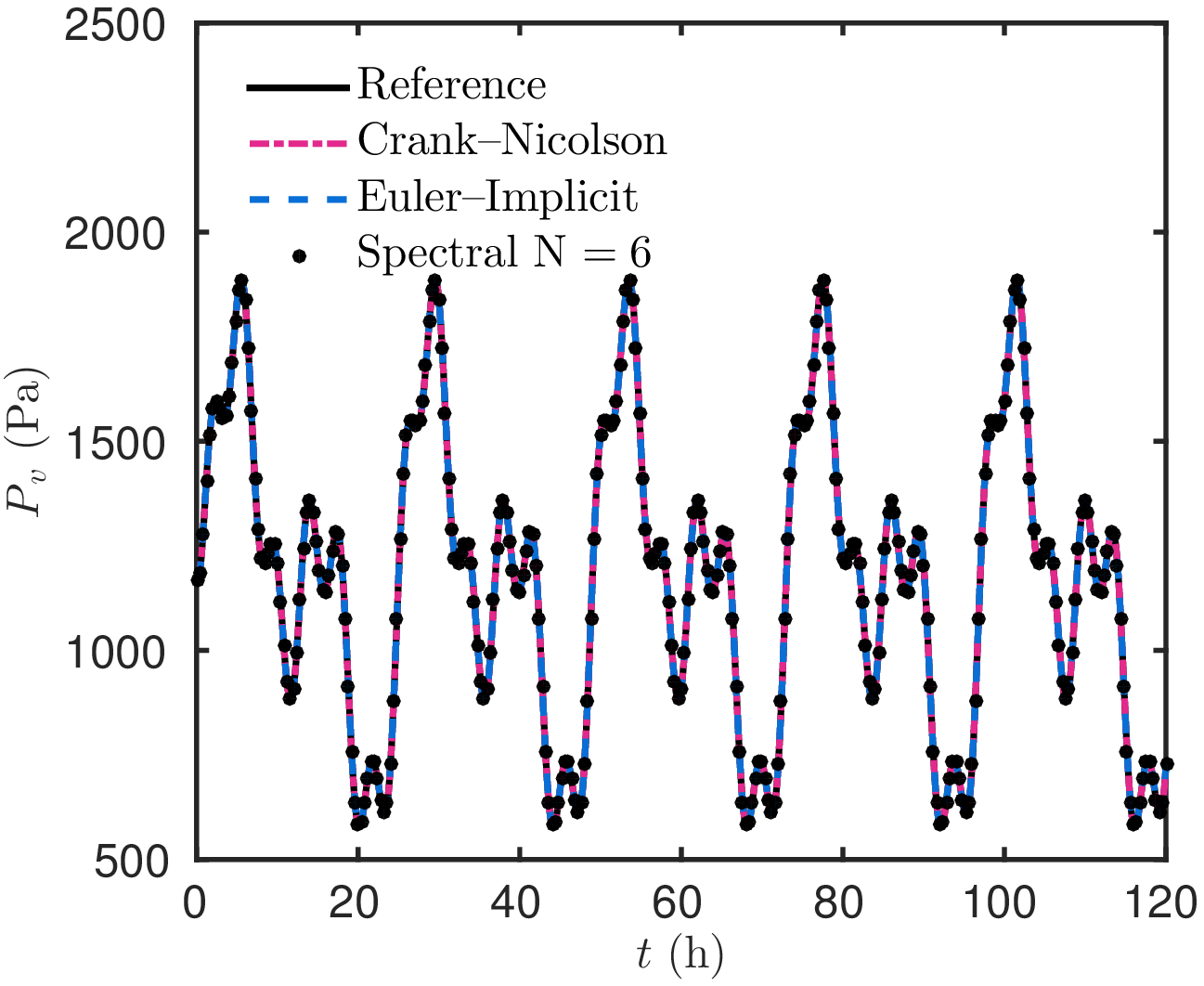}}
\subfigure[b][\label{fig_AN1:profil}]{\includegraphics[width=0.45\textwidth]{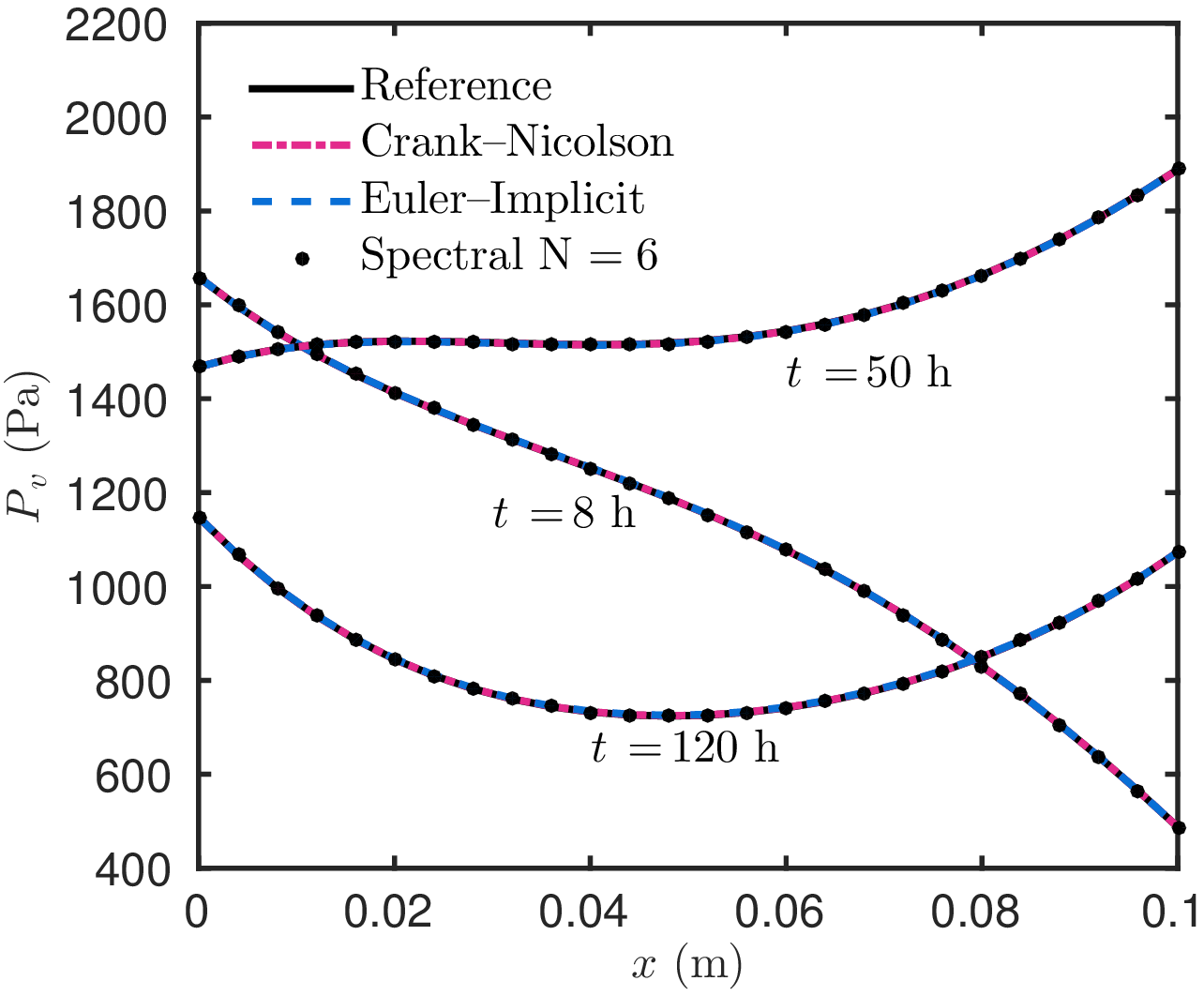}}
\caption{\small\em Evolution of the vapour pressure inside of the material, in $x \,=\, 0.04\,\mathsf{m}$ (a) and vapour pressure profiles at different times, for $t\, \in\, \left\lbrace 8 \,, 50 \,, 120 \right\rbrace\, \mathsf{h}$ (b).}
\end{figure}

It can be seen that the physical phenomena are well represented, as illustrated in Figure~\ref{fig_AN1:x0time} with the evolution of the vapour pressure at $x\,=\,0.04\, \mathsf{m}$. The variations follow the conditions of the left boundary and with the diffusion process going towards the periodic regime. It can be noted a good agreement between the Spectral--ROM and the other methods. Furthermore, the vapour pressure profile is shown in Figure~\ref{fig_AN1:profil} for the instants $t \,=\, \{ 8,\, 50,\,120\}\, \mathsf{h}$, enhancing the good accuracy of the solution to represent the physical phenomena.

\begin{figure}
\centering
\includegraphics[width=0.47\textwidth]{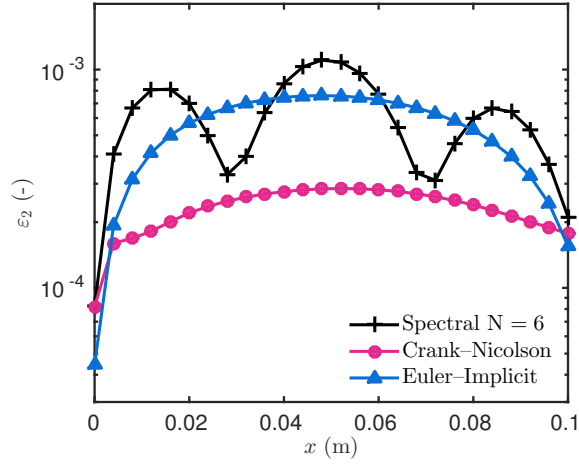}
\caption{\small\em Error $\varepsilon_{\,2}$ computed for the \CN ~method, for the \Eu ~implicit and for the Spectral with $N\ =\ 6$ modes.}
\label{fig_AN1:err_fx}
\end{figure}

The absolute error $\varepsilon_{\,2}$ of the numerical methods applied and the reference solution is of the order of $\O\,(\,10^{\,-4}\,)\,$, as illustrated in Figure~\ref{fig_AN1:err_fx}. The solutions of the problem have been computed for discretisation parameters of $\Delta \xs \,=\, 1 \cdot 10^{\,-2}$ and $\Delta \ts \,=\, 1 \cdot 10^{\,-1}$ for the Spectral--ROM and the \CN ~methods. However, the \Eu ~implicit scheme needed more refinement to reach the same order of accuracy, with $\Delta \xs \,=\, 1 \cdot 10^{\,-2} \text{ and } \Delta \ts \,=\, 1 \cdot 10^{\,-2}\,$.

\begin{figure}
\centering
\subfigure[a][\label{fig_AN1:profil_spect_fM}]{\includegraphics[width=0.45\textwidth]{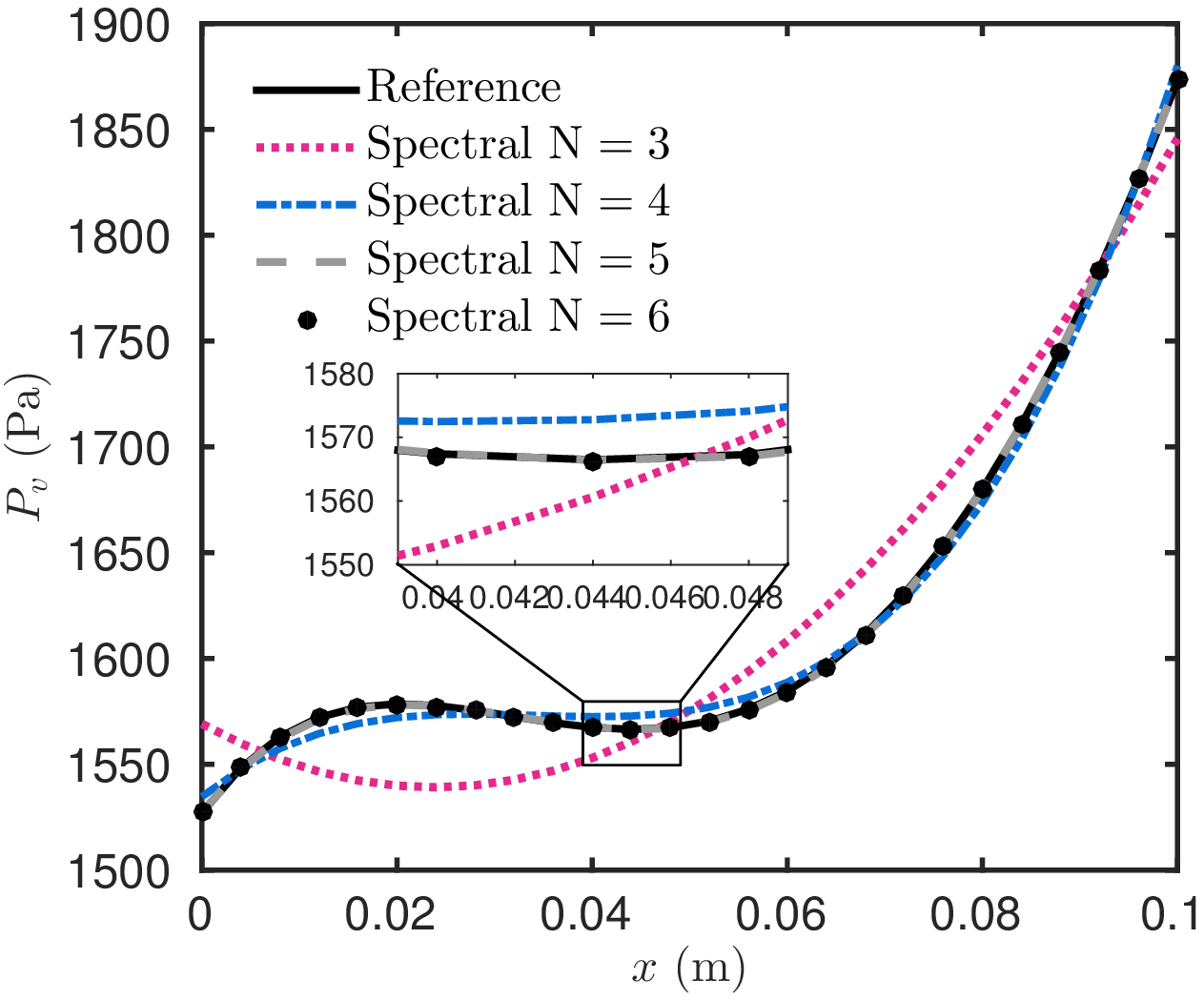}}
\subfigure[b][\label{fig_AN1:error_l2_all}]{\includegraphics[width=0.45\textwidth]{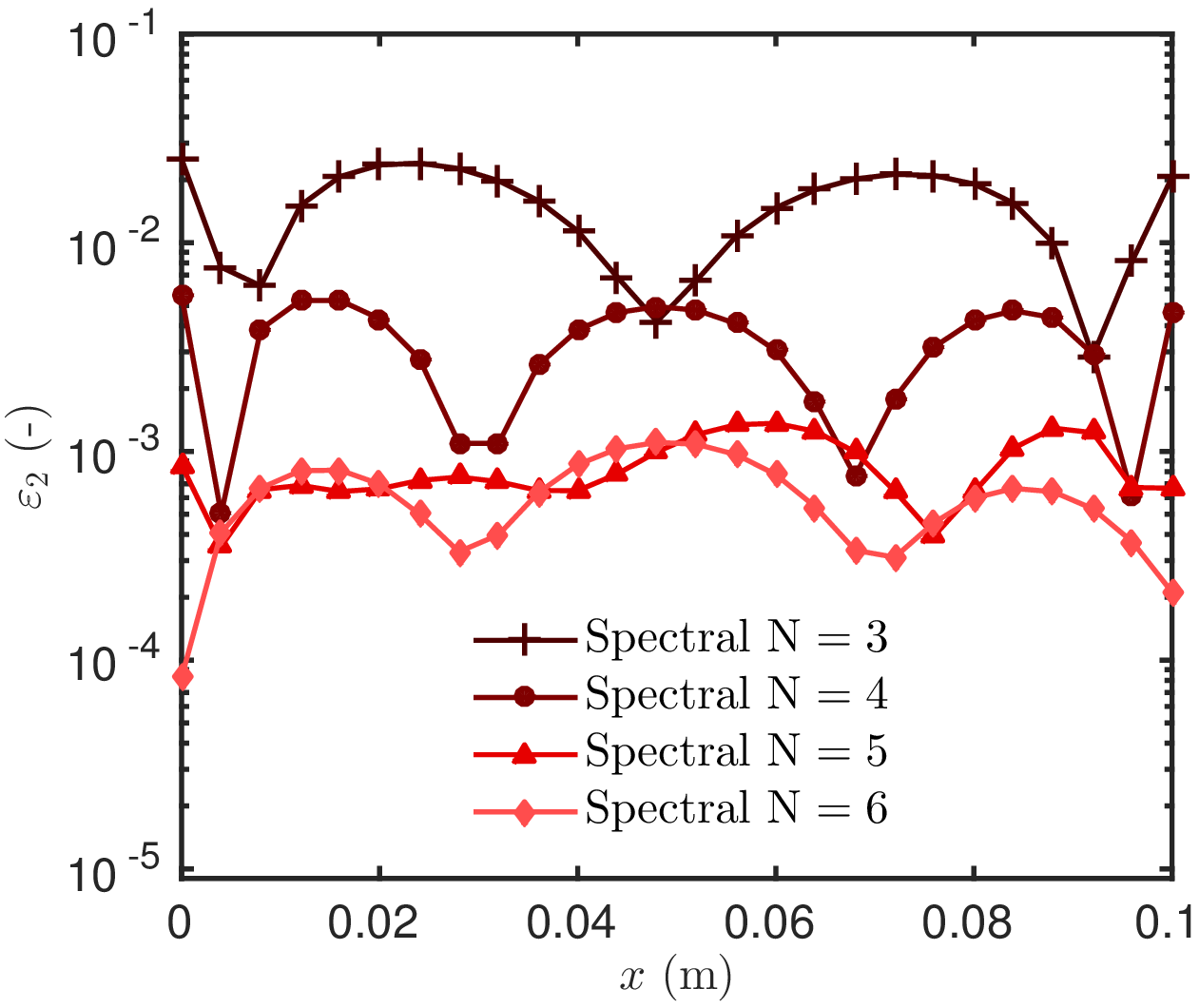}}
\caption{\small\em Vapour pressure profiles (at $t\,=\,2\, \mathsf{h}$) (a) computed with different numbers of spectral modes and the error $\varepsilon_{\,2}$ (b).}
\end{figure}

Figure~\ref{fig_AN1:error_l2_all} presents the absolute error $\varepsilon_{\,2}$ for the Spectral--ROM using different number of modes. As we increase the number of modes, the solution of the Spectral--ROM gets more accurate with the solution converging within few modes (less than 10). To illustrate the convergence of the solution, the profile of the vapour pressure for the last time instant of simulation is presented for a different number of modes in Figure~\ref{fig_AN1:profil_spect_fM}. In this case, if we compare the solution with $3$ modes to the solution with $5$ modes a significant difference can be noticed. With $5$ modes we already have a satisfactory solution to the problem, with the absolute error of the order of $\O\,(\,10^{-3}\,)\,$, while the solution with $3$ modes is still oscillating. The number of modes of the Spectral method is predetermined in order to build the system of ODEs. In this case, a number of six modes proved to be good enough.

\begin{figure}
\centering
\subfigure[a][\label{fig_AN1:coefA_spc_first3}]{\includegraphics[width=0.45\textwidth]{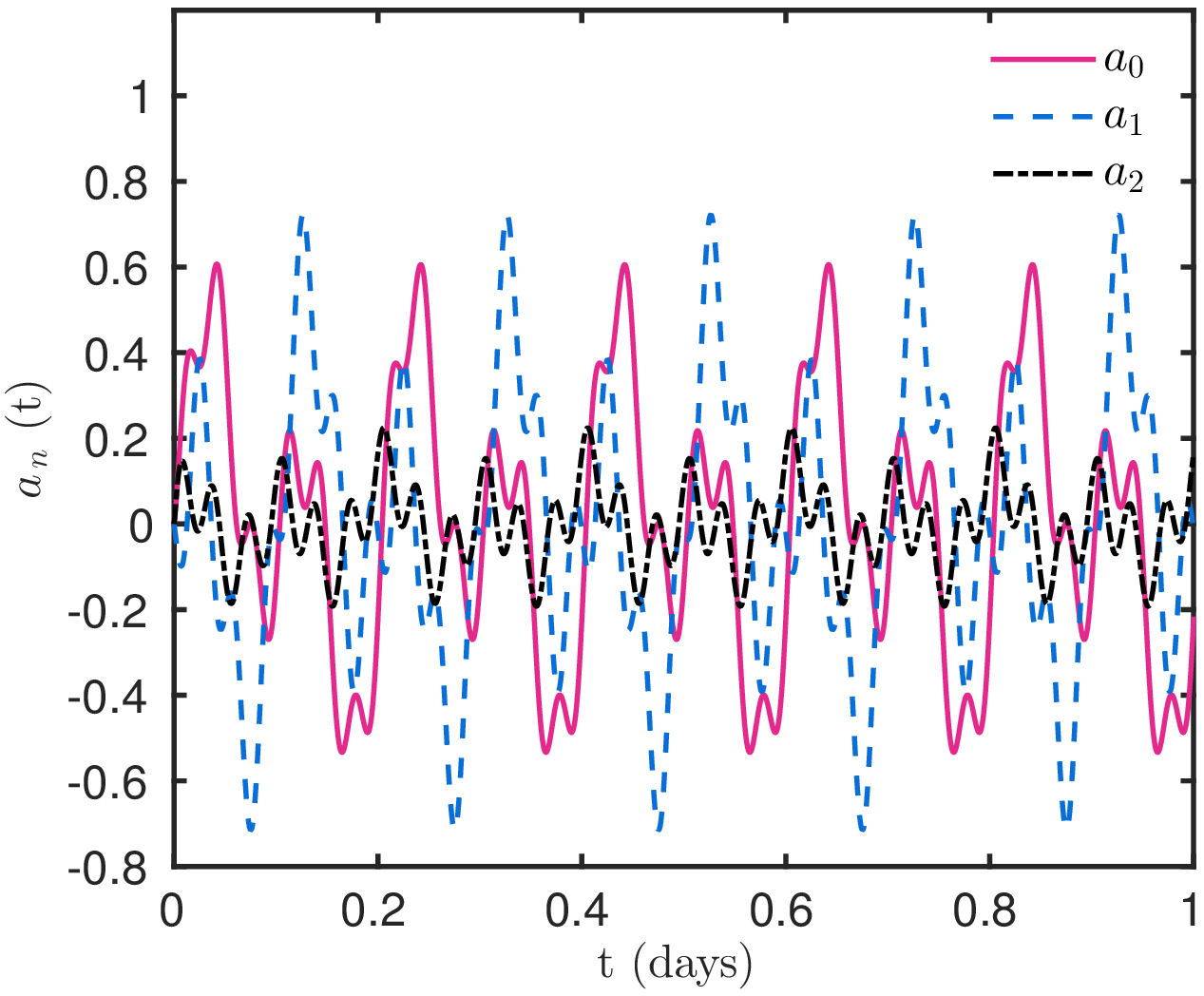}} \hspace{0.3cm}
\subfigure[b][\label{fig_AN1:coefA_spc_last3}]{\includegraphics[width=0.45\textwidth]{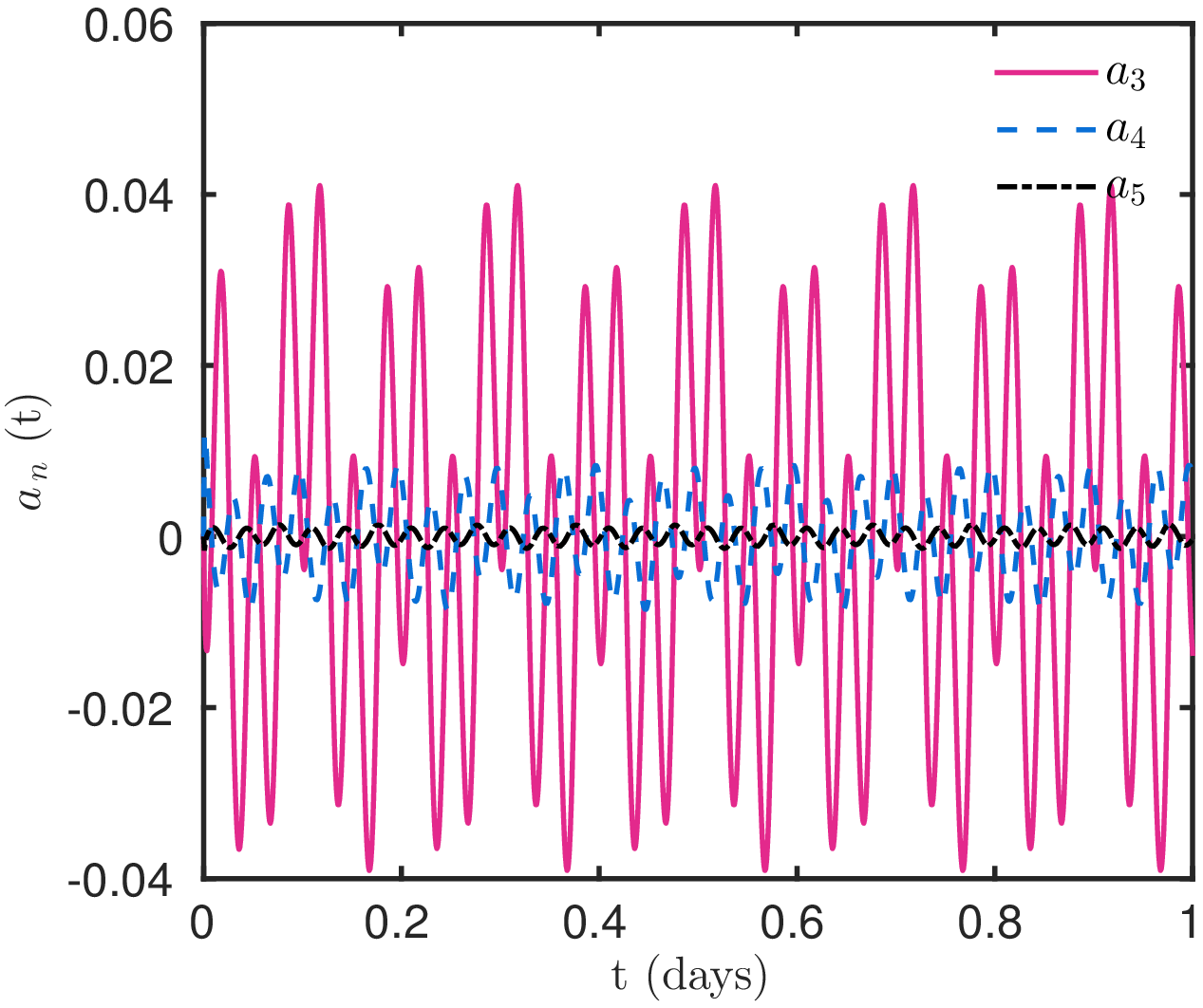}}
\caption{\small\em Evolution of the first three spectral coefficients $a_{\,n}$ (a) and of the last three coefficients (b).}
\end{figure}

\begin{figure}
\centering
\includegraphics[width=0.47\textwidth]{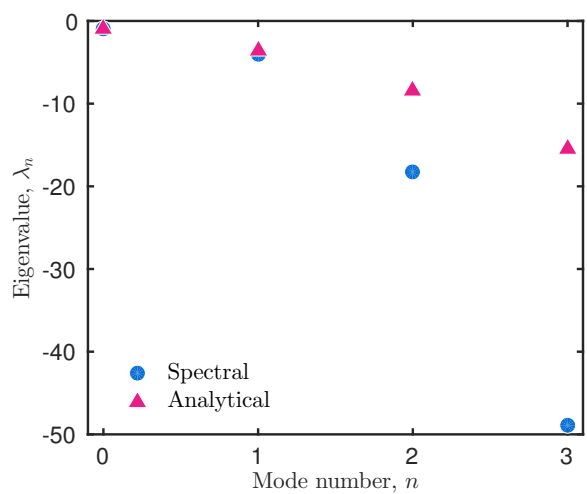}
\caption{\small\em Eigenvalues of the Analytical and of the Spectral solution corresponding to the first modes.}
\label{fig_AN1:eigenvalues}
\end{figure}

Spectral coefficients $a_{\,n}\,(\,t\,)$ are shown in Figures~\ref{fig_AN1:coefA_spc_first3} and \ref{fig_AN1:coefA_spc_last3}. It can be seen the first coefficients have the most significant values. For this reason, the Spectral method needs few modes to converge to the solution (an order of $10$) because its first modes have the highest magnitudes. A brief comparison with an analytical solution, built on \textsc{Fourier} decomposition \cite{Ozisik1993}, reveals that the eigenvalues of the Spectral method decrease faster, as shown in Figure~\ref{fig_AN1:eigenvalues}. Note that the eigenvalues of the analytical solution do not have to coincide with the ones of the Spectral method since the eigenfunctions are not the same for the \textsc{Chebyshev} polynomials and the trigonometric ones. Furthermore, the magnitude of the last spectral coefficient acts as an error estimator, determining the error upper limit.

\begin{figure}
\centering
\includegraphics[width=0.47\textwidth]{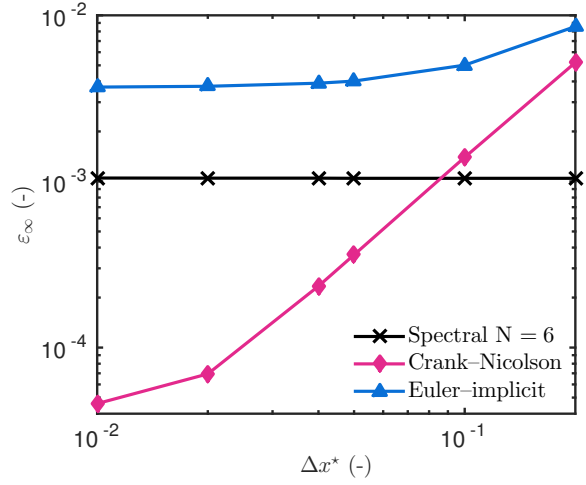}
\caption{\small\em Error $\varepsilon_{\, \infty}$ in function of the $\Delta \xs$ values.}
\label{fig_AN1:n_error_fNx}
\end{figure}

The global absolute error $\varepsilon_{\, \infty}$ for the conventional numerical methods applied is calculated as a function of spatial discretisation $\Delta \xs\,$. Fig.~\ref{fig_AN1:n_error_fNx} shows that the Spectral-ROM has the same accuracy for all values of $\Delta \xs$. It is due to the fact that the Spectral method is based on \textsc{Chebyshev} polynomials, which enables to calculate the solution in each spatial node, as an analytical solution. For this reason, the error of the Spectral solution is almost a straight line, not depending on the spatial discretisation. However, for the conventional methods, the solution gets inaccurate when the value of $\Delta \xs$ increases. It should be noted that the Spectral--ROM can provide even more accurate results, by increasing the number of modes or by decreasing the tolerance in the ODE \texttt{Matlab} solver to certain limits.

\subsection{Weakly nonlinear case}
\label{sec:case_weak_nonlinear}

This case is called weakly nonlinear because the boundary conditions remain linear and, only the diffusion coefficient has a slight dependency on the moisture field. Thus, the diffusion equation is written as:
\begin{align}\label{eq:diffusion_weak_nonlinear}
  \pd{u}{t} \egal \nu\,(\,u\,) \,  \pd{^{\,2} u}{x^{\,2}}  \,,
\end{align}
where, $\nu\,(\,u\,) \egal  \nu_{\,0} \plus \nu_{\,1} \cdot u\,$. Since we have the diffusion coefficient depending on the field $\nu\,(\,u\,)$, the diffusion equation --- Eq.~\eqref{eq:diffusion_weak_nonlinear} --- can be rewritten as:
\begin{align*}
  \pd{u}{t} \egal \nu_{\,0}\cdot \pd{^{\,2} u}{x^{\,2}} \plus \nu_{\,1} \cdot u \cdot \pd{^{\,2} u}{x^{\,2}}\,,
\end{align*}
where the residual has the following form:
\begin{align*}
  R \egal \underbrace{\sum_{i\egalb 0}^n \dot{a}_{\,i}\, (\,t\,)\, \T_{\,i}}_{\displaystyle{\pd{u}{t}}} \moins \nu_{\,0} \cdot \underbrace{\sum_{i\egalb 0}^n \Tilde{\Tilde{a}}_{\,i}\,(\,t\,)\, \T_{\,i}\,(\,x\,)}_{\displaystyle{\pd{^{\,2} u}{x^{\,2}}}} \moins \nu_{\,1} \cdot \underbrace{\sum_{i\egalb 0}^n \, a_{\,i}\, (\,t\,)\, \T_{\,i}\, (\,x\,)}_{\displaystyle{u}} \cdot \underbrace{\sum_{j\, = \, 0}^{n}\, \Tilde{\Tilde{a}}_{\,j}\, (\,t\,)\,  \T_{\,j}\,(\,x\,)}_{\displaystyle{\pd{^{\,2} u}{x^{\,2}}}} \,. 
\end{align*}
Then, by applying the \textsc{Tau}--\textsc{Galerkin} method, the residual is minimized by assuming it orthogonal to the basis functions $\langle\,R\,,\T_{\,k} \,\rangle \,=\, 0\,$, which is defined in Eq.~\eqref{eq:inner_product}, leading to the following equation:
\begin{align}
  \dot{a}_{\,i}\,(\,t\,) \egal \nu_{\,0}\cdot \tilde{\tilde{a}}_{\,i}\,(\,t\,)  \plus \nu_{\,1} \cdot \sum_{i\, = \, 0}^{n} \, \sum_{j\, = \, 0}^{n} \, c_{\, i,\,j,\,k}\ a_{\,i}\,(\,t\,) \, \tilde{\tilde{a}}_{\,j}\,(\,t\,) \,, \label{eq:AN2_residue}
\end{align}
where,
\begin{align*}
  c_{\, i,\,j,\,k} \egal \dfrac{2}{\pi} \, \int_{-1}^{1} \, \dfrac{\T_{\,i}\, (\,x\,)\,\T_{\,j}\, (\,x\,)\,\T_{\,k}\, (\,x\,)}{\sqrt{1 \moins x^{\,2}}}\, \mathrm{d}x \,.
\end{align*}

Eq.~\eqref{eq:AN2_residue} is a closed system of ODEs. Coefficients $c_{\, i,\,j,\,k} $ are calculated at once, and coefficient $\tilde{\tilde{a}}_{\,i}$ are related to $a_{\,i}$ though a linear transformation $\tilde{\tilde{a}} \egal \textsc{D}_{\,2} \cdot a$, in which $ \textsc{D}_{\,2}\in \Mat_{(n-2)\times (n-2)}(\R )$ is a second order derivative matrix.


\subsubsection{Case study}

\begin{figure}
\centering
\includegraphics[width=0.47\textwidth]{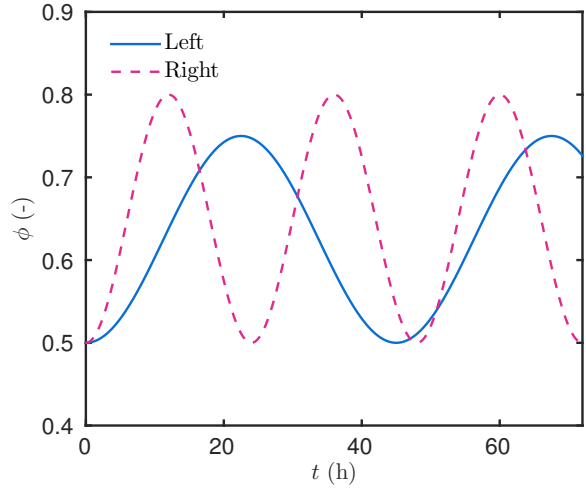}
\caption{\small\em Boundary conditions at the left side $(x \egal 0 \, \mathsf{m})$ and at the right side $(x \egal 0.1 \, \mathsf{m})$ of the domain.}
\label{fig_AN2:BC}
\end{figure}

This case considers that $\cm$ and $\dm$ have a slight dependency on the moisture. The material piece has a length of $0.1$ $\mathsf{m}$, with a relative humidity-dependent diffusion coefficient:
\begin{align*}
  \nu \egal \frac{\dm}{\cm} \egal 3.05\cdot 10^{\, -8} + 6.94\cdot 10^{\, -8} \cdot \phi \,.
\end{align*}
The initial vapour pressure in the material is considered uniform $\Pvi \egal 1.16 \cdot 10^{\,3}$ $\mathsf{Pa \,}$, corresponding to a relative humidity of $50\, \%$ and to temperature of $20^{\circ} \mathsf{C\,}$. Simulations are performed for a total time of $72\, \mathsf{h}$, the equivalent of three days. The boundary conditions, represented by the relative humidity $\phi$ are given in Figure~\ref{fig_AN2:BC}. The relative humidity oscillates sinusoidally between $50\, \%$ and $75\, \%$ on the left boundary and between $50\,\%$ and $80\,\%$ on the right boundary. The convective vapour coefficients are set to $\hvL \, =\, 3\cdot 10^{\,-8}\, \mathsf{s/m}$ and $\hvR \, =\, 2 \cdot 10^{\,-7}\, \mathsf{s/m}$ for the left and right boundaries, respectively. The dimensionless values of this case are also provided in Appendix~\ref{annexe:dimensionless}.

The Spectral reduced-order model is composed by $N \, = \, 6$ modes and its coefficients $\{a_{\,n}(\,t\,)\}$ are obtained through the use of the solver \texttt{ODE45}, with a tolerance set to $\mathsf{tol}\, =\,10^{\,-4}$. The discretisations used to compute the Spectral solution are $\Delta \xs \, =\, 10^{\,-2}$ and $\Delta \ts \, =\, 10^{\,-1}\,$. The reference solution is computed with the open source toolbox \texttt{Chebfun} in \texttt{Matlab}.

\begin{figure}
\centering
\subfigure[a][\label{fig_AN2:evolution}]{\includegraphics[width=0.45\textwidth]{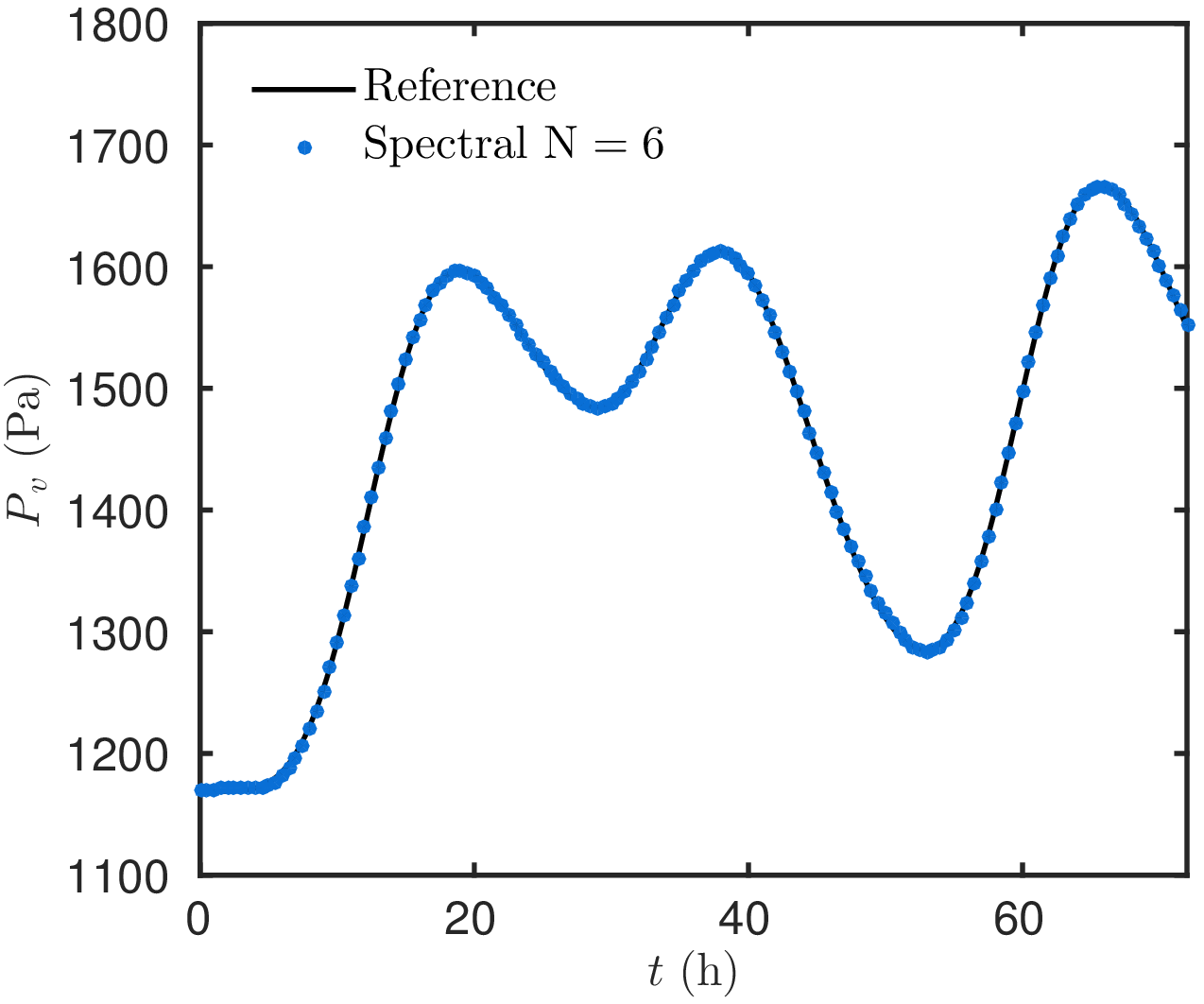}}
\subfigure[b][\label{fig_AN2:profile}]{\includegraphics[width=0.45\textwidth]{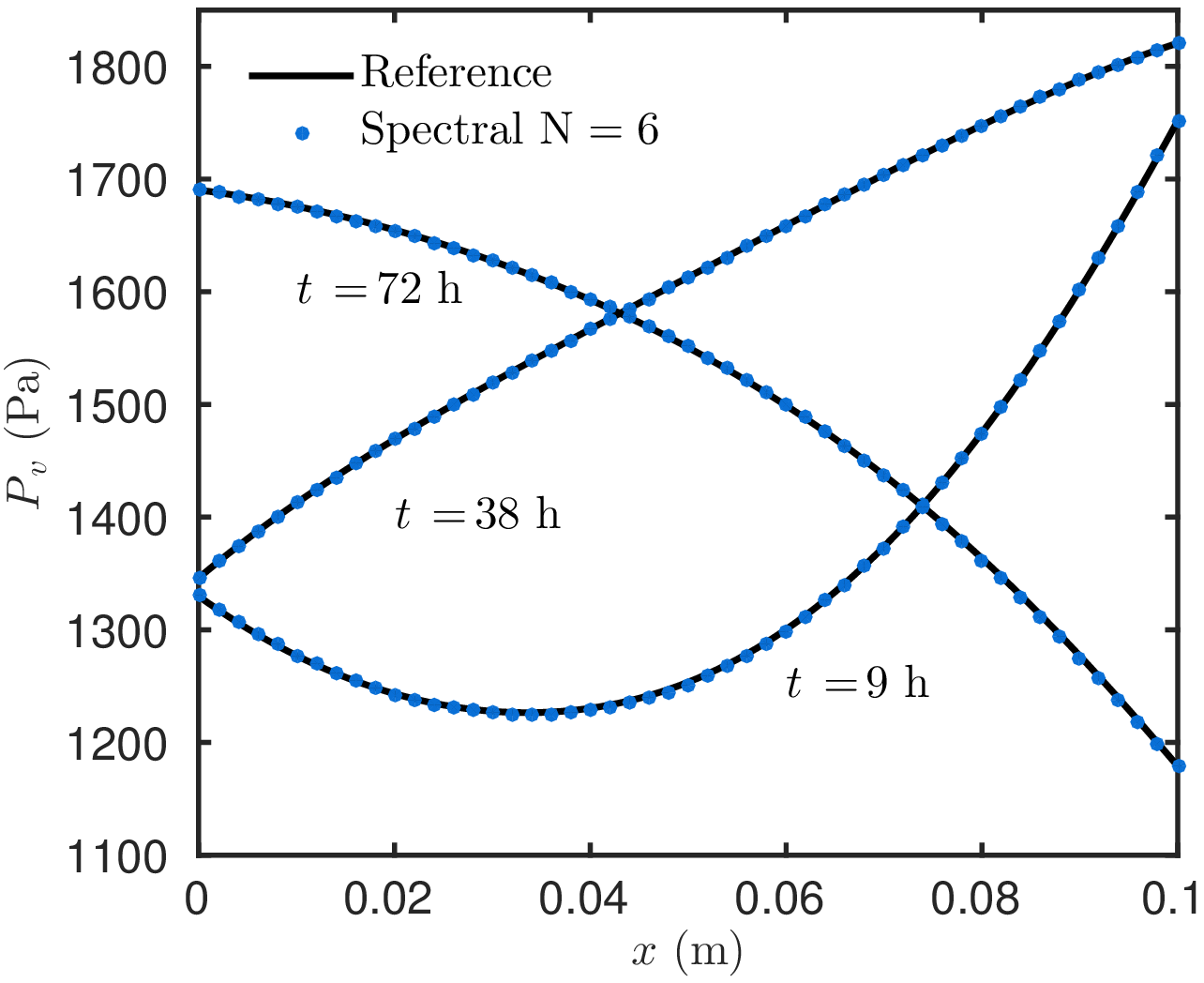}}
\caption{\small\em Evolution of the vapour pressure inside of the material, in $x \,=\, 0.05\, \mathsf{m}$ (a) and vapour pressure profiles at different times, for $t \in \left\lbrace 9 \,, 38 \,, 72 \right\rbrace$  $\mathsf{h}$ (b).}
\end{figure}

The evolution of the vapour pressure in the middle of the material, at $x \,=\, 0.05$ $\mathsf{m}$, is shown in Figure~\ref{fig_AN2:evolution}. The vapour pressure varies according to the sinusoidal fluctuations from both boundary conditions. The vapour pressure profiles at different times are illustrated in Figure~\ref{fig_AN2:profile} for $t \,=\, \{ 9,\, 38,\, 72\}\, \mathsf{h}$, highlighting the good agreement of the Spectral solution in representing the variations.

\begin{figure}
\centering
\includegraphics[width=0.47\textwidth]{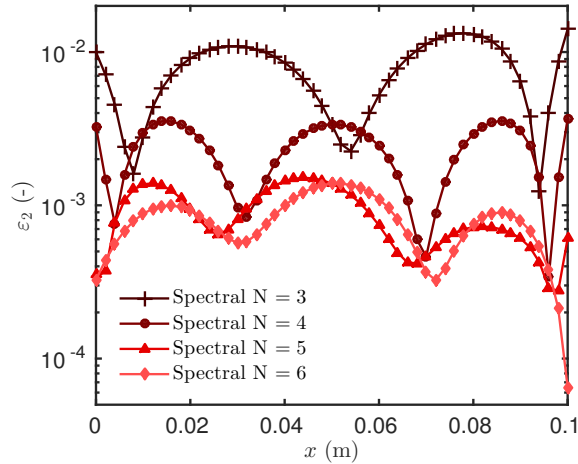}
\caption{\small\em Error $\varepsilon_{\,2}$ computed for the Spectral solution, varying the number of modes.}
\label{fig_AN2:n_error_l2_all}
\end{figure}

The absolute error $\varepsilon_{\,2}$ has been computed between the reference solution and the Spectral--ROM for a different number of modes, as illustrated in Figure~\ref{fig_AN2:n_error_l2_all}. For $N\, =\,6$ and $N\, =\,5$ modes the absolute error is of the same order, $\O\,(\,10^{\,-3}\,)\,$, proving the accuracy of the solution and showing that $5$ modes are good enough.

\begin{figure}
\centering
\subfigure[a][\label{fig_AN2:coefA_spc_first3}]{\includegraphics[width=0.45\textwidth]{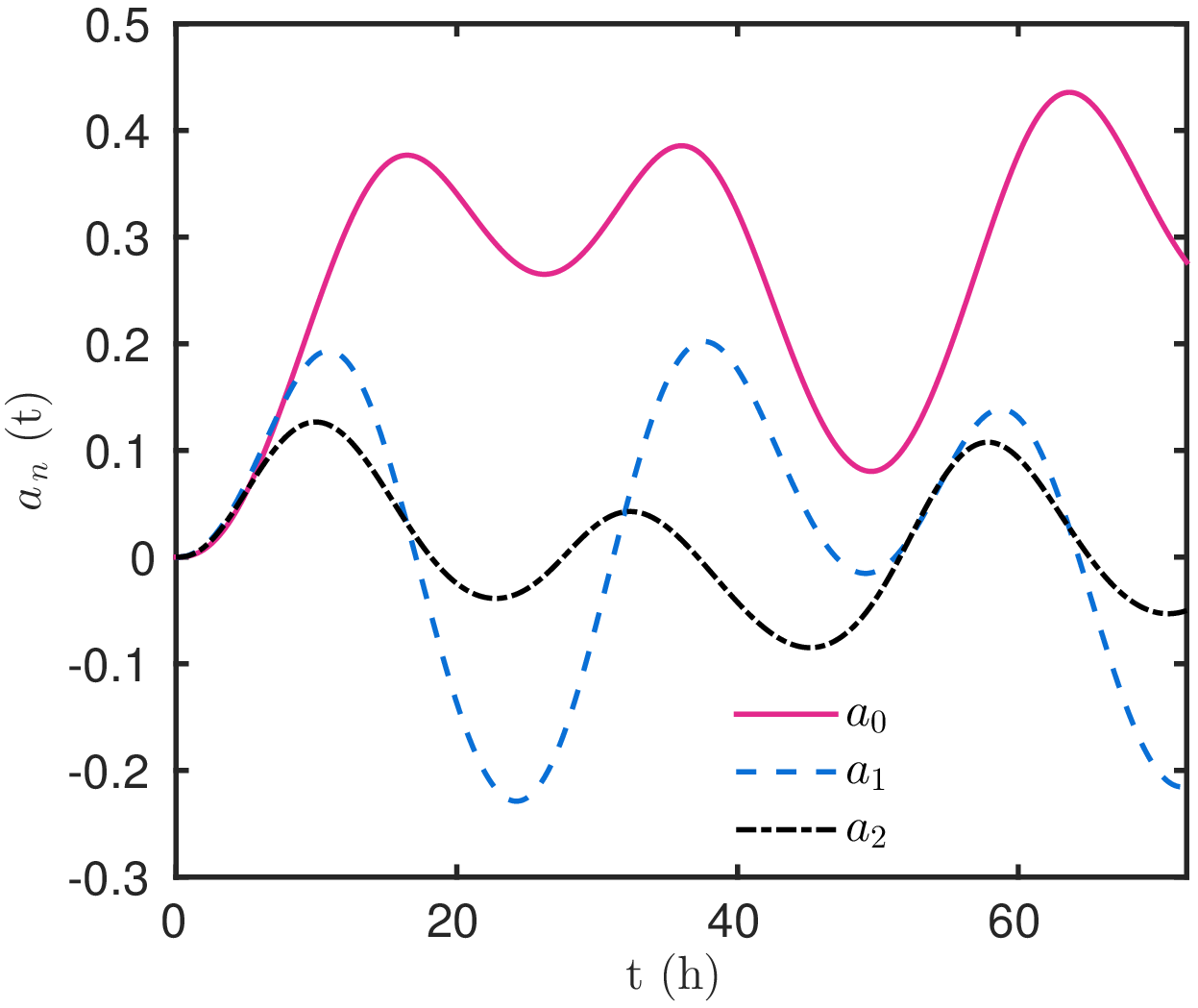}}
\subfigure[b][\label{fig_AN2:coefA_spc_last3}]{\includegraphics[width=0.45\textwidth]{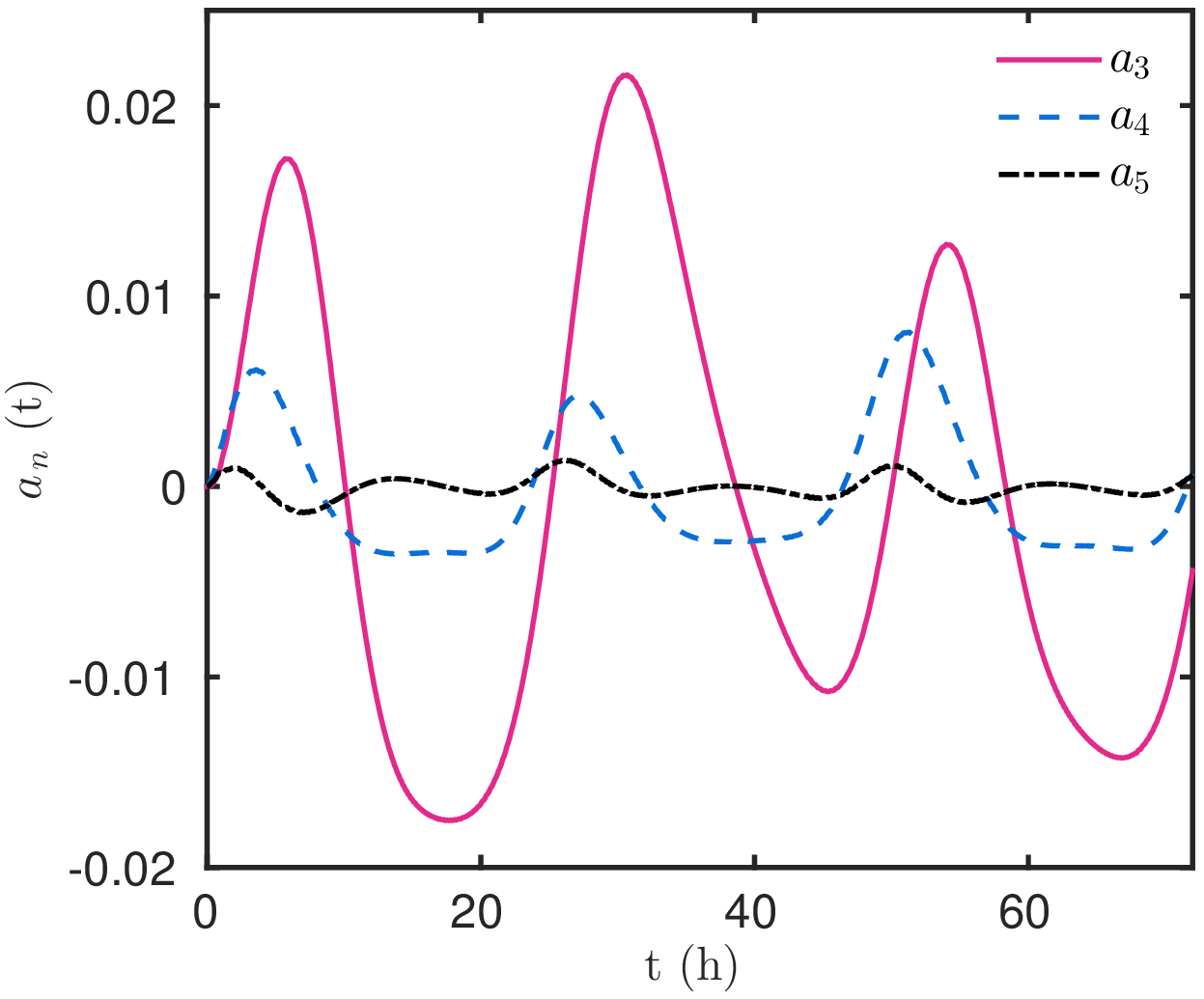}}
\caption{\small\em Evolution of the first three spectral coefficients $a_{\,n}$ (a) and of the last three coefficients (b).}
\end{figure}

Figures~\ref{fig_AN2:coefA_spc_first3} and \ref{fig_AN2:coefA_spc_last3} present the first and the last three coefficients $a_{\, n}\,$, respectively. The magnitude of the coefficient, in the total contribution of the solution, decreases with the order of the coefficient. The last coefficient determines the magnitude of the error, implying that the error will be lower than $a_{\,n}\,$. It is due to the truncation in the number of terms in the spectral representation of the solution and the fact that the solution is smooth. Thus, the higher the number of modes, the higher the accuracy. For this case, we cannot have a more precise solution than $\sup_{\, t \, \in \, \bigl[\, 0 \,, T \,\bigr]} \vert a_{\,6}\, (t) \vert \, = \,1.3\cdot 10^{\,-3}\,$.

\subsection{Numerical cost estimation}

The number of operations for each approach can be estimated. We denote by $N_{\,x}$ and $N_{\,t}$ the number of nodes according to the discretisation in both space and time domains. For explicit methods, it can be related to the CFL type conditions. A standard approach based on the \Eu ~implicit scheme requires $N_{\,x} \cdot N_{\,t}\,$, operations while the  \CN ~scheme requires at least twice as many, as it is built on both implicit and explicit parts. Considering the same discretisation parameters $N_{\,t} \, =\, 1200$ and $N_{\,x} \, =\, 100$ for both methods, the number of operations for the linear case scales with: 
\begin{align*}
  & \text{\Eu ~implicit:} && \O \Bigl(\, N_{\,x} \cdot N_{\,t} \, \Bigr)\ \simeq \  \O \Bigl(1.2 \cdot 10^{\,5}\Bigr)\,, \\
  & \text{\CN:} && \O \Bigl(\,2 \cdot N_{\,x} \cdot N_{\,t} \, \Bigr)\ \simeq \  \O \Bigl(2.4 \cdot 10^{\,5} \Bigr)\,.
\end{align*}
Considering these discretisation parameters, the order of accuracy is not the same for both methods, it is at the order of $\varepsilon_{\infty} \simeq \O\,(10^{-3}) $ for the \Eu ~implicit and of $\varepsilon_{\infty} \simeq \O\,(10^{-4}) $ for the \CN ~solution.

For the Spectral-ROM, the number is related to the solution of the system of ODEs (Eq.~\eqref{eq:system_ODE}), computed in this case with the \texttt{Matlab} solver \texttt{ODE45}. It is based on the iterative \textsc{Runge}--\textsc{Kutta} method to approximate the solution. The number of operation depends on the tolerance (\textsf{tol}) of the solver, which has a maximum tolerance of $ \sim 10^{\,-5}$ for \texttt{ODE45}. Thus, we have: 
\begin{align*}
  N_{\,t} \ \simeq \ \dfrac{\tau}{\Delta t} \ \simeq \ \dfrac{\tau}{(\mathsf{tol})^{\,1/5}} \,,
\end{align*}
where $\tau$ is the total time of simulation. At each time step, the \textsc{Runge}--\textsc{Kutta} needs to compute the vector product $\A_{\,s\times s}\,$, where $s$ depends on the degree of freedom $N$ of the solution ($s \, =\, N \, -\, 2$). Thus, it leads to $6\cdot s^{\,2}$ operations to perform, knowing that $s$ is of the order of $10\,$. Consequently, the total number of operations for the Spectral-ROM scales with approximately:
\begin{align*}
  & \O \biggl( \, \dfrac{6 \, (N \moins 2)^{\,2} \, \tau}{(\mathsf{tol})^{\,1/5}} \, \biggl) \,.
\end{align*}
Considering the first case, knowing that the tolerance was set to $10^{\, -4}\,$, with $N\, =\,6$ modes the number of operations performed by the Spectral--ROM is expressed by:
\begin{align*}
  & \text{Spectral--ROM:} && \O\biggl(\, \dfrac{6 \, (N \moins 2)^{\,2} \, \tau}{(\mathsf{10^{\, -4}})^{\, 1/5}} \, \biggr)\ \simeq\ \O\Bigl(\,  37 \cdot (6 \moins 2)^{\,2}  \cdot 120 \ \, \Bigr) \ \simeq\ \O\Bigl(\, 6.5 \cdot 10^{\, 4} \, \Bigr) \,.
\end{align*}

Comparing the number of operations of this case, we can already see that the Spectral--ROM is less costly than the other methods applied. Notice that the number of degrees of freedom necessary to solve the diffusion problem by means of the Spectral method is inferior to the ones necessary to solve the whole system of partial differential equations. Using \Eu ~or \CN ~methods, the computational complexity scales with $p \, =\, N_{\,x}\,$, whereas the one Spectral--ROM is $s \, =\, (\,N \, -\, 2\,)\,$. For this case, the numerical application gives $p \, \equiv \, 10^{\,2}$ and $s \, \equiv \, 4\,$. Moreover, we can note the reduction of the order of the solution, using the Spectral approach. According to the previous results, the fidelity of the model is not degraded but only the order of the solution. Besides, $N_{\,t}$ can not be reduced due to accuracy issues.


\subsubsection{Solving the system of ODEs}

The time spent on simulations is also related to the solver used to compute system of ordinary differential equations. For the weakly nonlinear case, different solvers were employed with different values of tolerances. Between all \texttt{Matlab} ODE solvers,  the \texttt{ODE15s} was the most efficient, combining accuracy and speed. By decreasing the tolerance of the solver, more accurate results can be obtained. But sometimes, if we improve the precision of the solver, the error can be limited by the magnitude of the last spectral coefficient. Thus, another way to get more precise solutions is through increasing the number of modes.

Therefore, depending on the accuracy sought on the results, several options are available. The choice of the ODE solver is related to the nature of the problem. For example, if the problem has two components which vary drastically on different time scales, then the problem is stiff, or difficult to evaluate. The solvers are then classified according to the problem type. For non-stiff problems, \texttt{ODE45}, \texttt{ODE23} and \texttt{ODE113} are the most appropriate, but for stiff problems, the other ODE solvers are recommended (\texttt{ODE15s}, \texttt{ODE23s}, \texttt{ODE23t} and \texttt{ODE23tb}). Further information can be found in \cite{Shampine1997}.


\section{Treating general nonlinearities}
\label{sec:spectral_nonlinear}

Problem~\eqref{eq:moisture_dimensionlesspb_1D} has an important difficulty in dealing with the nonlinearities of the moisture storage coefficient $\cm$ and of the diffusion coefficient $\dm\,$, both depending on the moisture content field. These coefficients are usually given by empirical functions from experimental data. Due to those nonlinearities, some modifications in the way of using the Spectral method have to be taken into account. For this reason, Eq.~\eqref{eq:moisture_governing_dimensionless} is recalled with a simplified notation:
\begin{align}\label{eq:diffusion_nonlinear}
  \cm \, (\, u \, ) \, \pd{u}{t} &\egal \pd{}{x} \left[ \, \dm \, (\, u \, ) \, \pd{u}{x} \, \right] \,.
\end{align}
In order to better apply the Spectral method, Eq.~\eqref{eq:diffusion_nonlinear} is rearranged as follows:
\begin{align}\label{eq:rearang_diff_nonlin}
  \pd{u}{t} & \egal \nu \, (\, u \, ) \, \pd{^{\,2}u}{x^{\,2}} \plus \lambda \, (\, u \, )\, \pd{u}{x}\,,
\end{align}
where,
\begin{align*}
  \nu \, \bigl(\, u \, \bigr) \eqdef \dfrac{\dm\, (\, u \,)}{\cm\, (\, u \,) } & & \text{and}   & &
  \lambda \, \bigl(\, u \, \bigr) \eqdef \dfrac{1}{\cm\, (\, u \,) }  \cdot  \dfrac{\mathrm{d} \Bigl(\, \dm\, (\,u\,)\, \Bigr)}{\mathrm{d} x}\,.
\end{align*}

By using Spectral methods the unknown $u\,(\,x\,,t\,)$ is approximated by the finite sum \eqref{eq:series_ap} with \textsc{Chebyshev} polynomials as basis functions. The derivatives are written as in the linear case, by Eqs.~\eqref{eq:derivative1}, \eqref{eq:derivative2} and \eqref{eq:derivative3}. Thus, substituting them into Eq.~\eqref{eq:rearang_diff_nonlin} gives:
\begin{align*}
  \sum_{i\, = \, 0}^{n} \,  \dot{a}_{\,i}\, (\,t\,)\, \T_{\,i}\, (\,x\,) \egal \nu \, \Biggl(\, \sum_{i\, = \, 0}^{n} \, a_{\,i} \,(\,t\,)\, \T_{\,i}\,(\,x\,) \, \Biggr) \, \sum_{i\, =\, 0}^{n} \, \Tilde{\Tilde{a}}_{\,i}\, (\,t\,)\, \T_{\,i}\, (\,x\,) \plus \nonumber \\ 
  \lambda \, \Biggl( \, \sum_{i\, = \, 0}^{n} \, a_{\,i}\, (\,t\,)\, \T_{\,i}\, (\,x\,)\,  \Biggr) \sum_{i\, =\, 0}^{n}\, \tilde{a}_{\,i}\,(\,t\,)\, \T_{\,i}\,(\,x\,) \,.
\end{align*}

By applying the \textsc{Galerkin} projection we have:
\begin{align}\label{eq:expres_simplif_res}
  \M \cdot \dot{a}_{\,i}\, (\,t\,) \egal \mathrm{G}_{\, i,\, j}\, \Bigl(\, \{a_{\,i}\} \,\Bigr) \cdot \Tilde{\Tilde{a}}_{\,i}\, (\,t\,) \plus \Lambda_{\, i,\, j} \, \Bigl(\, \{a_{\,i}\} \,\Bigr) \cdot \Tilde{a}_{\,i}\, (\,t\,) \,,
\end{align}
where,
\begin{align*}
  \mathrm{G}_{\,i,\, j}\, \Bigl(\, \{a_{\,i}\} \,\Bigr) &\egal \int_{-1}^{1} \, \dfrac{\nu\, (\, \sum\, )\, \T_{\,i}\, (\,x \,) \, \T_{\,j}\, (\,x \,) }{\sqrt{1 \moins x^{\,2}}} \, \mathrm{d}x \,, \\
  \Lambda_{\,i,\, j}\, \Bigl(\, \{a_{\,i}\} \,\Bigr) &\egal \int_{-1}^{1} \, \dfrac{\lambda\, (\, \sum\, )\, \T_{\,i}\, (\,x \,) \, \T_{\,j}\, (\,x \,) }{\sqrt{1 \moins x^{\,2}}} \,  \mathrm{d}x \,.
\end{align*}
Using the \textsc{Chebyshev--Gau}\ss ~quadrature, the integrals are also approximated by a finite sum:
\begin{align*}
  \mathrm{G}_{\,i,\, j}\, \Bigl(\, \{a_{\,i}\} \,\Bigr) &\ \approx\ \dfrac{\pi}{m} \ \sum_{k\, =\, 1}^{m}\, \nu_{\,k} \ \T_{\,i}\, (\,x_{\,k}\,)\, \T_{\,j}\, (\,x_{\,k}\,)\,, \\
  \Lambda_{\,i,\, j}\, \Bigl(\, \{a_{\,i}\} \,\Bigr) &\ \approx\ \dfrac{\pi}{m} \ \sum_{k\, =\, 1}^{m}\, \lambda_{\,k} \ \T_{\,i}\, (\,x_{\,k}\,)\, \T_{\,j}\, (\,x_{\,k}\,)\,,
\end{align*}
where, 
\begin{align*}
  \nu_{\,k}\ &\eqdef\ \nu \,\Biggl( \, \sum_{i\, =\, 0}^n \,a_{\,i}\,(\,t\,)\, \T_{\,i}\,(\,x_{\,k}\,) \, \Biggr) \,,\\
  \lambda_{\,k}\ &\eqdef\ \lambda \,\Biggl( \, \sum_{i\, =\, 0}^n \,a_{\,i}\,(\,t\,)\, \T_{\,i}\,(\,x_{\,k}\,) \, \Biggr) \,,
\end{align*}
and $x_{\,k}$ are the \textsc{Chebyshev} nodes:
\begin{align*}
  x_{\,k} \egal \cos\, \Biggl( \, \dfrac{2\, k \moins 1}{2\, m}\, \pi \, \Biggr) \,, & & k \egal 1, \,2, \,\ldots,\, m \,.
\end{align*}
The value of $m$ is determined according to numerical investigations and will be discussed for the next case study.

In addition, we have the expressions of the nonlinear boundary conditions:
\begin{subequations}
\begin{align}
  \dm \, \Biggl(\, \sum_{i\,=\,0}^{n} \,a_{\,i}\,(t)\, (-1)^{\,i}\, \Biggr)\, \sum_{i\, =\, 0}^{n}\, \tilde{a}_{\,i}\, (t)\, (-1)^{\,i} \moins \BivL \, \sum_{i\,=\,0}^{n}\, a_{\,i}\,(t)\, (-1)^{\,i} \plus \BivL \cdot \uL &\egal 0\,,\label{eq:spectr_bc1_1mat} \\
  -\ \dm \, \Biggl(\, \sum_{i\,=\,0}^{n}\, a_{\,i}\,(t)\, \Biggr)\,  \sum^n_{i\, =\, 0}\, \tilde{a}_{\,i}\, (t)\,  \moins \BivR \, \sum^n_{i\, =\, 0}\, a_{\,i}\,(t)\, \plus \BivR \cdot \uR &\egal 0 \,.  \label{eq:spectr_bc2_1mat} 
\end{align}
\end{subequations}
Different from the linear case, the boundary conditions cannot provide an explicit expression for the two last coefficients $a_{\,n}\,(\,t\,)$ and $a_{\,n-1}\,(\,t\,)\,$. Thus, it is not possible to compute the solution in the same way. Although, with all elements listed before, it is possible to set the system to be solved by composing a system of ODEs with two additional algebraic expressions for the boundary conditions. It results in a system of Differential--Algebraic Equations (DAEs) with the following form:
\begin{align}\label{eq:system_DAE}
  \M \, \dot{a}_{\,n}\,(\,t\,) \egal \A \, a_{\,n}\,(\,t\,) \plus \b\,(\,t\,) \,,
\end{align}
where, $\M$ is a diagonal and singular matrix ($\mathrm{rank}\,(\,\M\,)\,=\,n\ -\ 2$) containing the coefficients of the \textsc{Chebyshev} weighted orthogonal system, $\b\,(\,t\,)$ is a vector containing the boundary conditions and, $\A \cdot a_{\,n}\,(\,t\,)$ is composed by the right member of Eq.~\eqref{eq:expres_simplif_res}. The initial condition is given by Eq.~\eqref{eq:system_ODE_int} and the DAE system is solved by \texttt{ODE15s} or \texttt{ODE23t} from \texttt{Matlab}.


\subsection{A highly nonlinear case} \label{sec:case_strong_nonlinear}

\begin{figure}
\centering
\subfigure[a][\label{fig_AN3:cM}]{\includegraphics[width=0.42\textwidth]{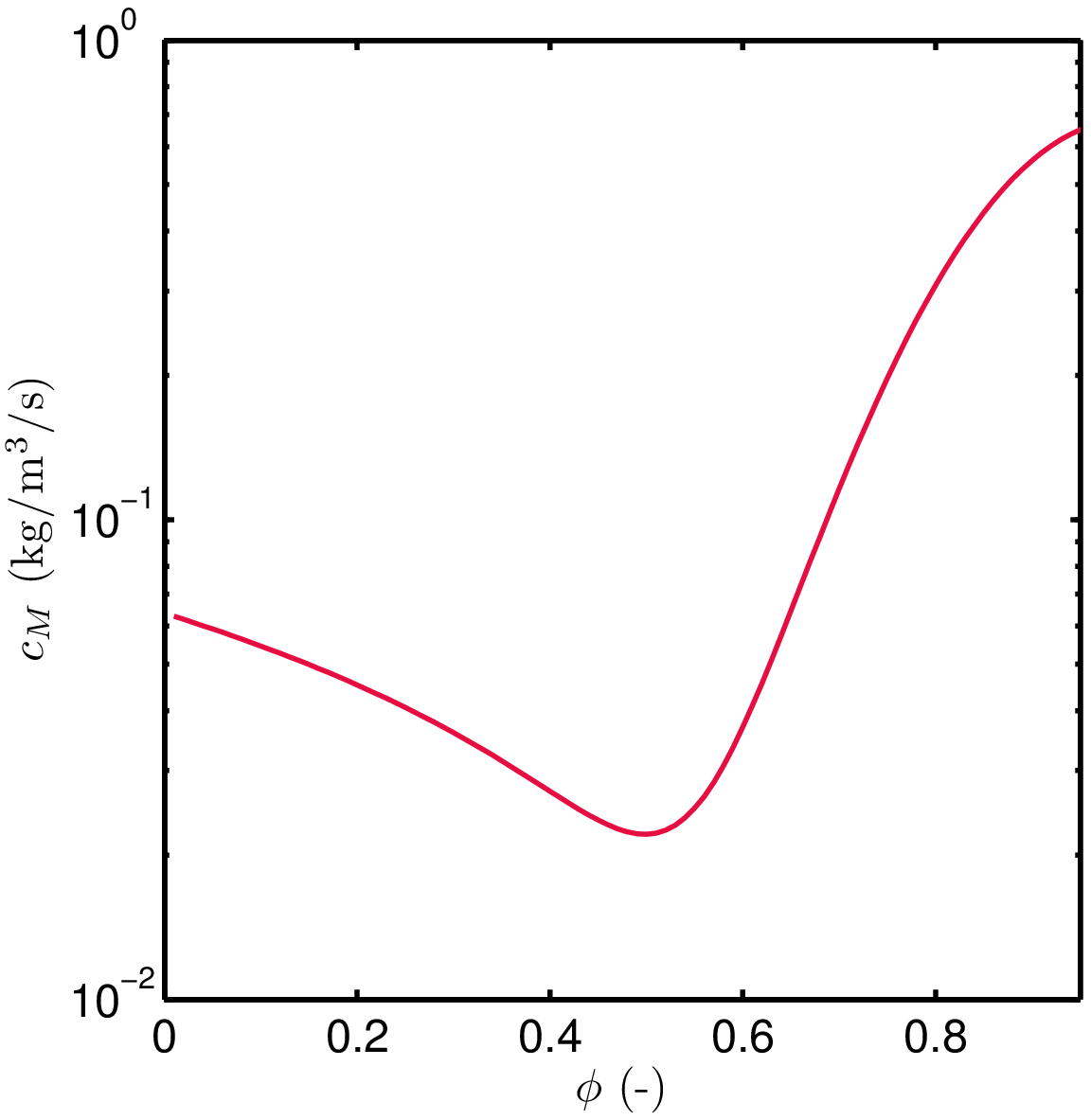}}
\subfigure[b][\label{fig_AN3:dM}]{\includegraphics[width=0.45\textwidth]{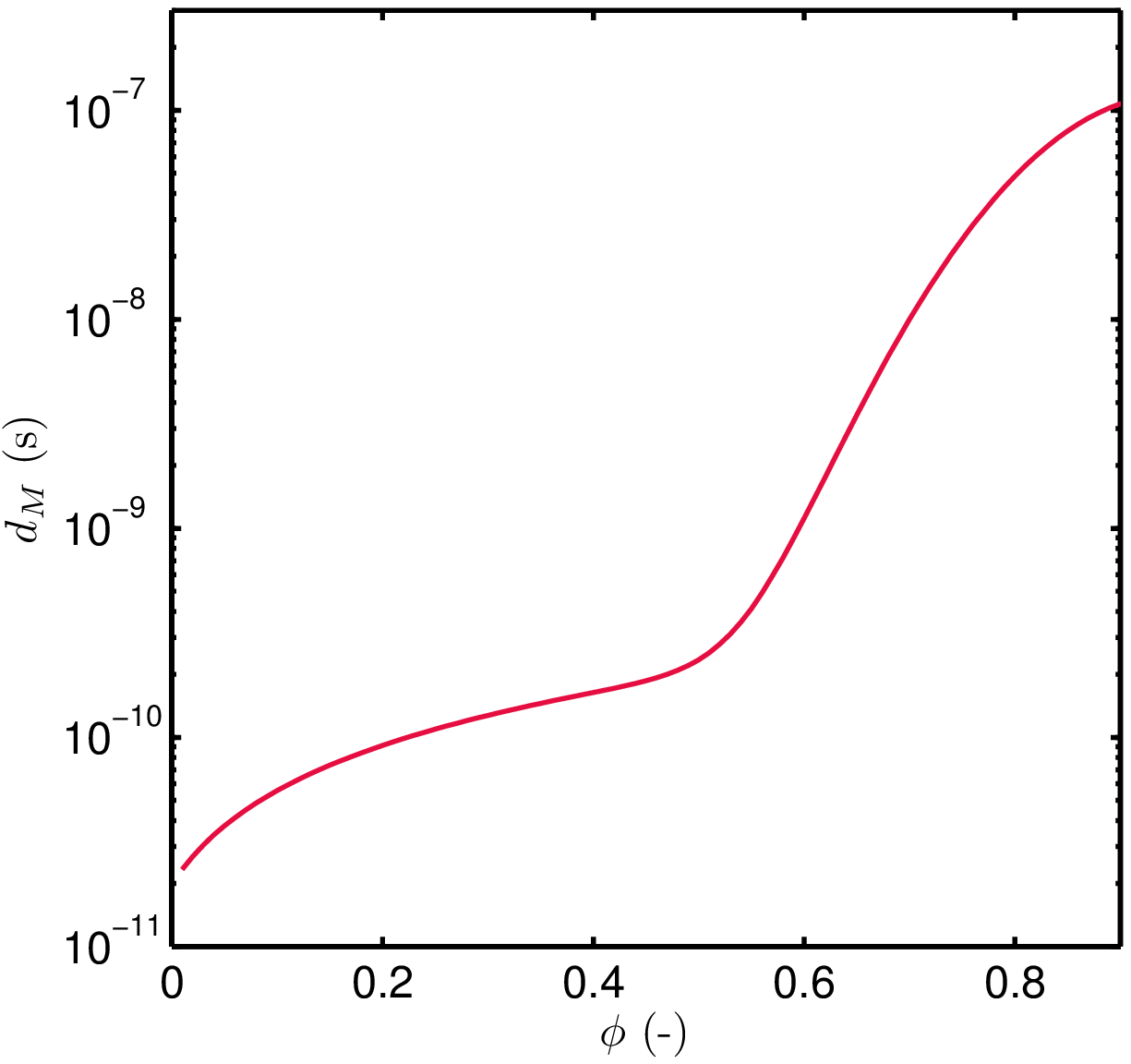}}
\caption{\small\em Variation of the moisture storage $\cm$ (a) and diffusion $\dm$ (b) as a function of the relative humidity $\phi\,$.}
\end{figure}

\begin{figure}
\centering
\includegraphics[width=0.47\textwidth]{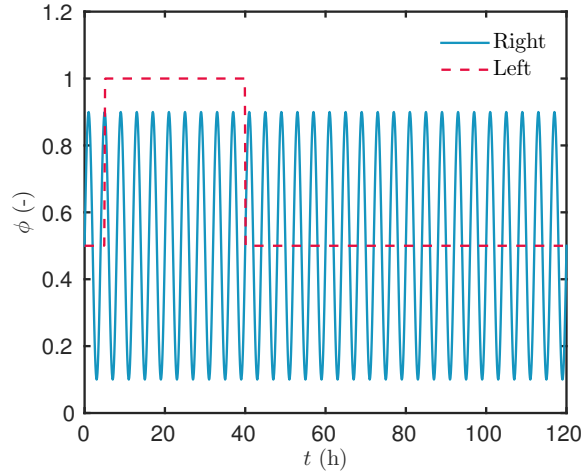}
\caption{\small\em Boundary conditions at the left side $(\,x \, = \, 0\,\mathsf{m}\,)$ and at the right side $(\,x \, = \, 0.1\,\mathsf{m}\,)\,$.}
\label{fig_AN3:BC}
\end{figure}

This case study considers moisture dependent coefficients $\cm$ and $\dm$, illustrated in Figures~\ref{fig_AN3:cM} and \ref{fig_AN3:dM}. Their variations are similar to the load bearing material from \cite{Janssen2014}. The initial vapour pressure is uniform $\Pvi \, =\, 1.16 \dix{\,3}$  $\mathsf{Pa}$. No moisture flow is taken into account at the boundaries. The ambient vapour pressures at the boundaries are illustrated in Figure~\ref{fig_AN3:BC}. At the left boundary, it has a fast drop until the saturation state and at the right boundary, it has a sinusoidal variation. The material is thus excited until the capillary state. The convective vapour transfer coefficients are set to $\hvL \, =\, 2 \cdot 10^{\,-7}$  $\mathsf{s/m}$ and $\hvR \, =\, 3 \cdot 10^{\,-8}$  $\mathsf{s/m}$ for the left and right boundary, respectively. The simulation is performed for $120\, \mathsf{h}\,$. As in the previous case study, the dimensionless values can be found in Appendix~\ref{annexe:dimensionless}.

The Spectral method is composed by $N \, = \, 9$ modes with $m \, = \, 7\,$. The \texttt{ODE15s} was used to solved System~\eqref{eq:system_DAE}, with a tolerance set to $10^{\,-4}\,$. For this case, the Spectral method was compared to the \CN ~\cite{Gasparin2017} and to a reference solution computed using the \texttt{Chebfun} \texttt{Matlab} toolbox \cite{Driscoll2014}. All solutions have been computed with the following discretisation parameters: $\Delta \ts \,=\, 10^{\,-2}$ and $\Delta \xs \,=\, 10^{\,-2}\,$.

\begin{figure}
\centering
\subfigure[a][\label{fig_AN3:Evolution}]{\includegraphics[width=0.45\textwidth]{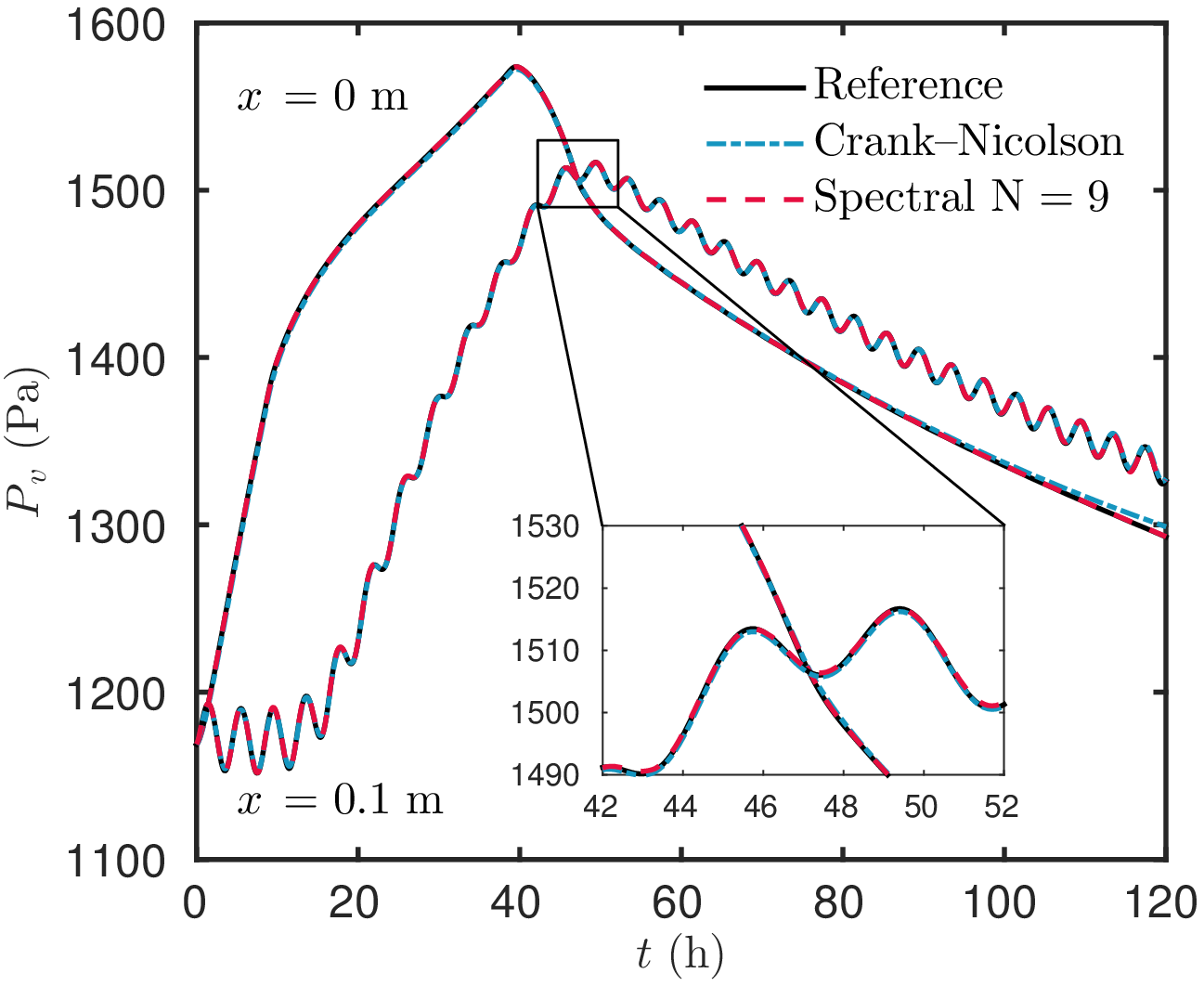}}
\subfigure[b][\label{fig_AN3:Profiles}]{\includegraphics[width=0.45\textwidth]{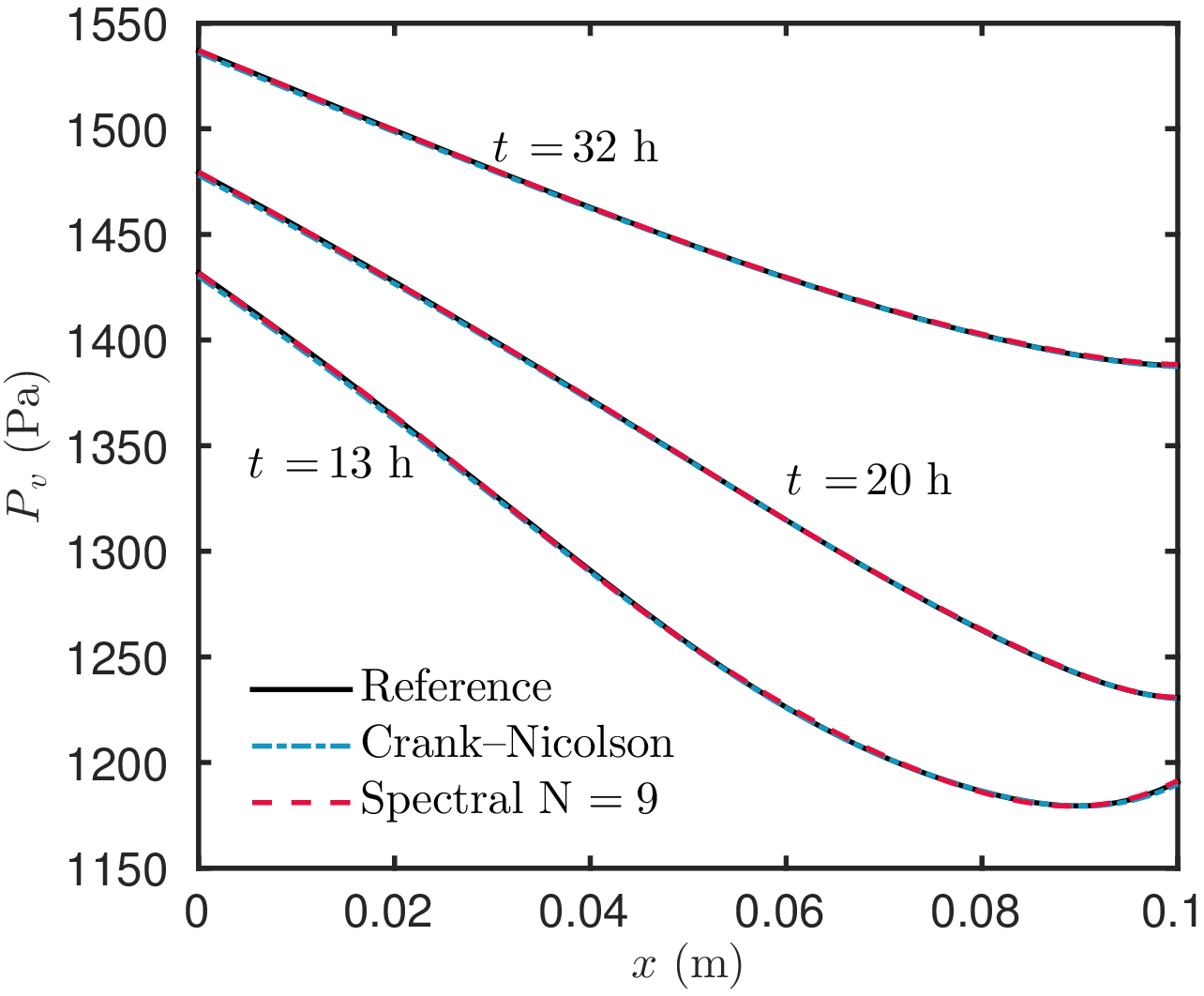}}
\caption{\small\em Evolution of the vapour pressure at the boundaries, $x\, \in\, \left\lbrace 0 \,, 0.1 \right\rbrace\, \mathsf{m}$ (a) and vapour pressure profiles for $t\, \in\, \left\lbrace 13 \,, 20 \,, 32 \right\rbrace\, \mathsf{h}$ (b).}
\end{figure}

Vapour pressure variations in the boundaries are shown in Figure~\ref{fig_AN3:Evolution}. The vapour pressure at $x \,=\, 0.1\, \mathsf{m}$ slowly oscillates according to the right boundary condition. It also increases within the material according to the step imposed at the left boundary $x \,=\, 0\, \mathsf{m}$. This increase can be also observed on three profiles of vapour pressure illustrated in Figure~\ref{fig_AN3:Profiles}, in which the diffusion process is represented going from left to right.

\begin{figure}
\centering
\subfigure[a][\label{fig_AN3:Error_fx}]{\includegraphics[width=0.45\textwidth]{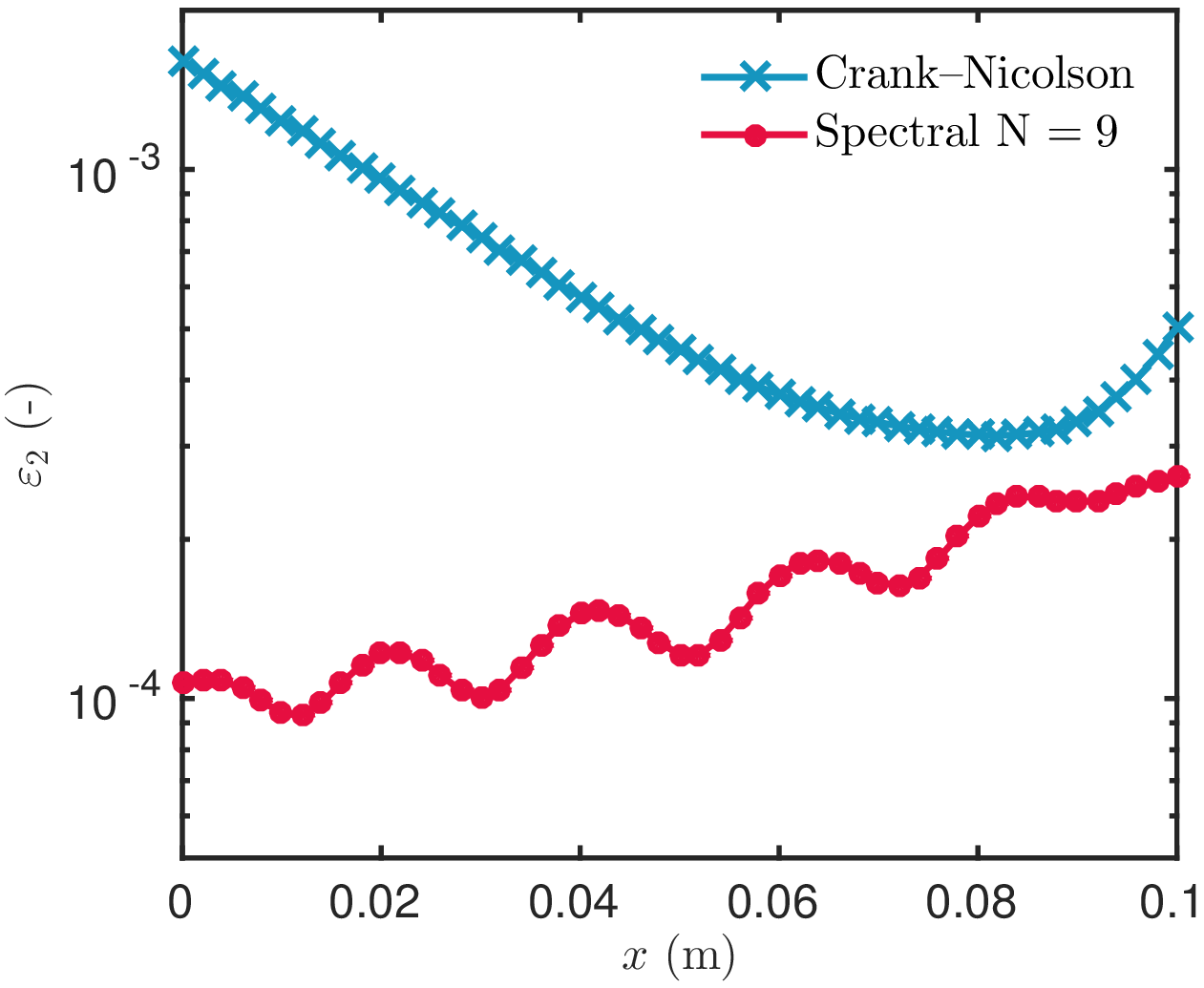}}
\subfigure[b][\label{fig_AN3:error_l2_all}]{\includegraphics[width=0.45\textwidth]{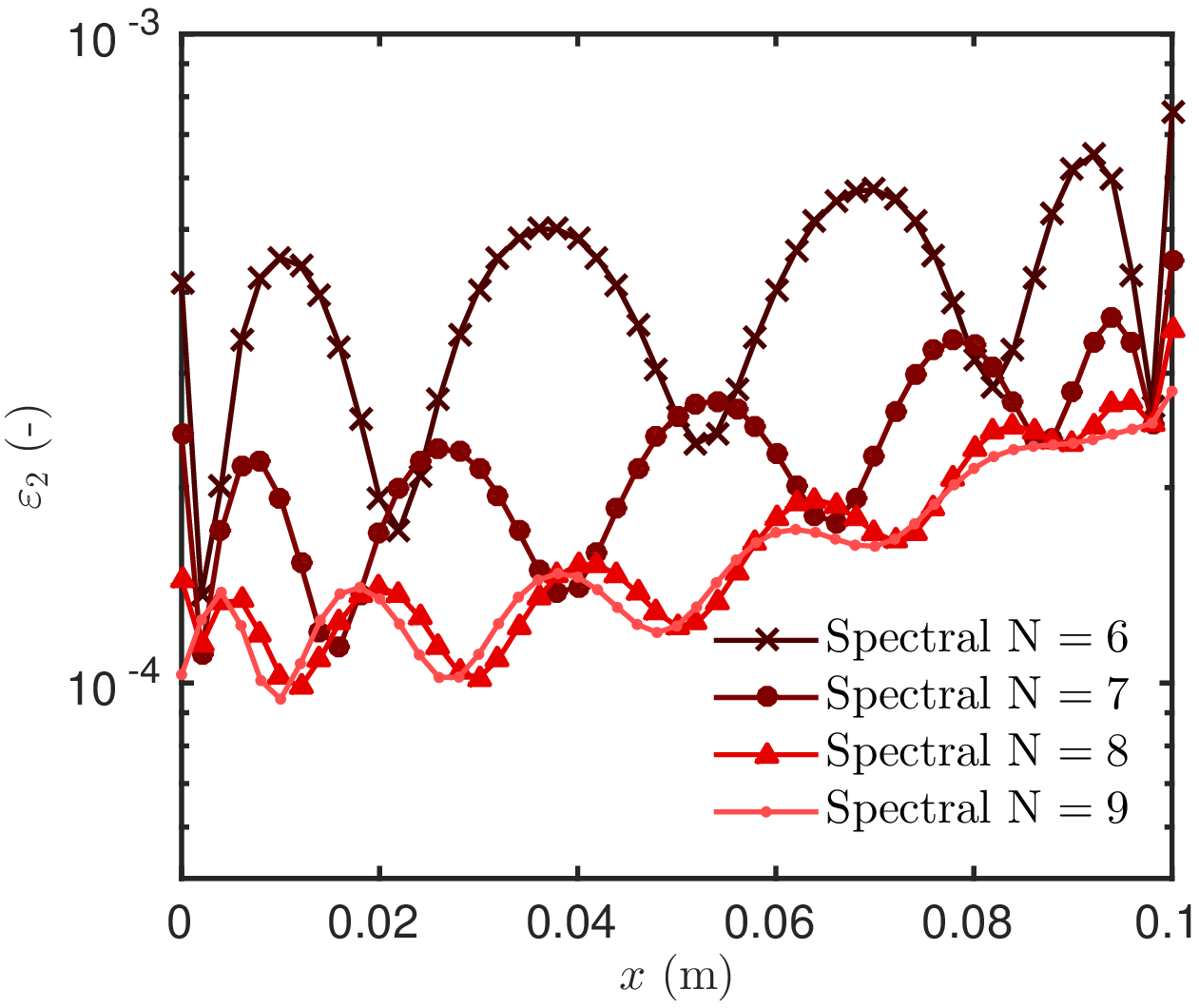}}
\caption{\small\em Error $\varepsilon_{\,2}$  computed for the \CN ~method and for the Spectral with $N\, =\,9$ modes (a), and the error $\varepsilon_{\,2}$ computed for the Spectral method with different number of modes (b).}
\end{figure}

All methods have demonstrated good agreement to represent the physical phenomenon. Again, the fidelity of the model does not deteriorate with the use of a Spectral approach. Results of the error $\varepsilon_{\,2}$ in function of $x$ are shown in Figure~\ref{fig_AN3:Error_fx}. The error of the \CN ~scheme is proportional to $\O\,(\,\Delta t^{\star \, 2}\,)\,$. The Spectral method with $N \,=\, 9$ modes is one order of magnitude more accurate and faster than the \CN ~method, even considering the same discretization parameters $\Delta \ts$ and $\Delta \xs\,$. Although, if we decrease the number of modes to $N \,=\, 6$ and maintaining the same discretization parameters $\Delta \ts$ and $\Delta \xs\,$, we reach the same order of accuracy of the \CN ~method, as observed in Figure~\ref{fig_AN3:error_l2_all}.

The solution of the Spectral methods becomes more accurate with the increase of the number of modes, as shown in Figure~\ref{fig_AN3:error_l2_all}. With $6$ modes, we have satisfactory results, with the error of the order of $\O\,(\,10^{\,-3}\,)\,$. As we increase only the number of modes, without changing other parameters, the error begins to stabilize, and with $8$ and $9$ modes the error remains the same.

As already observed in the linear case, the Spectral method does not depend on the number of spatial points, but on the order of the ODE solver tolerance and also on the number of modes. For the nonlinear case, the error also depends on the truncation of the sum $\sum_{k\, =\,1}^{m}\,$. For this reason, the error $\varepsilon_{\, \infty}$ in function of $m$ is shown in Table~\ref{tab:m_values}. The optimal $m$ number is approximated by numerical experimentation, and as can be seen in the Table~\ref{tab:m_values}, the best value for $m$ is the one equivalent to the number of modes.

\begin{table}
\centering
\small
\setlength{\extrarowheight}{.3em}
\begin{tabular}{lccccc}
\hline
& $N \egal 5$ & $N \egal 6$ & $N \egal 7$ & $N \egal 8$ & $N \egal 9$ \\ 
\hline
$m \egal 2$  & $7.0 \cdot 10^{\,-2}$  &   ---   &  $2.90\cdot 10^{\,-1}$  &   ---   &     ---     \\
$m \egal 3$  & $2.8\cdot 10^{\,-3}$  & $2.02\cdot 10^{\,-2}$  &  $1.94\cdot 10^{\,-2}$  &   ---   &     ---     \\
$m \egal 4$  & $2.7\cdot 10^{\,-3}$  & $1.43\cdot 10^{\,-3}$  &  $3.06\cdot 10^{\,-3}$  &   $2.46\cdot 10^{\,-2}$   &  $2.89\cdot 10^{\,-1}$  \\
$m \egal 5$  & $2.6\cdot 10^{\,-3}$  & $1.54\cdot 10^{\,-3}$  &  $9.30\cdot 10^{\,-4}$  &   $1.05\cdot 10^{\,-3}$   &  $4.13\cdot 10^{\,-3}$  \\
$m \egal 6$  & $2.6\cdot 10^{\,-3}$  & $1.39\cdot 10^{\,-3}$  &  $7.07\cdot 10^{\,-4}$  &   $3.40\cdot 10^{\,-4}$   &  $4.41\cdot 10^{\,-4}$  \\
$m \egal 7$  & $2.6\cdot 10^{\,-3}$  & $1.40\cdot 10^{\,-3}$  &  $6.59\cdot 10^{\,-4}$  &   $3.20\cdot 10^{\,-4}$   &  $2.60\cdot 10^{\,-4}$   \\
$m \egal 8$  & $2.6\cdot 10^{\,-3}$  & $1.39\cdot 10^{\,-3}$  &  $6.70\cdot 10^{\,-4}$  &   $3.20\cdot 10^{\,-4}$   &  $1.90\cdot 10^{\,-4}$   \\
$m \egal 9$  &  ---     &   ---   &     ---   &   $3.09\cdot 10^{\,-4}$   &  $2.40\cdot 10^{\,-4}$   \\
$m \egal 10$ &  ---     &   ---   &     ---   &   $3.09\cdot 10^{\,-4}$   &  $2.40\cdot 10^{\,-4}$   \\
\hline
\end{tabular}
\bigskip
\caption{\small\em Absolute error $\varepsilon_{\, \infty}$ for different number of modes $N$ and different truncations $m\,$.}\bigskip
\label{tab:m_values}
\end{table}

\begin{figure}
\centering
\subfigure[a][\label{fig_AN3:coefA_spc_first3}]{\includegraphics[width=0.45\textwidth]{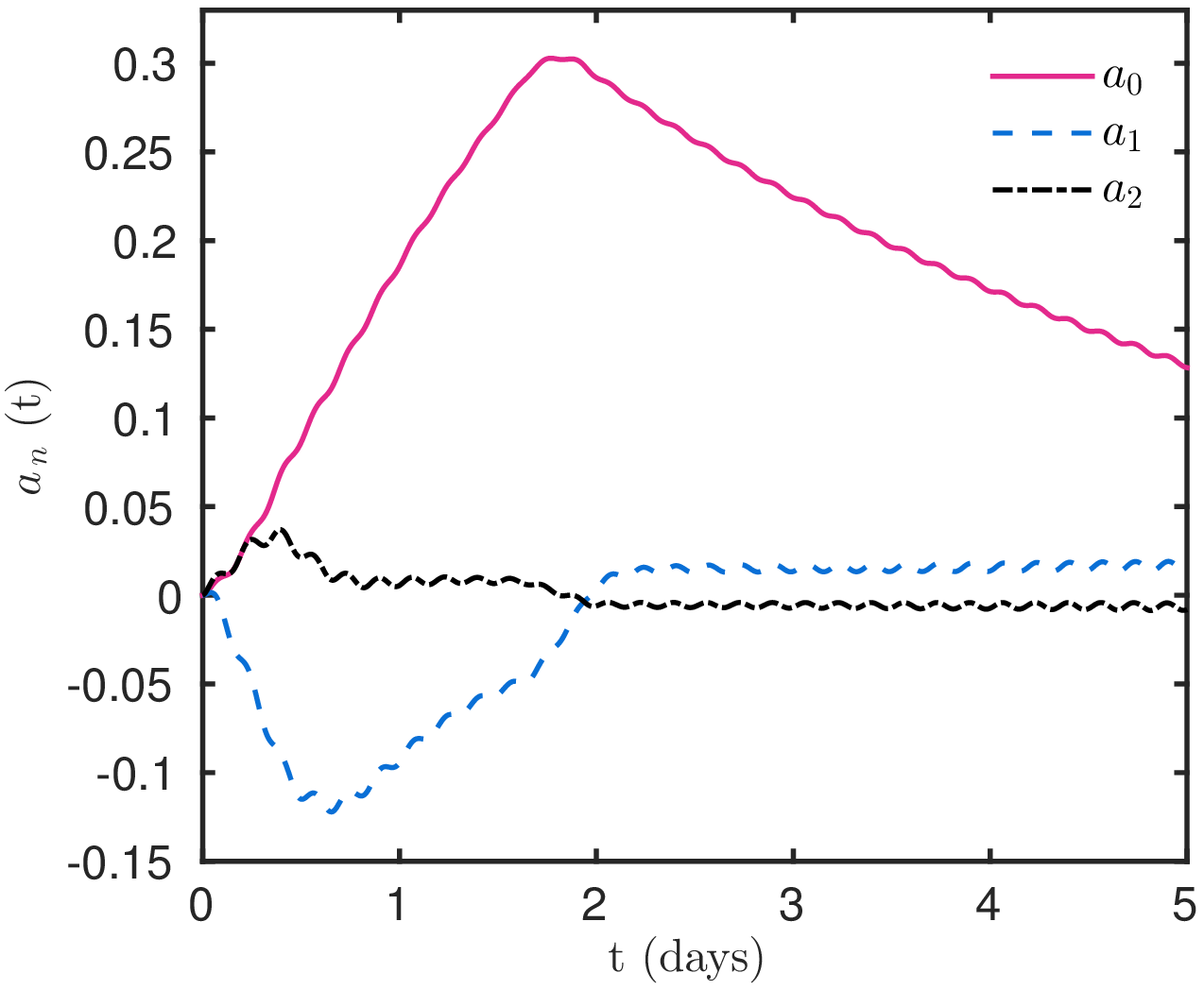}}
\subfigure[b][\label{fig_AN3:coefA_spc_last3}]{\includegraphics[width=0.45\textwidth]{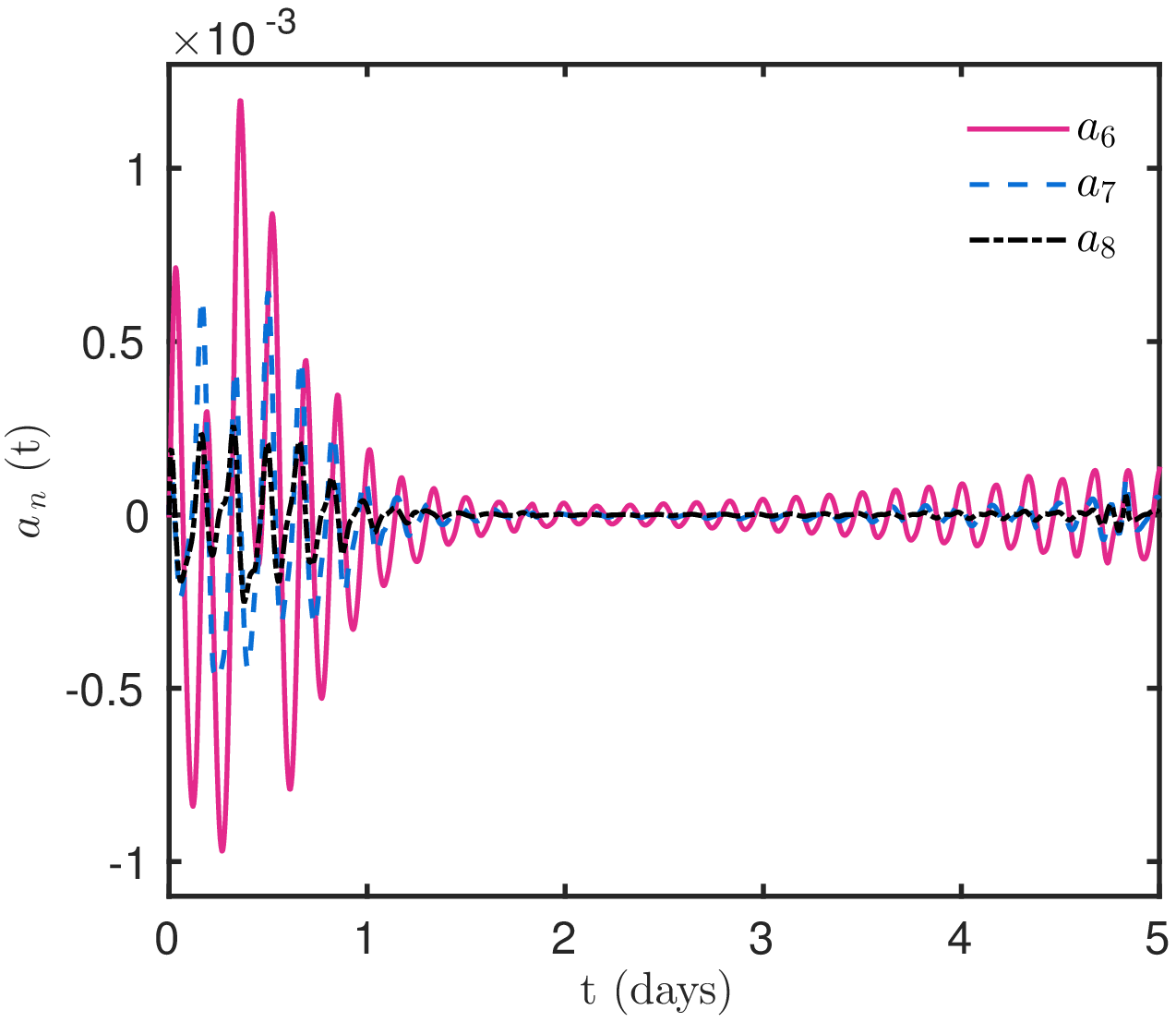}}
\caption{\small\em Evolution of the  first three (a) and of the last three (b) spectral coefficients $a_{\,n}\,$.}
\end{figure}

Figures~\ref{fig_AN3:coefA_spc_first3} and \ref{fig_AN3:coefA_spc_last3} represent the first and last three coefficients $a_{\, n}$ of the Spectral--ROM solution. The step in the left boundary can be also seen in these figures for the first days, and after that, the values tend to stabilize. It is possible to see the reduction in the magnitude of the coefficient with the increase of the number of coefficients. As for the previous cases, the last coefficients are always the smallest ones.

\begin{figure}
\centering
\subfigure[a][\label{fig_AN3:error_fmodes}]{\includegraphics[width=0.45\textwidth]{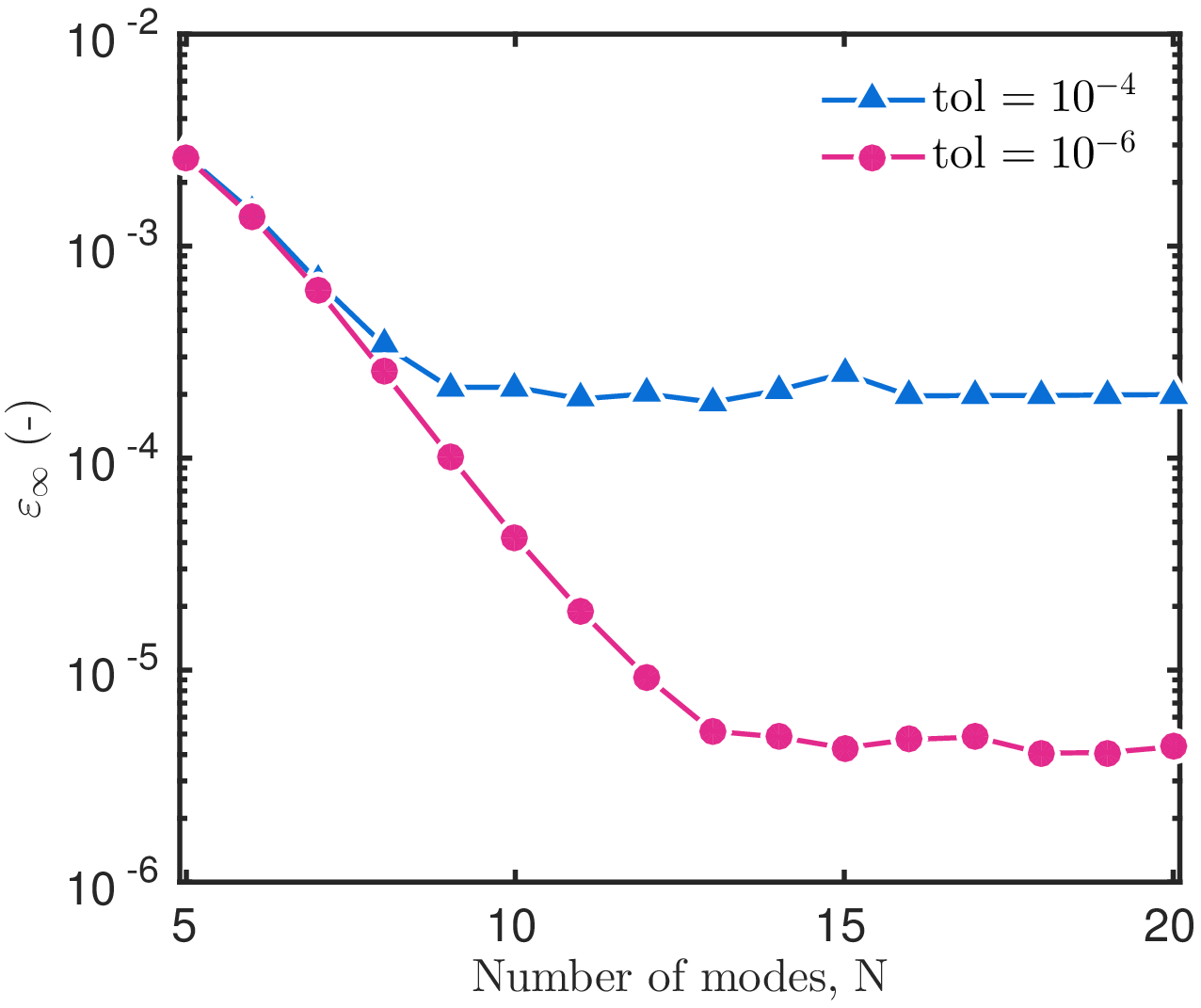}}
\subfigure[b][\label{fig_AN3:CPU_time}]{\includegraphics[width=0.432\textwidth]{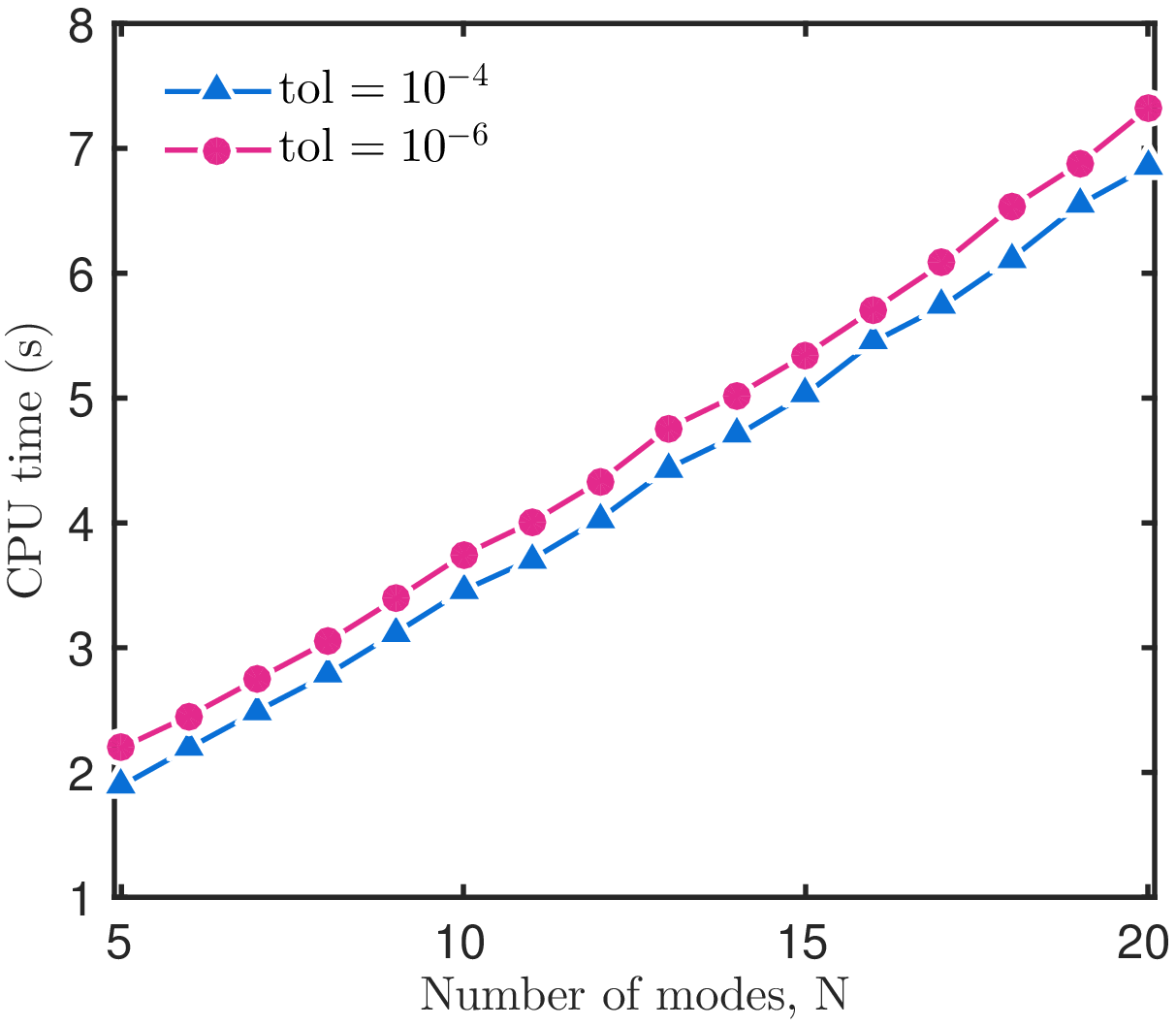}}
\caption{\small\em Maximum absolute error as a function of the number of spectral modes (a) and their respective CPU simulation time (b).}
\end{figure}

A parametric study is performed in order to verify the computational cost of the proposed method. The discretisation parameters are set to $\Delta \xs \,=\, 10^{\,-2}$ and $\Delta \ts \,=\, 10^{\,-2}\,$, while the number of modes $N$ of the Spectral solution and the tolerance of the solver vary. Figure~\ref{fig_AN3:error_fmodes} presents the maximum absolute error $\varepsilon_{\, \infty}$ in function of the number of spectral modes. As we increase the number of modes, the solution gets more accurate. Although, after a certain number of modes, the solution converges to a minimum value, that is related to the tolerance of the ODE solver. The time to perform each spectral simulation is presented in Figure~\ref{fig_AN3:CPU_time}. For this numerical application, the CPU time has been evaluated using \texttt{Matlab} platform on a computer with Intel i7 CPU and 8GB of RAM. The computational effort to perform the simulation increases linearly with the number of modes. However, it remains extremely low. To better appreciate the computational cost of each approach, Table~\ref{tab:CPU_time} provides the CPU time to compute the solution using the \CN ~scheme, the \texttt{Chebfun} toolbox for the same discretisation parameters. The Spectral solution has been computed with $N \, = \,9$ modes. It is preferable to focus on the ratio of computer run time rather than on absolute values, that is system-dependent. Even with an average number of sub-iterations is $\,N_{\,\mathrm{NL}}\, \,\simeq\, \O\,(1)$ of the \CN ~scheme, the Spectral method is substantially faster than the other methods. It represents only $1\,\%$ of the CPU time needed using the \CN ~approach.

\begin{table}
\center
\small
\setlength{\extrarowheight}{.3em}
\begin{tabular}[l]{@{}lccc}
\hline
Numerical Scheme & CPU time ($\mathsf{s}$) & CPU time ($\%$) & Average number of iterations \\
\hline
Spectral  $N \,= \,9$   &  $3$    & $1$     & --- \\
\texttt{Chebfun} (Reference)            &  $96$   & $29$    & --- \\
\CN             &  $327$  & $100$   & 1 \\
\hline
\end{tabular}
\bigskip
\caption{\small\em Computational time required for the numerical schemes perform the nonlinear case ($\Delta \xs \,=\, 10^{-2}$ and $\Delta \ts \,=\, 10^{-2}$).} \bigskip
\label{tab:CPU_time}
\end{table}


\section{Multilayer domain}

In constructions, multiple layers are commonly found. The configuration assumed at the interface between materials follows the hydraulic continuity \cite{DeFreitas1996}, which considers inter-penetration of both porous structure layers. Both materials are homogeneous and isotropic, and only moisture transfer is simulated, through a perfectly airtight structure. The hydraulic continuity establishes that there must be a continuous moisture flow through the interface and a continuous distributions of vapour content:
\begin{subequations}\label{eq:interf}
\begin{align}
  P_{\,v,\,1}\,(x_{\,\text{int}},t) & \egal P_{\,v,\,2}\,(x_{\,\text{int}},t) \,, \label{eq:interf_cont_field} \\
  d_{\,m,\,1} \ \pd{P_{\,v,\,1}}{x}\Bigg|_{x_{\,\text{int}}} & \egal d_{\,m,\,2}  \ \pd{P_{\,v,\,2}}{x}\Bigg|_{x_{\,\text{int}}} \,, \label{eq:interf_cont_flow}
\end{align}
\end{subequations}
where $x_{\,\text{int}}$ represents the location of the interface between materials and subscripts $1$ and $2$ stand for Material $1$ and Material $2\,$, respectively.


\subsection{Adaptation of the reduced Spectral Method}

\begin{figure}
\begin{center}
\includegraphics[width=0.55\textwidth]{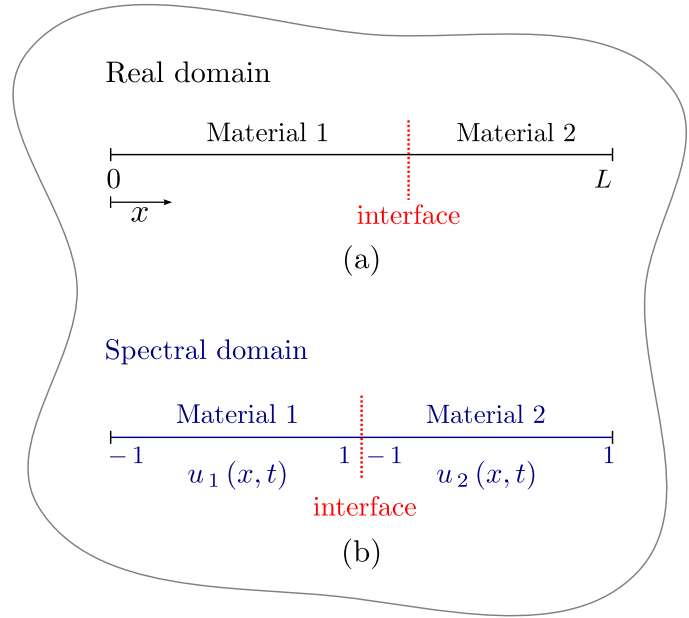}
\caption{\small\em Schematic representation of the domain division with the real domain (a) transformed linearly to obtain the spectral domain (b).}
\label{fig_AN4:layred_spectral}
\end{center}
\end{figure}

The original spatial domain $\Omega_{\,x} \egalb [\,0\,,\,L\,]$ is decomposed in two sub-domains $\Omega_{\,x,\,1} \egalb [\,0\,,x_{\,\text{int}}\,]$ and $\Omega_{\,x,\,2} \egalb [\,x_{\,\text{int}}\,,\,L\,]\,$, which represent each material surface. These sub-domains are linear transformed to the spectral domain $\bar{\Omega}_{\,x,\,1} \egalb [\,-1\,,1\,]$ and $\bar{\Omega}_{\,x,\,2} \egalb [\,-1\,,1\,]$ so they can fit within the interval of interest as illustrated in Figure~\ref{fig_AN4:layred_spectral}. From this, the unknown $u\,(\,x,\,t\,)$ is then defined as:
\begin{align*}
  u\, (\,x,\, t\,) \egal u_{\,1}\, (\,x,\, t\,)\ \cup\ u_{\,2}\, (\,x,\, t\,) \,,
\end{align*}
in which $u_{\,1}\, (x,\,t)$ is the solution defined over domain $\bar{\Omega}_{\,x,\,1}$ and $u_{\,2}\, (x,\,t)$ is the solution defined over domain $\bar{\Omega}_{\,x,\,2}\,$. Thus, $u_{\,1}$ and $u_{\,2}$ are written respectively as:
\begin{align*}
  u_{\,1}\, (\,x,\, t\,) \egal \sum_{i\, =\, 0}^{n} \, a_{\,i,\,1}\, (\,t\,)\, \T_{\,i}\, (\,x\,)  & & \text{and} & &
  u_{\,2}\, (\,x,\, t\,) \egal \sum_{i\, =\, 0}^{n} \, a_{\,i,\,2}\, (\,t\,)\, \T_{\,i}\, (\,x\,)\,,
\end{align*}
which represent the solution for Material $1$ and Material $2\,$, respectively. Note that the \textsc{Chebyshev} polynomials are always the same and the transformations always occur with the temporal coefficients.

The condition at the interface between two materials states the continuity of the fields and the flows. It implies that the derivative of the field $u$ is not continuous at the interface between two materials. This important remark has to be taken into account in the construction of the Spectral reduced order model. Indeed, the domain is decomposed in sub-domains to maintain a smooth solution and particularly a continuous derivative on each sub-domain. In this way, the model order reduction is optimal and ensure the error of the Spectral-ROM to decrease exponentially. It is totally possible to build the reduced order model considering the whole domain (without decomposition). However, the convergence is undermined since the solution and its derivatives are not smooth at the interface between two materials. More modes would be necessary to reach the same accuracy, as detailed in Theorem~$1$ of \cite[Chap. 4]{Trefethen2000}.

By considering the two materials, Eq.~\eqref{eq:expres_simplif_res} becomes:
\begin{align*}
\left( 
\begin{array}{cc}
\M & 0 \\
0 & \M
\end{array}
\right)
\left[ 
\begin{array}{c}
\dot{a}_{\,i,\,1} \\
\dot{a}_{\,i,\,2} 
\end{array}
\right] \egal
\left[ 
\begin{array}{c}
\mathrm{G}_{\, i,\, j,\,1}\, \Bigl( \{a_{\,i,\,1}\} \Bigr) \cdot \Tilde{\Tilde{a}}_{\,i,\,1}\, (t) + \Lambda_{\, i,\, j,\,1} \, \Bigl( \{a_{\,i,\,1}\} \Bigr) \cdot \Tilde{a}_{\,i,\,1}\, (t) \\
\mathrm{G}_{\, i,\, j,\,2}\, \Bigl( \{a_{\,i,\,2}\} \Bigr) \cdot \Tilde{\Tilde{a}}_{\,i,\,2}\, (t) + \Lambda_{\, i,\, j,\,2} \, \Bigl( \{a_{\,i,\,2}\}\Bigr) \cdot \Tilde{a}_{\,i,\,2}\, (\,t\,)
\end{array}
\right] +
\left[ 
\begin{array}{c}
b_{\,1}\, (t) \\
b_{\,2}\, (t) 
\end{array} 
\right]\,.
\end{align*}
The interface conditions --- Eqs.~\eqref{eq:interf_cont_field} and \eqref{eq:interf_cont_flow} --- are adimensionalized and written in the spectral form as:
\begin{subequations}
\begin{align*}
\beta_1 &\egal \sum^n_{i\, =\, 0}\, a_{\,i,\,1}\, (t)\,  \moins \sum^n_{i\, =\, 0} a_{\,i,\,2}\,(t)\, (-1)^{\,i}   \,, \\
\beta_2 &\egal d_{\,m,\,1} \Biggl(\, \sum_{i\,=\,0}^{n} \,a_{\,i,\,1}(t)\Biggr) \sum_{i\, =\, 0}^{n} \tilde{a}_{\,i,\,1}(t)\, \moins d_{\,m,\,2} \Biggl(\, \sum_{i\,=\,0}^{n} a_{\,i,\,2}(t)\, (-1)^{\,i}\Biggr) \sum_{i\, =\, 0}^{n} \tilde{a}_{\,i,\,2}(t)\, (-1)^{\,i} \,, 
\end{align*}
\end{subequations}
which are included in vector $b_{\,1}$ and set equal to zero. In the same way, the boundary conditions are written in the spectral form as:
\begin{subequations}
\begin{align*}
\alpha_1 &\egal -\ d_{\,m,\,2} \Biggl(\, \sum_{i\,=\,0}^{n}\, a_{\,i,\,2}(t) \Biggr) \sum^n_{i\, =\, 0} \tilde{a}_{\,i,\,2}\,(t) \moins \BivR \,\Biggl( \sum^n_{i\, =\, 0} a_{\,i,\,2}(t) \plus \uR\,(t)\Biggr) \,, \\ 
\alpha_2 &\egal d_{\,m,\,1} \Biggl(\, \sum_{i\,=\,0}^{n} a_{\,i,\,1}(t)(-1)^{\,i} \Biggr)\sum_{i\, =\, 0}^{n} \tilde{a}_{\,i,\,1} (t)(-1)^{\,i} \moins \BivL \Biggl(\, \sum_{i\,=\,0}^{n} a_{\,i,\,1}(t)(-1)^{\,i} + \uL\,(t)\Biggr) \,,
\end{align*}
\end{subequations}
which are included in $b_{\,2}$ and set equal to zero. Vectors $b_{\,1}$ and $b_{\,2}$ are column vectors of size $N\times 1$ with the form:
\begin{align*}
b_{\,1} \egal \left[
\begin{array}{c}
0\\
0\\
\vdots \\
0\\
\beta_1\\
\beta_2
\end{array} \right]
& & \text{and} &  &
b_{\,2} \egal \left[
\begin{array}{c}
0\\
0\\
\vdots \\
0\\
\alpha_1\\
\alpha_2
\end{array} \right] \,.
\end{align*}

With all elements listed before, it is possible to set the system to be solved. Different from the previous case, here the  system of ODEs has the double of the size $2\,N$ and it has \textbf{four} additional algebraic expressions for the boundary and interface conditions. The initial condition is also given by Eq.~\eqref{eq:system_ODE_int} and the DAE system is solved by \texttt{ODE15s} from \texttt{Matlab}. In this work, the approach was presented for a wall with two layers for the sake of clarity, knowing that it can be extended to any number of layers.


\subsection{A multilayer case}
\label{sec:case_multilayer}

\begin{figure}
\begin{center}
\includegraphics[width=.55\textwidth]{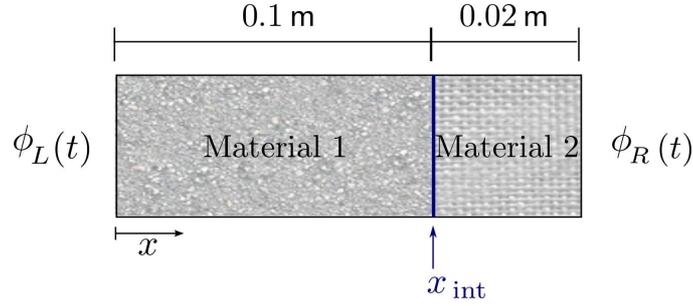}
\caption{\small\em Schematic representation of the two-layer wall.}
\label{fig_AN3:layred_configuration}
\end{center}
\end{figure}

\begin{figure}
\centering
\subfigure[a][\label{fig_AN4:cm}]{\includegraphics[width=0.45\textwidth]{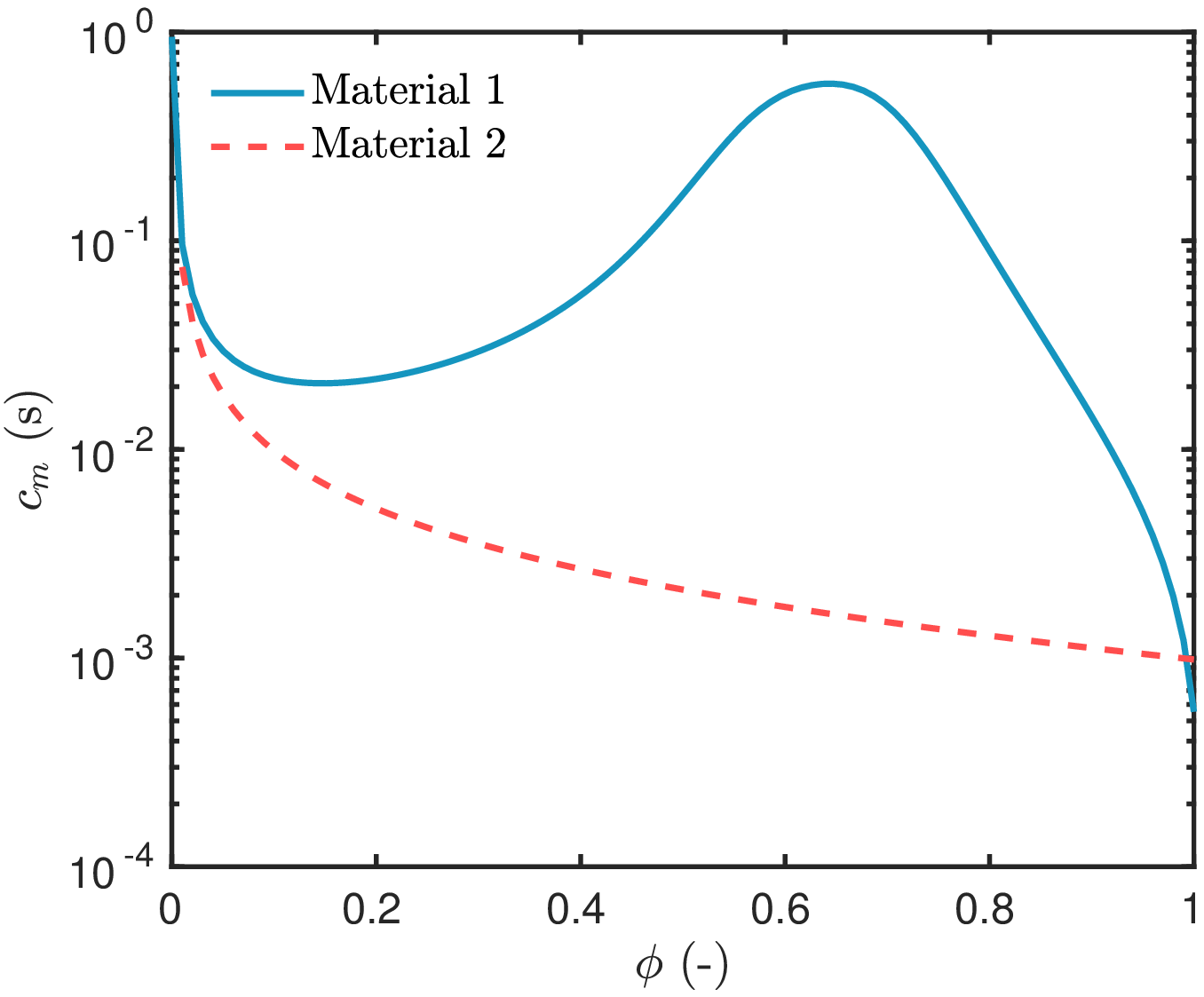}}
\subfigure[b][\label{fig_AN4:dm}]{\includegraphics[width=0.45\textwidth]{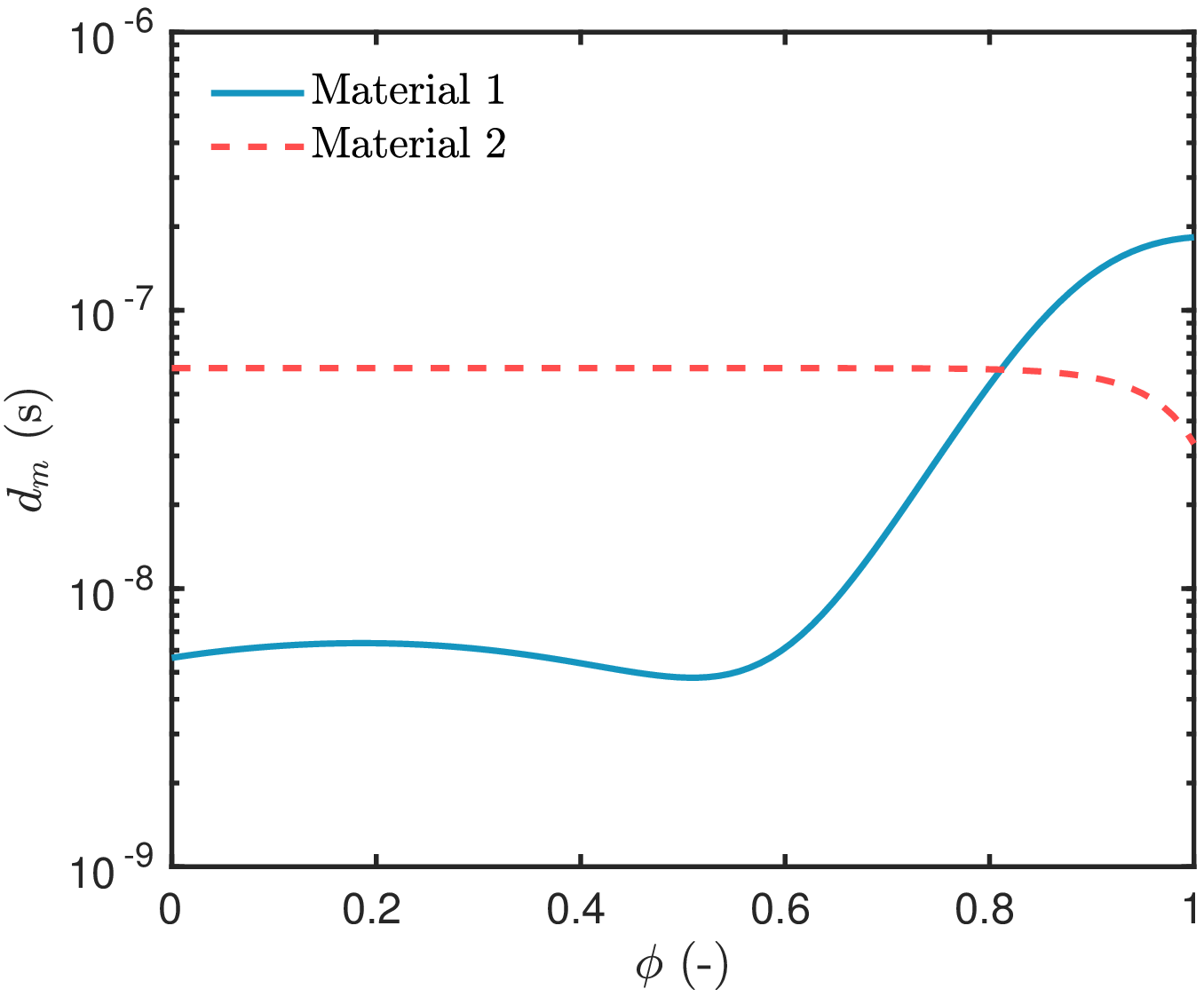}}
\caption{\small\em Variation of the moisture storage $\cm$ (a) and diffusion $\dm$ (b) as a function of the relative humidity $\phi\,$.}
\end{figure}

This case study considers a porous wall formed by $2$ layers: $10\ \mathsf{cm}$ of a load bearing material and $2\ \mathsf{cm}$ of a finishing material, as illustrated in Figure~\ref{fig_AN3:layred_configuration}. The selected materials complicate the case, with the first layer having a faster liquid transfer while the second layer acts as an hygroscopic finish. The properties used for these materials were obtained from \cite{Hagentoft2004} and are presented in Figures~\ref{fig_AN4:cm} and \ref{fig_AN4:dm}. Temperature dependence was neglected and transport coefficients modeled as a function of moisture content. Boundary and initial conditions are set with the same values as in the previous case study: initial vapour pressure of $\Pvi \egalb 1.16\cdot 10^{\,3}\ \mathsf{Pa}$ on both materials and boundary conditions represented in Figure~\ref{fig_AN3:BC}.

By using the Spectral approach, it is assumed that at the interface, the solution and the flow of the problem are continuous. In this way, the method will search for the solution that can satisfy both conditions. Simulations were performed using the \texttt{ODE15s}, with a tolerance set to $\mathsf{tol} \egalb 10^{\,-4}$ and with $N \egalb 10$ modes. These values were chosen based on the previous numerical study. The time is incremented with a discretization of $\Delta \ts \egalb 10^{\,-2}\,$, which is equivalent to $36\ \mathsf{s}\,$.

\begin{figure}
\begin{center}
\subfigure[][\label{fig_AN4:Evolution}]{\includegraphics[width=.45\textwidth]{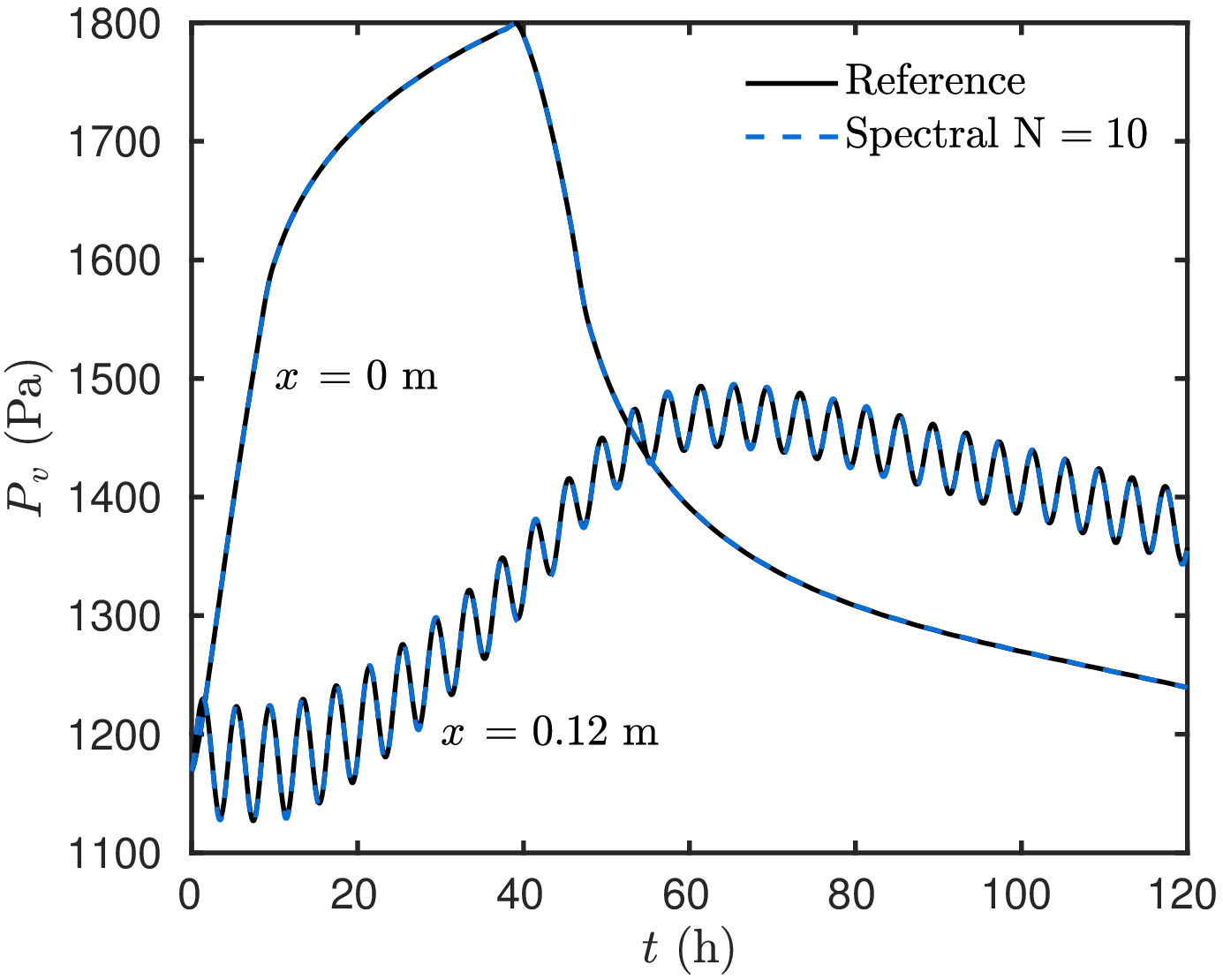}}
\subfigure[][\label{fig_AN4:Profiles}]{\includegraphics[width=.45\textwidth]{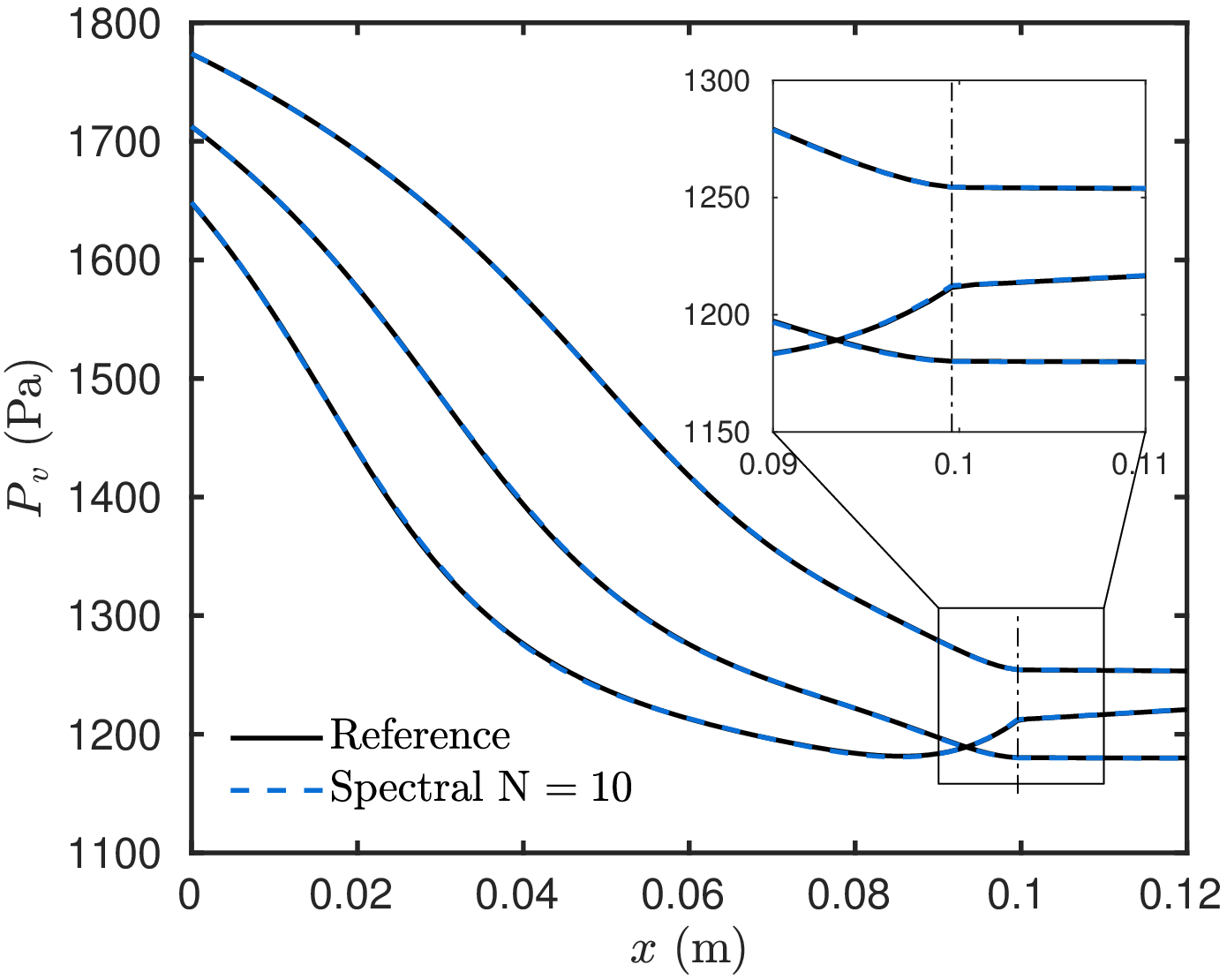}} 
\caption{\small\em Evolution of the vapour pressure at the boundaries, $x\, \in\, \left\lbrace 0 \,, 0.12 \right\rbrace\, \mathsf{m}$ (a) and vapour pressure profiles for $t\, \in\, \left\lbrace 13 \,, 20 \,, 32 \right\rbrace\, \mathsf{h}$ (b).}
\end{center}
\end{figure}

The evolution of vapour pressure at the boundary surfaces ($x \egalb 0\ \mathsf{m}$ and $x \egalb 0.12\ \mathsf{m}$) is shown in Figure~\ref{fig_AN4:Evolution}. At $x \egalb 0\ \mathsf{m}$\,, the vapour pressure suddenly increases due to the step imposed at the surface. The moisture from the vapour pressure step diffuses through both layers. Although, as the second layer is composed with a less hygroscopic material, the vapour pressure completely reaches this surface by $65\ \mathsf{h}\,$. At $x \egalb 0.12\ \mathsf{m}\,$, the vapour pressure varies according to the sinusoidal fluctuations of the boundary conditions until the flow arrives. This increasing can also be observed on three profiles of vapour pressure illustrated in Figure~\ref{fig_AN4:Profiles}. Different from the previous case, the moisture flow takes more time to reach the right boundary due to the material properties of the second layer. Finally, at $t \egalb 120\ \mathsf{h}$, it is still possible to observe the influence of the step on the vapour pressure.

\begin{figure}
\begin{center}
\includegraphics[width=.45\textwidth]{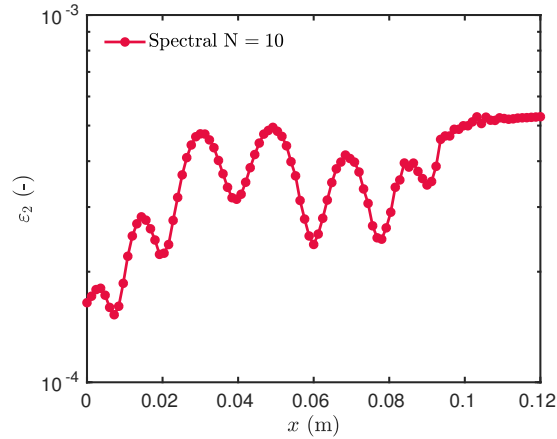}
\caption{\small\em Error $\varepsilon_{\,2}$ of the vapour pressure solution on the multilayered case.}
\label{fig_AN4:Error_fx}
\end{center}
\end{figure}

The Spectral--ROM has demonstrated a good agreement with the reference to represent the moisture diffusion trough composed walls. Distribution of the error $\varepsilon_{\,2}$ on function of $x$ is given in Figure~\ref{fig_AN4:Error_fx}. The order of the error is the same as in the previous case $\varepsilon_{\,2}\simeq \O (10^{\,-4})\,$, but the error $\varepsilon_{\,\infty}$ is higher here. This is explained since we keep the same numerical configurations of the other case but the nonlinearities increase compared to the previous configuration. Nonetheless, results provided by the Spectral reduced-order model are still acceptable.

Regarding the CPU time, this case has also presented competitive outputs. The way in which the spatial domain was split, makes the reduced system to have the double of the size, if compared with a single layer simulation. Now, the matrix $\A$ has the double of its size, making the computer run time twice as high $(7\ \mathsf{s}\,)$, as shown in Figure~\ref{fig_AN3:CPU_time}.


\section{Conclusions}

Most of the numerical methods applied to mathematical models used in building physics are commonly based on implicit schemes to compute the solution of diffusive problems. Its main advantage is due to the stability conditions for the choice of the time step $\Delta \ts\,$. However, implicit schemes require important sub-iterations when treating nonlinear problems. This work was therefore devoted to exploring the use of an innovative reduced-order approach based on the Spectral method. Spectral methods are well-known in other applications, such as meteorology and wave propagation, although it was not used before as a reduced-order model. Thus, in this work, we showed that they can be applied in some one-dimensional building physics problems to compute a reduced-order model.

The first case study considered a linear diffusive moisture transfer through a porous material. The Spectral--ROM was compared to the classical \Eu ~implicit scheme, to the \CN ~scheme and to a reference solution obtained using \texttt{Chebfun} toolbox for \texttt{Matlab}. Results have shown the dynamics and amplitude of hygrothermal fields are perfectly represented by the Spectral-ROM solution. The fidelity of the physical model is totally conserved by the Spectral-ROM. Only the order of the solution is highly reduced. Using standard approaches, the order of the solution rises with $10^{\,2}$ whereas with the Spectral method, the order of the solution scales with $6\,$. In the second case, a weak nonlinear problem was treated, which has a field dependent diffusion coefficient. To build the reduced system of ODEs, the same features of the linear case were used. Its reduced system was written with an explicit formulation and then implemented in \texttt{Matlab}. In the highly nonlinear case, the reduced system is numerically obtained as the system of ODEs cannot be explicitly expressed. The third case study focused on such general highly nonlinear transfer model, with material properties strongly dependent on the relative humidity field. To treat the nonlinearities, the \textsc{Chebyshev--Gau}\ss ~quadrature was employed to solve the integrals. Again, the accuracy of the approach has been demonstrated by representing accurately the physical phenomenon, with an absolute error of the order of $\varepsilon_{\,2} \simeq \O \,(10^{\,-4}) $ comparing to the reference solution. A parametric study on the number of modes and the tolerance of the ODE solver has also been carried out. Moreover, when comparing the CPU time of the different approaches, the \CN ~is one hundred times longer than the Spectral method to compute the solution. To bring applications closer to building physics problems, a wall with two materials is used for the last case study. By using the Spectral reduced-order model the spatial domain is decomposed and the interface conditions can be easily imposed. As the complexity of the problems rises, the Spectral method needs more modes, with still a very low computational effort compared to standard approaches, and yet it does not mean that the Spectral method loses its efficiency.    

The application of Spectral methods is not straightforward, neither intuitive as for example for the finite-difference method. Although, the efforts used in its implementation are compensated by the results, which showed to be very promising. In other domains, Spectral methods have also demonstrated their great potential for solving more complex problems \cite{Ma2017, Pasban2017, Liu2016}, which instigates the development of further work in the building physics field on the solution of combined heat and moisture transfer and through multidimensional geometries.


\bigskip

\subsection*{Acknowledgments}
\addcontentsline{toc}{subsection}{Acknowledgments}

The authors acknowledge the \textsc{Brazilian} Agencies CAPES of the Ministry of Education and CNPQ of the Ministry of Science, Technology, and Innovation, for the financial support. The authors also acknowledge the Junior Chair Research program ``\textit{Building performance assessment, evaluation and enhancement}'' from the University of \textsc{Savoie Mont Blanc} in collaboration with The French Atomic and Alternative Energy Center (CEA) and Scientific and Technical Center for Buildings (CSTB).

\bigskip


\section*{Nomenclature}
\addcontentsline{toc}{section}{Nomenclature}

\begin{tabular*}{0.7\textwidth}{@{\extracolsep{\fill}} |cll| }
\hline
\multicolumn{3}{|c|}{\emph{Latin letters}} \\
$c_{\,m}$ & moisture storage capacity & $[\mathsf{kg/m^3/Pa}]$ \\
$d_{\,m}$ & moisture diffusion & $[\mathsf{s}]$ \\
$g$ & liquid flux & $[\mathsf{kg/m^{\,2}/s}]$ \\
$h_{\,v}$ & vapour convective transfer coefficient & $[\mathsf{s/m}]$ \\
$k$ & permeability & $[\mathsf{s}]$ \\
$L$ & length & $[\mathsf{m}]$ \\
$\Pc$ & capillary pressure & $[\mathsf{Pa}]$ \\
$\Ps$ & saturation pressure & $[\mathsf{Pa}]$ \\
$\Pv$ & vapour pressure & $[\mathsf{Pa}]$ \\
$R_v$ & water gas constant & $[\mathsf{J/kg/K}]$\\
$T$ & temperature & $[\mathsf{K}]$ \\
\multicolumn{3}{|c|}{\emph{Greek letters}} \\
$\phi$ & relative humidity & $[-]$ \\
$\rho$ & specific mass & $[\mathsf{kg/m^3}]$ \\ 
\multicolumn{3}{|c|}{\emph{Abbreviations}} \\
ODE & Ordinary Differential Equation & \\
ROM & Reduced-Order Model &  \\
\hline
\end{tabular*}


\appendix
\section{Dimensionless values}
\label{annexe:dimensionless}

\subsection{Case from Section~\ref{sec:case_linear}}

Problem~\eqref{eq:moisture_dimensionlesspb_1D} is considered with $\glsL \,=\, \glsR \egal 0$ and the dimensionless properties of the material are equal to $\dms \,=\, 1$ and $\cms \,=\, 8.6\,$. The reference time is $\tref \,=\, 1$  $\mathsf{h}$, thus the final simulation time is fixed to $\tau^{\,\star} \,=\, 120\,$. The \textsc{Biot} numbers are $\BivL \,=\, 101.5$ and $\BivR \,=\, 15.2\,$. The boundary conditions are expressed as:
\begin{align*}
  & \uL \,(\,\ts\,) \egal 1 \plus 0.5 \cdot \sin \left(\, 2\,\pi \, \ts /24\,\right) \plus 0.4 \cdot \sin \left(\, 2\pi \, \ts/4\,\right) \,,\\
  & \uR \,(\,\ts\,) \egal 1 \plus 0.8 \cdot \sin \left(\, 2\,\pi \, \ts/12\,\right) \,.
\end{align*}


\subsection{Case from Section~\ref{sec:case_weak_nonlinear}}

Simplification of the problem \eqref{eq:moisture_dimensionlesspb_1D} are carried out, considering $\glsL \,=\, \glsR \,=\, 0\,$ and $\nu\,(\,u\,) \,=\, \dms\,(\,u\,)/ \cms\,(\,u\,)$. 
In this way, the problem is written as:
\begin{align*}
  \pd{u}{\ts} &\egal \nu\,(\,u\,) \;  \pd{^{\,2} u}{x^{\,\star\, 2} }  \,,
  & \ts & \ > \ 0\,, \;&  \xs & \ \in \ \big[ \, 0, \, 1 \, \big] \,, \\[3pt]
  \pd{u}{\xs} &\egal \BivL \cdot \Bigl( \, u \moins \uL \,(\, \ts\,) \, \Bigr) \,,
  & \ts & \ > \ 0\,, \,&  \xs & \egal 0 \,, \\[3pt]
  -\ \pd{u}{\xs} &\egal \BivR \cdot \Bigl( \, u \moins \uR \,(\, \ts\,) \, \Bigr) \,,
  & \ts & \ > \ 0\,, \,&   \xs & \egal 1 \,, \\[3pt]
  u &\egal 1 \,,
  & \ts & \egal 0\,, \,&  \xs & \ \in \ \big[ \, 0, \, 1 \, \big] \,.
\end{align*}
The reference time is $\tref \,=\, 1$  $\mathsf{h}\,$, thus the final simulation time is fixed to $\tau^{\,\star} \,=\, 72\,$. The \textsc{Biot} numbers are $\BivL \,=\, 15.2$ and $\BivR \,=\, 101.5\,$. The boundary conditions are expressed as:
\begin{align*}
  & \uL\, (\,\ts\,) \egal 1 \plus 0.5 \cdot \sin^{\,2} \left(\, 2\pi \, \ts/90\,\right) \,,\\
  & \uR\, (\,\ts\,) \egal 1 \plus 0.6 \cdot \sin^{\,2} \left(\, 2\pi \, \ts/48\,\right) \,.
\end{align*}
and, the dimensionless property of the material is:
\begin{align*}
  \nu \, \Bigl( u\,(\,\xs\,,\ts\,) \Bigr) \egal 1.1\cdot 10^{\,-2} \plus 5\cdot 10^{\,-2} \cdot u\,(\,\xs\,,\ts\,) \,.
\end{align*}


\subsection{Case from Section~\ref{sec:case_strong_nonlinear}}

Problem \eqref{eq:moisture_dimensionlesspb_1D} is considered with $\glsL \,=\, \glsR \,=\, 0\,$. 
In this way, the dimensionless governing equations are written as:
\begin{align*}
 \cms\,(\,u\,) \; \pd{u}{\ts} &\egal \pd{}{\xs} \left( \, \dms\,(\,u\,) \; \pd{u}{\xs} \, \right) \,,
& \ts & \ > \ 0\,, \;&  \xs & \ \in \ \big[ \, 0, \, 1 \, \big] \,, \\[3pt]
 \dms\,(\,u\,) \; \pd{u}{\xs} &\egal \BivL \cdot \Bigl( \, u \moins \uL \,(\, \ts\,) \, \Bigr) \,,
& \ts & \ > \ 0\,, \,&  \xs & \egal 0 \,, \\[3pt]
\moins \dms\,(\,u\,) \; \pd{u}{\xs} &\egal \BivR \cdot \Bigl( \, u \moins \uR \,(\, \ts\,) \, \Bigr) \,,
& \ts & \ > \ 0\,, \,&   \xs & \egal 1 \,, \\[3pt]
 u &\egal 1 \,,
& \ts & \egal 0\,, \,&  \xs & \ \in \ \big[ \, 0, \, 1 \, \big] \,.
\end{align*}
in which, the dimensionless properties of the material are:
\begin{align*}
& \dms\,(\,u\,) \egal 0.1 \plus 0.91 \, u \plus 600 \cdot \exp \Bigl[ \,-10 \, \bigl(\, u \moins 1.5 \, \bigr)^{\,2} \, \Bigr] \,,\\
& \cms\,(\,u\,) \egal 900 \moins 656 \, u \plus 10^4 \cdot \exp \Bigl[ \,-5 \, \bigl(\, u \moins 1.3 \, \bigr)^{\,2} \,  \Bigr] \,.
\end{align*}

Simulations are performed for a total time of $\tau^{\,\star} \,=\, 120\,$. The ambient water vapour pressure at the boundaries are different from the previous case study. At the left boundary, $\uL$ has a fast jump until the saturation state $\uL \,=\, 2, \; \forall t \in \, \bigr[ \, 10  \,, 40 \, \bigl] $ and at the right boundary, $\uR\,(\,\ts\,) \,=\, 1 \plus 0.8 \; \sin \left(\, \dfrac{2\pi \, \ts}{4} \,\right)\,$, with $\BivL \,=\, 101.5$ and $\BivR \,=\, 15.2 \,$. Reference values are $\dmref \egalb 1.98 \cdot 10^{\,-10}\ [\mathsf{s}]\,,$  $\tref  \egalb 3600 \ [\mathsf{s}]$ and $L  \egalb 0.1 \ [\mathsf{m}]\,$.


\subsection{Case from Section~\ref{sec:case_multilayer}}

The dimensionless properties of Material 1 are:
\begin{align*}
& d_{\,m,\,1}^{\, \star} \,(\,u\,) \egal \frac{2.723\cdot u^{\,3} \moins 4.16\cdot u^{\,2} \moins 1.383\cdot u \plus 3.515}{u^{\,2} \moins 3.618\cdot u \plus 3.412 }\,,\\
& c_{\,m,\,1}^{\, \star} \,(\,u\,) \egal \frac{\moins 5.541\cdot u^{\,4} \plus 22.05\cdot u^{\,3} \moins 26.85\cdot u^{\,2} \plus 8.032\cdot u \plus 3.837}{u^{\,5} \moins 5.101\cdot u^{\,4} \plus 9.803\cdot u^{\,3} \moins 8.409\cdot u^{\,2} \plus 2.714\cdot u \plus 0.005535}
\end{align*}
and of Material 2 are:
\begin{align*}
& d_{\,m,\,2}^{\, \star} \,(\,u\,) \egal -2.98 \cdot 10^{-5} \cdot u^{17.43} \plus 11.33 \,,\\
& c_{\,m,\,2}^{\, \star} \,(\,u\,) \egal 1.848 \cdot u^{-0.8696} \moins 0.2912 \,.
\end{align*}
with $u \egalb [0,\,2]$. The ambient water vapour pressure at the boundaries are the same from the previous case study, with $\BivL \,=\, 4.4$ and $\BivR \,=\, 0.65 \,.$ Reference values are $\dmref \egalb 5.5 \cdot 10^{\,-9}\ [\mathsf{s}]\,,$  $\tref \egalb 3600 \ [\mathsf{s}]$ and $L \egalb 0.12 \ [\mathsf{m}]\,$.


\bigskip
\addcontentsline{toc}{section}{References}
\bibliographystyle{abbrv}
\bibliography{biblio}
\bigskip\bigskip

\end{document}